\documentclass[10pt,a4paper,twocolumn]{article}
\usepackage[utf8]{inputenc}
\usepackage[english]{babel}
\pdfoutput=1
\usepackage{amsmath}
\usepackage{amsfonts}
\usepackage{lipsum}
\usepackage{amssymb}
\usepackage{graphicx,dblfloatfix}
\usepackage{geometry}
\usepackage{abstract}
\usepackage{mathtools}
\usepackage{float}
\usepackage{subfig}
\usepackage{booktabs}
\usepackage{xcolor}
\usepackage{amsbsy}
\usepackage{hyperref}
\usepackage[export]{adjustbox}
\usepackage{setspace}
\usepackage{hyphenat}
\usepackage{multirow}
\usepackage[artemisia]{textgreek}
\usepackage[keeplastbox]{flushend}
\newtheorem{definition}{Definition}[section]
\usepackage{bm}
\makeatletter
\renewcommand*{\@opargbegintheorem}[3]{\trivlist
	\item[\hskip \labelsep{\bfseries #1\ #2}] \textbf{(#3)}\ \itshape}
\makeatother
\DeclarePairedDelimiter{\ket}{\lvert}{\rangle}
\title{\textbf{Effective Hamiltonians for interacting superconducting qubits: local basis reduction and the Schrieffer-Wolff transformation}}
\author{Gioele Consani${}^\dagger$ and Paul A. Warburton${}^\dagger$\\${}^\dagger$\textit{University College London}}
\begin{document}
	\twocolumn[
	\maketitle
	\begin{onecolabstract}
	An open question in designing superconducting quantum circuits is how best to reduce the full circuit Hamiltonian which describes their dynamics to an effective two-level qubit Hamiltonian which is appropriate for manipulation of quantum information. Despite advances in numerical methods to simulate the spectral properties of multi-element superconducting circuits\cite{yurke1984quantum,reiter2012effective,amin2012approximate}, the literature lacks a consistent and effective method of determining the effective qubit Hamiltonian. Here we address this problem by introducing a novel \textit{local basis reduction method}. This method does not require any \textit{ad hoc} assumption on the structure of the Hamiltonian such as its linear response to applied fields. We numerically benchmark the local basis reduction method against other Hamiltonian reduction methods in the literature and report specific examples of superconducting qubits, including the capacitively-shunted flux qubit, where the standard reduction approaches fail. By combining the local basis reduction method with the Schrieffer-Wolff transformation we further extend its applicability to systems of interacting qubits and use it to extract both non-stoquastic two-qubit Hamiltonians and three-local interaction terms in three-qubit Hamiltonians.
	\vspace{1cm}
	\end{onecolabstract}
	]
\section{Introduction}
Since their first appearance, superconducting (SC) circuits including Josephson junctions have proved to be one of the most promising platforms for quantum information processing applications\cite{wendin2017quantum,devoret2013superconducting,clarke2008superconducting,krantz2019quantum,paraoanu2014recent}. The lithographic fabrication process allows fine tuning of the physical properties of each superconducting circuit, thus resulting in qubits with different spectral properties. Individual qubits can be manufactured in large arrays, with electrostatic and magnetic interactions coupling pairs of them. The strength of the local fields on each qubit and of the two-qubit interactions can further be adjusted dynamically by applying external electrostatic and magnetic fields, making for a flexible and scalable architecture for both gate-based quantum computation (GBQC) and quantum annealing (QA)\cite{orlando1999superconducting,devoret2013superconducting,wendin2017quantum,arute2019quantum,hauke2019perspectives}.
\par A two decades quest to improve the coherence metrics of superconducting qubits, by materials and circuit engineering, has led to a number of SC qubit designs, such  as capacitively-shunted flux qubits and transmons, having $T_1$ and $T_2$ times in the 100 \textmu s range\cite{koch2007charge,yan2016flux}.
These circuits, as much as the earlier designs, including rf-SQUID qubits\cite{harris2010experimental}, persistent-current qubits\cite{orlando1999superconducting} and single-Cooper-pair boxes\cite{nakamura1999coherent} are, by construction, characterised by the fact that, under specific operation conditions, they can be regarded as two-level systems (in the sense that any additional stationary state of the system has a substantially higher energy and a small probability of being populated)\cite{makhlin2001quantum}.
\par The fundamental theory describing SC circuits, i.e. \textit{quantum network theory}, is well established and can be used, at least in some approximate form, to numerically determine the energy spectrum of an arbitrary SC qubit circuit\cite{yurke1984quantum,reiter2012effective,amin2012approximate}. The literature seems, however, to be missing an agreed and consistent way of connecting the electromagnetic Hamiltonian $\hat{H}_{e.m.}$ of an arbitrary system of \textit{n} SC qubits to the corresponding \textit{effective qubit Hamiltonian} $\hat{H}_1$, or, equivalently, of numerically determining the parameters of an \textit{n}-spin Hamiltonian which reproduces the low-energy spectrum of $\hat{H}_{e.m.}$, as well as the computational state probabilities and the expectation values of the system observables. As we will see below, where such mapping methods do exist (see, for instance, supplementary materials of Ref. \cite{boixo2016computational,ozfidan2019demonstration}), they are not guaranteed to reproduce the correct low-energy spectrum of the circuit.
\par A general scheme for reducing the circuit Hamiltonian of an arbitrary SC qubit system to the correct effective qubit Hamiltonian would serve several purposes. Firstly, the effective qubit Hamiltonian can be used to model and interpret quantum state evolution experiments, since it contains all the necessary information to describe the dynamical evolution of the qubit system, as long as this is not excited outside of the computational space (\textit{leakage})\cite{krantz2019quantum,ozfidan2019demonstration}, while being much more compact than the full circuit Hamiltonian.
Secondly, specifically in the context of adiabatic quantum computing (AQC), identification of non-stoquastic and multi-local terms in the qubit Hamiltonian could help the engineering of such terms, which are fundamental to implement non-stoquastic AQC (which is thought to be more powerful than its stoquastic counterpart\cite{albash2018adiabatic}) and error suppression protocols based on stabiliser codes\cite{jordan2006error}, respectively. In experiments involving such non-stoquastic and multi-body interaction terms, the analysis of spectroscopic data can also be made substantially easier by the availability of the reduced Hamiltonian\cite{ozfidan2019demonstration}. Lastly, in the case of single qubits, the calculation of the effective qubit Hamiltonian represents an improved way of estimating the tunnelling amplitudes between semi-classical potential minima that is an alternative to instanton-based approaches and therefore potentially more accurate, especially in the limit of large tunnelling amplitudes\cite{friedman2002aharonov,rajaraman1989solitons}.
\par In this paper we propose a method of implementing \textit{Hamiltonian reduction} based on a natural local definition of the computational basis. Our method does not require any \textit{ad hoc} assumption on the structure of the Hamiltonian, such as its linear response to the applied electrostatic and magnetic fields, which is at the core of standard perturbative reduction methods\cite{boixo2016computational}. Additionally the scheme can be applied to individual SC qubits of any kind, as well as to systems of qubits and coupler circuits, interacting magnetically or electrostatically. In the interacting case the scheme makes use of the Schrieffer-Wolff transformation to separate the low energy subspace from the rest of the Hilbert space\cite{bravyi2011schrieffer}.
\par The article is structured as follows: in the next section we revise how to write a general electromagnetic Hamiltonian for isolated and coupled superconducting circuits. In section \ref{hamred} we introduce some of the state-of-the-art reduction methods in the literature and then present our novel approach to the problem, in the context of both single and interacting qubits. In section \ref{numres} we present the numerical results of Hamiltonian reduction applied to systems of superconducting qubits, with a specific reference to recent publications. Finally we summarise our conclusions.
\section{Circuit Hamiltonians from Quantum Circuit Analysis}
\label{sec1}
Since we want to establish a way to numerically derive an effective qubit Hamiltonian from the full Hamiltonian describing the superconducting circuit, we begin this paper by revising how to write down the circuit Hamiltonian for a generic non-dissipative circuit. We start with isolated circuits and later consider the presence of interactions. The framework which we use is that of \textit{quantum network theory}, which is the quantum version of Lagrangian mechanics applied to electrical circuits\cite{yurke1984quantum,zagoskin2011quantum}. Following the standard procedure we will first write the classical Hamiltonian and then quantise it by replacing the variables with the corresponding operators. The reader who is familiar with these concepts may wish to skip to the next section.
\subsection{Isolated circuits}
\begin{figure}[b!]
	\centering
	\includegraphics[height=.35\textwidth]{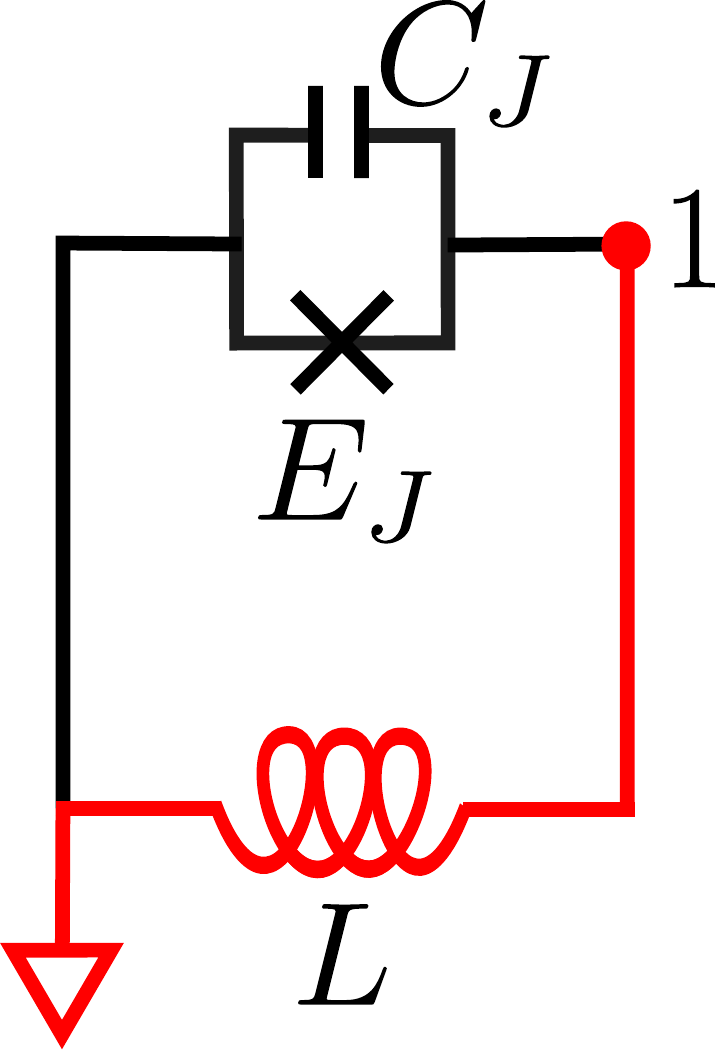}
	\caption{Equivalent lumped-element circuit of an rf-SQUID flux qubit\cite{harris2010experimental}. One of the two possible choices of the spanning tree is highlighted in red.}
	\label{fig:rfsquida}
\end{figure}
The first key assumption we make in order to apply quantum network theory is that the size of our qubit is sufficiently small relative to microwave wavelengths, such that it is appropriate to use a lumped-element description of the circuit\cite{devoret2004superconducting}. Because the system is superconductive, this circuit will consist of nodes connected by branches containing only non-dissipative elements, namely inductors, capacitors and Josephson junctions. An example representing the equivalent circuit of an rf-SQUID flux qubit is shown in figure \ref{fig:rfsquida}. Then, without loss of generality, we can arbitrarily assign one of the circuit nodes to ground. (For a floating qubit there will be a capacitor between the ground node and the rest of the circuit.)
\par At this point, in order to later take into account the effect of external magnetic fields, we need to choose a \textit{spanning tree}, i.e. a path of connected branches going from the ground node to every other node, without generating loops. The specific choice of the spanning tree will not affect our final results\cite{zagoskin2011quantum}. A possible choice of the spanning tree for the rf-SQUID flux qubit in figure \ref{fig:rfsquida} is highlighted in red. We will indicate the set of branches in the spanning tree by $\mathcal{T}$ and the complementary set of \textit{closure branches} by $\mathcal{C}$. Every closure branch is associated with an irreducible loop in the circuit, which is the smallest loop formed by that closure branch and by other branches in the spanning tree. For instance, in the flux qubit in Fig. \ref{fig:rfsquida} the closure branch $b_{01}$ is associated with the single loop in the circuit\cite{zagoskin2011quantum}.
\par Every state of our circuit is defined by specifying the instantaneous voltages at each of the nodes. Alternatively, we can define, for every node \textit{j} (excluding ground), a \textit{node flux} variable $\Phi_j$, representing the integral over time of its voltage, \textit{i.e.}
\begin{equation}
\Phi_j(t)=\int_{0}^{t}V_j(t')dt'.
\end{equation}
The ground node acts as the voltage reference, so its associated voltage and flux are set to be identically equal to 0\cite{zagoskin2011quantum}. The node fluxes can be used, together with the voltages, to write down the circuit Lagrangian $\mathcal{L}_{e.m.}(\{\Phi_i\},\{\dot{\Phi}_i\})$, which in turn allows to define the variables canonically conjugate to the node fluxes, \textit{i.e.} the \textit{node charges}\cite{zagoskin2011quantum}:
\begin{equation}
Q_j=\frac{\partial}{\partial \dot{\Phi}_j}\mathcal{L}_{e.m.}(\{\Phi_i\},\{\dot{\Phi}_i\}).
\end{equation}
For brevity we omit here the derivation of the system Lagrangian (which can be found, for instance, in \cite{zagoskin2011quantum}) and we simply report the final form we obtain for the circuit Hamiltonian,
\begin{equation}
\begin{gathered}
H_{e.m.}(\{\Phi_i\},\{Q_i\}):=\sum_{i=1}Q_i\dot{\Phi}_i|_{\dot{\Phi}_i=\dot{\Phi}_i(\{Q_i\})}+\\
-\mathcal{L}_{e.m.}(\{\Phi_i\},\{\dot{\Phi}_i\}).
\end{gathered}
\end{equation}
If we take care to define the spanning tree so as not to leave any inductive branch in the closure set $\mathcal{C}$, this takes a particularly simple and general form:
\begin{equation}
H_{e.m.}=H_{LC}+H_J,
\label{eq:circuitham}
\end{equation}
where
\begin{equation}
H_{LC}=\frac{1}{2}\sum_{i,j=1}^N\bigg[(\mathbf{C}^{-1})_{ij}Q_iQ_j+(\mathbf{L}^{-1})_{ij}\Phi_i\Phi_j\bigg]
\end{equation}
is its linear part, with  $N$ the number of circuit nodes, $\mathbf{C}$ and $\mathbf{L}$ are the $(N\times N)$ capacitance and inductance matrices of the circuit, respectively, (see appendix \ref{appendix1} for their definition) and where
\begin{equation}
H_{J}=\sum_{i=0}^N \sum_{j=i+1}^N E_{J,b_{ij}}\cdot\bigg[1-\cos\bigg(\frac{2\pi}{\Phi_0}\Phi_{b_{ij}}\bigg)\bigg]
\label{eq:josham}
\end{equation}
is the Josephson energy component. Here $E_{J,b_{ij}}$ is the \textit{Josephson energy} of the Josephson junction in the branch $b_{ij}$ connecting nodes \textit{i} and \textit{j} (the index 0 refers to the ground node here) and $\Phi_0=h/(2e)\simeq2.0678\cdot10^{-15}$Wb is the magnetic flux quantum. The \textit{branch fluxes} $\{\Phi_{b_{ij}}\}_{j>i=0,\dots,N}$  appearing inside the expression are defined as
\begin{equation}
\Phi_{b_{ij}}=
\begin{cases}
\Phi_i-\Phi_j,\hspace{4.1em}\textrm{if }b_{ij}\in\mathcal{T},\\
\Phi_i-\Phi_j+\Phi_{ij}^{ext},\hspace{1em}\textrm{if } b_{ij}\in\mathcal{C},
\end{cases}
\label{eq:branchflux}
\end{equation}
where $\Phi_{ij}^{ext}$ is the external magnetic flux threading the irreducible loop associated with $b_{ij}$.
\par Our definition of the branch fluxes includes the effect of external magnetic fields on the energy of the system. Current and voltage biases, however, may also be applied to the circuit and each of them will contribute with its own term to the Hamiltonian. In the case of current bias, this is applied through a dangling inductive branch. Let \textit{a} be the origin node of this branch, $L_a$ its inductance and $I_{ext}$ the bias current; the corresponding Hamiltonian term is\cite{zagoskin2011quantum}:
\begin{equation}
\Delta H_{e.m.}=\frac{(\Phi_a-L_aI_{ext})^2}{2L_a}.
\end{equation}
In order to apply a voltage bias, a voltage source $V_g$ is connected to the desired circuit node \textit{a} through a gate capacitor $C_g$. The resulting effect on the Hamiltonian is to change the capacitance matrix $\mathbf{C}\rightarrow\tilde{\mathbf{C}}$ (to take into account that the total capacitance attached to node \textit{a} has increased by $C_g$) and to introduce the additional term\cite{zagoskin2011quantum}:
\begin{equation}
\Delta H_{e.m.}=C_gV_g\cdot\sum_{i\neq a}(\tilde{\mathbf{C}}^{-1})_{ai}Q_i+\frac{1}{2}(\tilde{\mathbf{C}}^{-1})_{aa}(C_gV_g)^2.
\end{equation}
\par Now that we have put together all the necessary Hamiltonian terms, we can finally obtain the quantum Hamiltonian of the circuit $\hat{H}_{e.m.}$ by simply replacing the variables $\{\Phi_j,Q_j\}_{i=1,\dots,N}$ with the corresponding Hermitian operators. These will obey the canonical commutation relations\cite{yurke1984quantum}:
\begin{equation}
\left[\hat{\Phi}_j,\hat{Q}_k\right]=i\hbar\delta_{jk}.
\end{equation}
\subsection{Interacting circuits}
\label{intqsec}
Let us now consider a system of $N$ superconducting circuits of the kind just considered which are interacting with each other. The total electromagnetic Hamiltonian of the system will have the general form $\hat{H}_{e.m.}=\hat{H}_0+\hat{H}_{int}$, where:
\begin{equation}
\hat{H}_0=\sum_{i=1}^N\hat{H}_i
\label{eq:multiqham}
\end{equation}
is the unperturbed part, with $\hat{H}_i$ the Hamiltonian of the \textit{i}-th circuit, in the form of Eq. \eqref{eq:circuitham}, and
\begin{equation}
\hat{H}_{int}=\sum_{i=1}^N\sum_{j=i+1}^N\sum_{k,l}\alpha_{i_k,j_l}\hat{O}_{i_k}\hat{O}_{j_l}
\label{eq:multiqhamint}
\end{equation}
describes the interactions between pairs of different circuits. Here, $\{\hat{O}_{i_k}\}_{k=1,2,...}$ is a set of operators (either node or branch operators) acting on the \textit{i}-th circuit and the $\alpha_{i_k,j_l}$'s are the interaction constants.
\par In practice the interactions can be electrostatic, mediated by the charge operators, and magnetostatic, involving the flux operators. (In principle, there could also be additional interactions mediated by Josephson junctions shared between two circuits, but, for simplicity, we will not consider these here.) The electrostatic interaction is achieved by connecting the $k$-th node of circuit \textit{i} with the $l$-th node of circuit $j\neq i$ with a coupling capacitor $C_{i_k,j_l}$. This has two effects on the system Hamiltonian: it rescales the inverse capacitance matrices of the two circuits (known as \textit{capacitive loading}),
\begin{equation}
\mathbf{C}^{-1}_i\rightarrow\widetilde{\mathbf{C}}^{-1}_i,\quad \mathbf{C}^{-1}_j\rightarrow\widetilde{\mathbf{C}}^{-1}_j,
\label{eq:crescale}
\end{equation}
as shown explicitly in appendix \ref{appendix5}, and introduces the interaction term
\begin{equation}
\alpha_{i_k,j_l}\hat{O}_{i_k}\hat{O}_{j_l}=(\mathbf{C}_{m}^{-1})_{i_k,j_l}\hat{Q}_{i_k}\hat{Q}_{j_l},
\label{eq:cinteract}
\end{equation}
where $\mathbf{C}_{m}^{-1}$ is a suitable inverse mutual capacitance matrix (see appendix \ref{appendix5})\cite{wendin2005superconducting}.
\par The magnetostatic interactions are the result of the mutual inductive coupling between pairs of branches belonging to two different circuits, say $b_{i_k}$ and $b_{j_l}$. The effect of this mutual inductance is again twofold: it rescales the inverse inductance matrices of the circuits (\textit{inductive loading}),
\begin{equation}
\mathbf{L}^{-1}_i\rightarrow\widetilde{\mathbf{L}}^{-1}_i,\quad \mathbf{L}^{-1}_j\rightarrow\widetilde{\mathbf{L}}^{-1}_j,
\label{eq:lrescale}
\end{equation}
and introduces in the Hamiltonian the interaction term
\begin{equation}
\alpha_{i_k,j_l}\hat{O}_{i_k}\hat{O}_{j_l}=(\mathbf{M}^{-1})_{i_k,j_l}\hat{\Phi}_{b_{i_k}}\hat{\Phi}_{b_{j_l}},
\label{eq:minteract}
\end{equation}
where $\hat{\Phi}_{b_i}$ is the branch-flux operator associated with the branch $b_i$ (see appendix \ref{appendix5} for the definitions of $\widetilde{\mathbf{L}}^{-1}_{i}$, $\widetilde{\mathbf{L}}^{-1}_{j}$ and $\mathbf{M}^{-1}$)\cite{wendin2005superconducting}. Notice that the uncoupled Hamiltonians $\{\hat{H}_i\}$ in equation \eqref{eq:multiqham} are intended to be corrected for capacitive and inductive loading.
\section{Hamiltonian reduction methods}
\label{hamred}
In this section we review some of the state-of-the-art numerical Hamiltonian reduction approaches and successively introduce two novel protocols, one for single qubits (subsection \ref{sqc}) and one for multiple interacting qubits (subsection \ref{miq}). We also point out the key differences between the standard methods and our new method and demonstrate how the latter improves the range of applicability of the reduction. The standard reduction protocols described here will be used in numerical simulations (section \ref{numres}) for a comparison against the new protocols.
\subsection{Single qubits}
\label{sqc}
Let us begin by introducing a formal definition of the reduction process. In the case of one isolated qubit, this amounts to finding an effective single-spin Hamiltonian, that is:
\begin{definition}[Effective Single-Qubit Hamiltonian:]
A Hermitian operator $\hat{H}_q$ acting on a Hilbert space with dimension 2, whose spectrum matches the two lowest energy eigenstates ($E_0$ and $E_1$) of the SC qubit circuit Hamiltonian $\hat{H}_{e.m.}$.
\end{definition}
\par Assuming that the SC qubit is at thermal equilibrium with an environment at temperature \textit{T}, then, if $k_BT$ is small compared to the transition energy to the second excited state, $E_2-E_0$, the probability that this state, or any further excited state, is occupied at any given time is exponentially small. In fact, in the absence of any resonant drive term in the Hamiltonian, the higher excited states of the qubit circuit can only be occupied as a result of environment-induced relaxation. The stationary probability that the system occupies a state with energy $E_i$ at the end of this process is $p_i\propto\exp\left[(E_i-E_0)/(k_BT)\right]$\cite{breuer2002theory}. Under this hypothesis, the dynamics of the qubit are effectively restricted to the eigenspace associated with the two lowest energy eigenstates of the (potentially time-dependent) circuit Hamiltonian $\hat{H}_{e.m.}(t)$ (\textit{i.e.} the \textit{qubit subspace} $\mathcal{H}_q=\textrm{Span}\{|E_0(t)\rangle,|E_1(t)\rangle\}$) and can be described in terms of an (instantaneous) effective single qubit Hamiltonian\cite{makhlin2001quantum}.
\par Let us now consider the spectral decomposition of the circuit Hamiltonian,
\begin{equation}
\hat{H}_{e.m.}=E_0|E_0\rangle\langle E_0|+E_1|E_1\rangle\langle E_1|+\sum_{i=2}^{+\infty}E_i|E_i\rangle\langle E_i|,
\end{equation}
where we have sorted the energy eigenvalues in increasing order. By considering the definition of the qubit Hamiltonian, we see immediately that a good candidate for $\hat{H}_q$ is the restriction of $\hat{H}_{e.m.}$ to the qubit subspace, that is:
\begin{equation}
\hat{H}_q=\hat{P}_0\hat{H}_{e.m.}\hat{P}_0=E_0|E_0\rangle\langle E_0|+E_1|E_1\rangle\langle E_1|,
\label{eq:qubitham}
\end{equation}
where $\hat{P}_0=|E_0\rangle\langle E_0|+|E_1\rangle\langle E_1|$ is the projector on $\mathcal{H}_q$.
\par This expression, however, is not particularly useful to describe the evolution of the qubit in a quantum computation process. In fact, the \textit{computational basis} used to encode the information on the quantum computer does not correspond, in general, to the system energy eigenbasis. (Note that, in this basis, the Hamiltonian is diagonal, and therefore \textit{classical}\cite{cubitt2018universal}.) It is therefore necessary to define the two computational states and their relationship to the energy eigenstates\cite{makhlin2001quantum}.
\par The computational basis for a superconducting qubit is defined in terms of two eigenstates of an observable which is used in practice to measure the qubit state. This \textit{operational definition} distinguishes, therefore, between the two main categories of SC qubit design. For circuits of the \textit{flux-qubit} type (including rf-SQUID qubits\cite{friedman2000quantum}, three and four-Josephson-junction persistent current qubits\cite{orlando1999superconducting,shimazu2009four} and C-shunt flux qubits\cite{yan2016flux}), the computational states are identified with two states with opposite and well-defined values of persistent current in the qubit loop. For \textit{charge-qubit}-type designs (including single Cooper-pair box qubits\cite{nakamura1999coherent} and transmons\cite{koch2007charge}), $|0\rangle$ and $|1\rangle$ are instead identified with states with a different number of Cooper pairs on the superconducting island\cite{wendin2017quantum}.
\subsubsection{Perturbative reduction (\textit{PR}) method}
The usual approach to identifying the computational basis states for theory and simulations, which is extensively used in the literature (cf. for example \cite{boixo2016computational,yan2016flux,nakamura1999coherent,vinci2017non}), is based on a series expansion of the circuit Hamiltonian around a fixed value of one of its bias parameters (voltage or magnetic flux bias). For clarity, let us consider the specific case of the rf-SQUID qubit, whose circuit is shown in figure \ref{fig:rfsquida}. Following the method introduced in section \ref{sec1}, we can write its circuit Hamiltonian (up to an additive constant) as\cite{harris2010experimental}:
\begin{equation}
\hat{H}_{e.m.}(f_z)=\frac{\hat{Q}^2}{2C_J}+\frac{\hat{\Phi}^2}{2L}-E_J\cos\left[2\pi\left(\frac{\hat{\Phi}}{\Phi_0}+f_z\right)\right],
\end{equation}
where $f_z=\Phi_z/\Phi_0:=\Phi_{01}^{ext}/\Phi_0$ is the magnetic flux applied externally to the rf-SQUID loop, in units of $\Phi_0$. When $f_z\simeq0.5$, we can rewrite the previous equation as:
\begin{equation}
\begin{aligned}
\hat{H}_{e.m.}(f_z)\simeq&\hat{H}_0+\delta\hat{H}:=\\=&\hat{H}_{e.m.}(0.5)+\delta f_z\frac{\partial\hat{H}_{e.m.}(f_z)}{\partial f_z}\bigg\vert_{f_z=0.5},
\end{aligned}
\label{eq:seriesexp}
\end{equation}
where $\delta f_z=f_z-0.5$ and
\begin{equation}
\begin{gathered}
\frac{\partial\hat{H}_{e.m.}(f_z)}{\partial f_z}=E_J\cdot\sin\left[\frac{2\pi}{\Phi_o}\left(\hat{\Phi}-\Phi_z\right)\right]\equiv\\\equiv-\Phi_0\frac{\hat{\Phi}}{L}:=-\Phi_0\hat{I},
\end{gathered}
\end{equation}
with $\hat{I}$ the loop current operator, which will define our computational basis. Notice that we used Kirchhoff's current law to go from the first to the second line in the last equation\cite{yurke1984quantum}.
\par At this point, we can invoke \textit{stationary perturbation theory} to write the $n$-th eigenstate of $\hat{H}_{e.m.}(f_z)$, up to first order in $\delta f_z$ as\cite{shankar2012principles}:
\begin{equation}
|E_n\rangle=|E_n^{(0)}\rangle+\sum_{m\neq n}\frac{\langle E_m^{(0)}|\delta\hat{H}|E_n^{(0)}\rangle}{E_n^{(0)}-E_m^{(0)}}|E_m^{(0)}\rangle,
\label{eq:perturb}
\end{equation}
where $|E_n^{(0)}\rangle:\hat{H}_0|E_n^{(0)}\rangle=E_n^{(0)}|E_n^{(0)}\rangle$ is the $n$-th eigenstate of $\hat{H}_0$. As we can see, the various terms of the first order correction $|E_n\rangle-|E_n^{(0)}\rangle$ scale with the inverse of the differences between the unperturbed energies. Then, since for the rf-SQUID the spectrum of $\hat{H}_0=\hat{H}_{e.m.}(f_z=0.5)$ is largely \textit{anharmonic}, i.e. $E_2^{(0)}\gg E_1^{(0)}$, the two lowest-energy perturbed eigenstates are approximately linear combinations of their unperturbed counterparts only.
\par Since we are only interested in the two lowest eigenstates of the system (the qubit subspace), we can now project Eq. \eqref{eq:seriesexp} on $|E_0^{(0)}\rangle$ and $|E_1^{(0)}\rangle$ and use the fact that $\langle E_0^{(0)}|\hat{I}|E_0^{(0)}\rangle=\langle E_1^{(0)}|\hat{I}|E_1^{(0)}\rangle=0$ (due to the symmetry of the Hamiltonian under magnetic field inversion about the point $f=f_z=0.5$) to get:
\begin{equation}
\begin{gathered}
\mathbf{H}_q(f_z)\simeq \frac{E_0^{(0)}+E_1^{(0)}}{2}\pmb{\sigma}_I+\delta f_z\Phi_0I_p\pmb{\sigma}_x+\\-\frac{E_0^{(0)}-E_1^{(0)}}{2}\pmb{\sigma}_z,
\end{gathered}
\end{equation}
where $\pmb{\sigma}_x$ and $\pmb{\sigma}_z$ are two of the standard Pauli matrices, $\pmb{\sigma}_I$ is the $2\times2$ identity matrix and $I_p:=\langle E_0^{(0)}|\hat{I}|E_1^{(0)}\rangle=\langle E_1^{(0)}|\hat{I}|E_0^{(0)}\rangle>0$ (notice that we can always ensure these two conditions by multiplying $|E_0^{(0)}\rangle$ and $|E_1^{(0)}\rangle$ by appropriate phase factors). At this point we can diagonalise the current operator part of the Hamiltonian simply by introducing the two following computational states:
\begin{definition}[Computational basis states (perturbative)]
\begin{equation}
\begin{aligned}
|0\rangle=&\frac{|E_0^{(0)}\rangle+|E_1^{(0)}\rangle}{\sqrt{2}},\\
|1\rangle=&\frac{|E_0^{(0)}\rangle-|E_1^{(0)}\rangle}{\sqrt{2}}
\end{aligned}
\label{eq:computs}
\end{equation}
\end{definition}
Using the results above, it is trivial to show that these are actually eigenstates of $\hat{I}$ with opposite eigenvalues: $\langle0|\hat{I}|0\rangle=-\langle 1|\hat{I}|1\rangle=I_p$\cite{vinci2017non}. In this basis, the effective Hamiltonian reads
\begin{equation}
\begin{gathered}
\mathbf{H}_q(f_z)\simeq \frac{E_0^{(0)}+E_1^{(0)}}{2}\pmb{\sigma}_I-\frac{E_0^{(0)}-E_1^{(0)}}{2}\pmb{\sigma}_x+\\+\delta f_z\Phi_0I_p\pmb{\sigma}_z.
\end{gathered}
\label{eq:qham}
\end{equation}
Notice that every single-qubit Hamiltonian can be written in the general form
\begin{equation}
\mathbf{H}_q=\sum_{i=I,x,y,z}h_i\pmb{\sigma}_i,
\label{eq:qubithamgen}
\end{equation}
where $\pmb{\sigma}_I\equiv\pmb{\mathbb{I}}_2$, $\pmb{\sigma}_{i=x,y,z}$ are the three standard Pauli matrices and the $h_i$'s are real coefficients. In the following we will call these \textit{Pauli coefficients}, with specific reference to their values in the computational basis.
\par Although equation \eqref{eq:qham} already contains the analytic expressions of the Pauli coefficients (which apply to the rf-SQUID qubit), it is useful to consider the following equivalent derivation, which has a straightforward extension to the interacting qubit case. Once we have found the computational states according to \eqref{eq:computs}, we can use the homomorphism between $\mathbb{C}^2$ and qubit subspace ($\mathcal{H}_q=\textrm{Span}\{|0\rangle,|1\rangle\}$) to introduce the following four operators, which represent the action of the Pauli matrices on $\mathcal{H}_q$:
\begin{equation}
\begin{gathered}
\hat{\sigma}_I=|0\rangle\langle 0|+|1\rangle\langle 1|,\qquad \hat{\sigma}_x=|0\rangle\langle 1|+|1\rangle\langle 0|,\\
\hat{\sigma}_y=-i|0\rangle\langle 1|+i|1\rangle\langle 0|,\qquad \hat{\sigma}_z=|0\rangle\langle 0|-|1\rangle\langle 1|.
\end{gathered}
\end{equation}
Then, using the following property of the Pauli matrices,
\begin{equation}
\textrm{Tr}\left(\pmb{\sigma}_i\cdot\pmb{\sigma}_j\right)=2\delta_{ij},
\label{eq:trace}
\end{equation}
we find that:
\begin{equation}
\begin{gathered}
h_i=\frac{1}{2}\textrm{Tr}\left(\mathbf{H}_{q}\cdot\pmb{\sigma}_i\right)=\frac{1}{2}\textrm{Tr}\left(\hat{H}_{q}\cdot\hat{\sigma}_i\right)\equiv\\\equiv\frac{1}{2}\textrm{Tr}\left(\hat{H}_{e.m.}\cdot\hat{\sigma}_i\right).
\end{gathered}
\label{eq:paulicoeffs}
\end{equation}
Notice that here, both $\hat{H}_{e.m.}$ and $\hat{\sigma}_{I,x,y,z}$ are conveniently expressed in whatever basis we initially choose for $\hat{H}_{e.m.}$.
\par The perturbative reduction approach has a clear disadvantage: the effective Hamiltonian \eqref{eq:qham} reproduces the two lowest energy levels of the full circuit Hamiltonian only in the limit in which the first order perturbative expansion \eqref{eq:perturb} holds. This entails two requirements. Firstly that the spectrum of the unperturbed Hamiltonian (in other words, the circuit Hamiltonian at the point of the expansion) is highly anharmonic, which is only true for some SC qubit designs and not for others (such as the capacitively-shunted flux qubit and the transmon)\cite{yan2016flux,koch2007charge}. Secondly, the perturbation to the bias parameter must be small, for instance $|\delta f_z|\ll1$ for the rf-SQUID qubit\cite{boixo2016computational}.
\subsubsection{Instanton approach}
A second common approach to the numerical calculation of the Pauli coefficients is the use of \textit{semi-classical theory}. In this case the quantum state of the system is approximated by one that minimises its semi-classical potential, which is the part of the classical Hamiltonian depending on the coordinate variable (\textit{i.e.} the flux in a flux qubit and the charge in a charge qubit). At the operational point the semi-classical potential of qubit circuits assumes a general double-well shape (or, more generally, that of a system of wells in more than one dimension), with two local minima very close in energy, such that quantum tunnelling can occur between them.
\par In this picture, the \textit{longitudinal} Pauli coefficient $h_z$ is identified with the difference in energy between the two potential minima, whereas the effective \textit{transverse} field $h_x$ corresponds to the tunnelling energy. This is calculated using the semi-classical \textit{instanton method} (or equivalently the \textit{WKB approximation})\cite{garg2000tunnel}. These calculations are only accurate in the limit in which the tunnelling action across the potential barrier is very large, which implies that the tunnelling energy has to be exponentially small\cite{landau2013quantum}. The instanton calculation of the transverse field for the rf-SQUID qubit is described in detail in appendix \ref{appendix4}.
\subsubsection{Local basis reduction (\textit{LR}) method}
\label{localb}
In order to overcome the difficulties of the standard reduction approaches outlined above, we propose an alternative reduction method which relies on a local definition of the computational basis, \textit{i.e.} one that explicitly depends on all of the circuit bias parameters. In other words, in this case the computational basis states are built as a linear combination of the two local circuit low-energy states:
\begin{equation}
\begin{aligned}
|0\rangle=u_{00}|E_0\rangle+u_{01}|E_1\rangle,\\
|1\rangle=u_{10}|E_0\rangle+u_{11}|E_1\rangle,
\label{eq:lincomb}
\end{aligned}
\end{equation}
where $\hat{H}_{e.m.}|E_i\rangle=E_i|E_i\rangle$ and $\hat{H}_{e.m.}$ is the local circuit Hamiltonian. In order for these two states to be appropriately orthonormal, the $u_{ij}$'s have to be the elements of a unitary matrix,
\begin{equation}
\mathbf{U}=\begin{pmatrix}
u_{00} & u_{10}\\
u_{01} & u_{11}
\end{pmatrix},
\label{eq:u}
\end{equation}
which we will have to find. The unitarity condition ensures that when we transform from the energy eigenbasis $\{|E_0\rangle,|E_1\rangle\}$ to the local computational basis $\{|0\rangle,|1\rangle\}$ the spectrum of the effective qubit Hamiltonian \eqref{eq:qubitham} is unchanged and the two lowest-energy levels of the circuit Hamiltonian are preserved.
\par Owing to the orthonormality of $\mathbf{U}$ columns, we can always rewrite $\mathbf{U}$, up to an irrelevant global phase multiplication factor, as:
\begin{equation}
\mathbf{U}=\begin{pmatrix}
e^{i\varphi_1}\cos\theta & e^{i\varphi_2}\sin\theta\\
-e^{-i\varphi_2}\sin\theta & e^{-i\varphi_1}\cos\theta
\end{pmatrix},
\end{equation}
where
\begin{gather}
\theta=\textrm{acos}\vert u_{00}\vert=\textrm{acos}\vert u_{11}\vert,\label{eq:theta}\\
\varphi_1=\frac{1}{2}\textrm{acos}\left(\frac{u_{00}u_{11}^*+u_{00}^*u_{11}}{2\cos^2\theta}\right),\\
\varphi_2=\frac{1}{2}\left[\pi-\textrm{acos}\left(\frac{u_{01}u_{10}^*+u_{01}^*u_{10}}{2\sin^2\theta}\right)\right].
\end{gather}
so that $\theta\in[0,\pi]$ and $\varphi_1,\varphi_2\in[0,\pi/2]$.
\par Now we consider again the operational definition of the computational states. This specifies that these should be eigenstates of a certain observable $\hat{O}$. For a flux qubit $\hat{O}=\hat{I}$, the current operator associated with the qubit SC loop, whereas for a charge qubit $\hat{O}=\hat{Q}$ represents the charge on the qubit SC island. One can easily see that imposing this condition on the states \eqref{eq:lincomb} is equivalent to finding the two eigenstates of the operator
\begin{equation}
\begin{aligned}
\hat{O}_p=&\hat{P}_0\hat{O}\hat{P}_0=\\
=&\langle E_0|\hat{O}|E_0\rangle|E_0\rangle\langle E_0|+\langle E_0|\hat{O}|E_1\rangle|E_0\rangle\langle E_1|+\\
&+\langle E_1|\hat{O}|E_0\rangle|E_1\rangle\langle E_0|+\langle E_1|\hat{O}|E_1\rangle|E_1\rangle\langle E_1|,
\end{aligned}
\label{eq:Op}
\end{equation}
associated with a non-zero eigenvalue\footnote{One can easily show that $\hat{O}_p$ achieves its maximum rank of two as long as $\hat{O}|E_0\rangle$ and $\hat{O}|E_1\rangle$ are linearly independent.}, that is
\begin{definition}[Computational basis states (local)]
$|0\rangle$ and $|1\rangle$ such that
\begin{equation}
\begin{gathered}
\hat{O}_p|0\rangle=u_0|0\rangle,\\
\hat{O}_p|1\rangle=u_1|1\rangle,
\end{gathered}
\end{equation}
with $u_0\neq u_1$ and $|u_0|,|u_1|>0$.
\label{localdef}
\end{definition}
Notice that this definition coincides with the one used in the perturbative method at the specific bias point at which the Hamiltonian expansion is performed (for instance at $f_z=0.5$ in the the rf-SQUID qubit case).
\par By identifying $|E_0\rangle$ with the vector $(1,0)$ and $|E_1\rangle$ with $(0,1)$, we can rewrite $\hat{O}_p$ in the matrix form
\begin{equation}
\mathbf{O}_p=\begin{pmatrix}
\langle E_0|\hat{O}|E_0\rangle & \langle E_0|\hat{O}|E_1\rangle\\
\langle E_1|\hat{O}|E_0\rangle & \langle E_1|\hat{O}|E_1\rangle
\end{pmatrix}.
\end{equation}
Finding the eigenvalues and the eigenvectors of this $2\times2$ matrix is straightforward. In particular, for the eigenvalues, we have:
\begin{equation}
u_{0,1}=\frac{t\pm\sqrt{t^2-4d}}{2},
\end{equation}
where $t=\textrm{Tr}(\mathbf{O}_p)$ and $d=\det(\mathbf{O}_p)$. In accordance with the operational definitions given above, we need to enforce one condition on these eigenvalues. For a flux-qubit type circuit, we need to have $u_1<0<u_0$, which implies $\det(\mathbf{I}_p)<0$, or, more explicitly
\begin{equation}
\langle E_0|\hat{I}|E_0\rangle\langle E_1|\hat{I}|E_1\rangle<|\langle E_0|\hat{I}|E_1\rangle|^2.
\end{equation}
For a qubit of the charge type, instead, we will require $u_1=u_0\pm2e$ (up to some suitably small numerical error). If this condition is not satisfied, then the circuit cannot be operated as a qubit with the desired computational states and the reduction protocol fails. Note, however, that since we are not making use of a perturbative expansion or a semi-classical approximation here, the range of applicability of this local reduction method should be wider than that of the standard methods presented before.
\par If we now write the eigenvectors of $\mathbf{O}_p$ as $\vec{u}_0=(u_{00},u_{01})$ and $\vec{u}_1=(u_{10},u_{11})$, then equation \eqref{eq:lincomb} returns our desired computational basis states, which make $\hat{O}_p$ diagonal. Armed with $\vec{u}_0$ and $\vec{u}_1$, we can easily calculate the general expression of the effective qubit Hamiltonian in the computational basis. If we keep working with $2\times2$ matrices, the qubit Hamiltonian in the energy eigenbasis \eqref{eq:qubitham} takes the obvious diagonal form
\begin{equation}
\mathbf{H}'_q=\begin{pmatrix}
E_0&0\\
0&E_1
\end{pmatrix}.
\end{equation}
Going from this basis to the computational basis amounts to applying the unitary transformation $\mathbf{U}$ defined above; this gives the effective qubit Hamiltonian in the computational basis as:
\begin{equation}
\begin{gathered}
\mathbf{H}_q=\mathbf{U}^\dagger\mathbf{H}_q'\mathbf{U}=\frac{E_0+E_1}{2}\pmb{\mathbb{I}}_2+\\ +\frac{E_0-E_1}{2}\begin{pmatrix}
\vert u_{00}\vert^2-\vert u_{01}\vert^2 & u_{00}^*u_{10}-u_{01}^*u_{11}\\
u_{10}^*u_{00}-u_{11}^*u_{01} & \vert u_{10}\vert^2-\vert u_{11}\vert^2
\end{pmatrix}=\\
\frac{E_0+E_1}{2}\pmb{\mathbb{I}}_2+\frac{E_0-E_1}{2}\begin{pmatrix}
\cos 2\theta & e^{-i\varphi}\sin 2\theta\\
e^{i\varphi}\sin 2\theta & -\cos 2\theta\end{pmatrix}=\\
\frac{E_0+E_1}{2}\pmb{\mathbb{I}}_2+\frac{E_0-E_1}{2}\big[\sin 2\theta \cdot(\cos\varphi\pmb{\sigma}_x+\sin\varphi\pmb{\sigma}_y)+\\+\cos 2\theta\pmb{\sigma}_z\big],
\end{gathered}
\label{eq:qubitham2}
\end{equation}
where $\pmb{\mathbb{I}}_2$ is the $2\times 2$ identity matrix and $\varphi=\varphi_1-\varphi_2\in[-\pi/2,\pi/2]$.
\par We observe that, by rescaling the computational states $\vec{u}_0$ and $\vec{u}_1$ by two phase factors, say $e^{i\phi_0}$ and $e^{i\phi_1}$, \textit{i.e.} by applying some \textit{local gauge transformation} in the qubit subspace, we can always remove the imaginary component $h_y\pmb{\sigma}_y$ of $\mathbf{H}_q$. In fact such a gauge transformation $\mathbf{G}(\phi_0,\phi_1)$ corresponds to a spin rotation around the \textit{z} axis, multiplied by a global phase:
\begin{equation}
\begin{gathered}
\mathbf{G}(\phi_0,\phi_1)=\begin{pmatrix}
e^{i\phi_0}&0\\
0& e^{i\phi_1}
\end{pmatrix}=\\
=e^{i\frac{\phi_0+\phi_1}{2}}\begin{pmatrix}
e^{i\frac{\phi_0-\phi_1}{2}}&0\\
0& e^{-i\frac{\phi_0-\phi_1}{2}}
\end{pmatrix}=e^{i\frac{\phi_0+\phi_1}{2}}\cdot e^{i\frac{\phi_0-\phi_1}{2}\pmb{\sigma}_z}.
\end{gathered}
\end{equation}
Hence $\mathbf{G}(-\varphi_1,-\varphi_2)$, which represents a rotation around \textit{z} by the angle $\varphi_2-\varphi_1=-\varphi$ (followed by a rescaling by $e^{-i(\varphi_1+\varphi_2)/2}$), transforms $\cos\varphi\pmb{\sigma}_x+\sin\varphi\pmb{\sigma}_y$ into $\pmb{\sigma}_x$, and makes the effective qubit Hamiltonian real, that is:
\begin{equation}
\mathbf{H}_q=\frac{E_0+E_1}{2}\pmb{\mathbb{I}}_2-\frac{\Delta}{2}\pmb{\sigma}_x-\frac{\varepsilon}{2}\pmb{\sigma}_z,
\label{eq:qubitham3}
\end{equation}
where $\Delta=(E_1-E_0)\sin 2\theta$ and $\varepsilon=(E_1-E_0)\cos 2\theta$. Notice that this gauge transformation can equivalently be written as:
\begin{equation}
\begin{aligned}
\vec{u}_0&\rightarrow\frac{\vert u_{00}\vert}{u_{00}}\cdot\vec{u}_0,\\
\vec{u}_1&\rightarrow\frac{\vert u_{10}\vert}{u_{10}}\cdot\vec{u}_1.
\end{aligned}
\label{eq:rescale}
\end{equation}
Expression \eqref{eq:qubitham3} for the effective qubit Hamiltonian, is the one adopted by most of the literature on SC qubits\cite{devoret2004superconducting,boixo2016computational,bouchiat1998quantum}. (Note that the coefficient $\Delta$ is usually further assumed to be positive, a condition which can also always be achieved with a $\pi$ rotation about \textit{z}.)
\par An equivalent and more convenient way of calculating the four \textit{Pauli coefficients} $h_i$, $i=I,x,y,z$ than using equations \eqref{eq:theta}, \eqref{eq:qubitham2} and \eqref{eq:qubithamgen} together is again to use the computational states to build the Pauli operators and then to apply equation \eqref{eq:paulicoeffs}.\\
\linebreak
In section \ref{numres} we will present numerical simulations which benchmark the performance of the local reduction method against the standard methods and demonstrate the increased accuracy of the former relative to the latter ones.
\subsection{Multiple qubits}
\label{miq}
Let us now consider the Hamiltonian reduction process in the case of multiple interacting superconducting qubits. Given a system of $N$ qubits and $M$ additional coupling circuits, coupled inductively and/or capacitively, its effective qubit Hamiltonian is one that reproduces the lowest $2^N$ energy levels of the total system Hamiltonian, as well as the expectation values of the qubit operators. Notice that any such Hamiltonian can be written in the general form
\begin{equation}
\mathbf{H}_q=\sum_{\vec{\eta}}h_{\vec{\eta}}\pmb{\sigma}_{\vec{\eta}},
\label{eq:mqham}
\end{equation}
where $\vec{\eta}=(\eta_1,\dots,\eta_N)$, $\eta_i\in\{I,x,y,z\}$ and $\pmb{\sigma}_{\vec{\eta}}=\pmb{\sigma}_{\eta_1}\otimes\cdots\otimes\pmb{\sigma}_{\eta_N}$ is a $2^N\times2^N$ matrix in the Pauli group $\textrm{G}_N$. Recalling the equality \eqref{eq:trace} and using the following property of the trace
\begin{equation}
\textrm{Tr}(\mathbf{A}_1\otimes\cdots\otimes\mathbf{A}_N)=\textrm{Tr}(\mathbf{A}_1)\times\cdots\times\textrm{Tr}(\mathbf{A}_N),
\end{equation}
we can see that the real Pauli coefficients $h_{\vec{\eta}}$ obey the equation 
\begin{equation}
h_{\vec{\eta}}=\frac{1}{2^N}\textrm{Tr}\left(\mathbf{H}_q\cdot\pmb{\sigma}_{\vec{\eta}}\right).
\label{eq:paulis}
\end{equation}
\par According to section \ref{intqsec}, the circuit Hamiltonian of the system can be written as
\begin{equation}
\begin{aligned}
\hat{H}_{e.m.}=&\hat{H}_0+\hat{H}_{int}=\\
=&\sum_{i=1}^N\hat{H}_i+\sum_{i=1}^M\hat{H}_{c,i}+\hat{H}_{int},
\end{aligned}
\label{eq:csysham}
\end{equation}
with $\hat{H}_i$ ($\hat{H}_{c,i}$) the unperturbed Hamiltonian of the \textit{i}-th qubit (coupler) circuit and where $\hat{H}_{int}$ includes all the interaction terms. Notice that the unperturbed Hamiltonians are assumed to be corrected for capacitive and inductive loading (cf. section \ref{intqsec}).
Now we can define the qubit subspace, in analogy with the single-qubit case, to be the one spanned by the lowest two eigenstates the unperturbed Hamiltonian of each qubit. Since the couplers are designed to be classical elements which always remain in their ground state, while adiabatically following the qubits, the qubit subspace will at the same time be the one spanned by the ground state of each coupler circuit Hamiltonian\cite{hutter2006tunable}. We therefore have, in symbolic form:
\begin{equation}
\mathcal{H}_q=\bigotimes_{i=1}^{N}\textrm{Span}\{|E_{i,0}\rangle,|E_{i,1}\rangle\}\otimes\bigotimes_{j=1}^{M}\textrm{Span}\{|E_{cj,0}\rangle\},
\label{eq:multispace}
\end{equation}
where $|E_{i,j}\rangle$ ($|E_{ci,j}\rangle$) is the \textit{j}-th eigenstate of $\hat{H}_i$ ($\hat{H}_{c,i}$).
\par One could then think of defining the qubit Hamiltonian for this $N$-qubit system simply as in Eq. \eqref{eq:qubitham}: $\hat{H}_q=\hat{P}_0\cdot\hat{H}_{e.m.}\cdot\hat{P}_0$, where again $\hat{P}_0$ is the projector on $\mathcal{H}_q$. This operator $\hat{H}_q$, however, does not have the correct spectrum, matching the lowest $2^N$ energy levels of $\hat{H}_{e.m.}$, and therefore does not satisfy our initial definition of qubit Hamiltonian. The reason for this is that the interaction described by $\hat{H}_{int}$ mixes the states in $\mathcal{H}_q$ with those outside it, i.e. the higher excited states of the individual circuits. Such mixed states become the new low-energy eigenstates of $\hat{H}_{e.m.}$\cite{bravyi2011schrieffer,ozfidan2019demonstration}.
\par Contrary to the single-qubit case, the literature concerning Hamiltonian reduction for multiple interacting SC qubits is relatively scarce. In the following subsections we present two protocols adopted in recent publications and later present a new alternative reduction method, which overcomes some of their limitations and explicitly addresses the problem of the mixing of the qubit subspace with the rest of the Hilbert space by using the \textit{Schrieffer-Wolff transformation} theory\cite{bravyi2011schrieffer}.
\subsubsection{Approximate rotation method}
\label{approt}
\par In this subsection we briefly review the reduction method outlined in a recent work by Ozfidan et al. \cite{ozfidan2019demonstration}. This method starts by writing the low-energy part of the total circuit Hamiltonian $\hat{H}_{e.m.}$, \textit{i.e.} the component associated with its lowest $2^N$ eigenvalues, in its diagonal form: $\mathbf{H}'_q=\textrm{diag}(E_0,\dots,E_{2^N-1})$. Then a sequence of two rotations, say $\mathbf{R}_1,\mathbf{R}_2\in \textrm{SO}(2^N)$, is applied to it, producing the effective qubit Hamiltonian $\mathbf{H}_q=\mathbf{R}_2^T\mathbf{R}_1^T\mathbf{H}_q'\mathbf{R}_1\mathbf{R}_2$. Since orthogonal operations do not change the spectrum of an operator, this protocol guarantees by construction that the spectrum of $\mathbf{H}_q$ matches the low-energy spectrum of the circuit Hamiltonian.
\par The first rotation applied in this protocol, $\mathbf{R}_1$, maps from the low-energy eigenbasis of the total Hamiltonian $\hat{H}_{e.m.}$, $\{|E_0\rangle,\dots,|E_{2^N-1}\rangle\}$ to that of the unperturbed Hamiltonian $\hat{H}_0$, i.e. $\{|E_0^{(0)}\rangle,\dots,|E_{2^N-1}^{(0)}\rangle\}$, and is initially calculated as
\begin{equation}
(\mathbf{R}_1)_{ij}=\langle E_i|E_j^{(0)}\rangle.
\end{equation}
However, as we pointed out before, $|E_i\rangle$ also has components outside of the subspace $\textrm{Span}\{E_0^{(0)}\rangle,\dots,|E_{2^N-1}^{(0)}\rangle\}$, which implies that this matrix is not orthogonal.
$\mathbf{R}_1$ must therefore be explicitly orthonormalised, for instance using the Gram-Schmidt procedure. This step is only justified if the columns of $\mathbf{R}_1$ are already approximately orthonormal\cite{ozfidan2019demonstration}. Since in our case orthogonality follows from normalisation, it suffices to check that
\begin{equation}
\sum_{i=0}^{2^N-1}(\mathbf{R}_1)_{ij}^2\approx 1,\;\forall j=0,\dots,2^N-1,
\label{eq:approx}
\end{equation}
before we apply the Gram-Schmidt procedure.
\par To obtain the qubit Hamiltonian we now need the second rotation $\mathbf{R}_2$ to map from the basis of the energy eigenstates $|E_0^{(0)}\rangle,\dots$ to the computational basis. We then take
\begin{equation}
(\mathbf{R}_2)_{ij}=\langle i|E^{(0)}_j\rangle,
\end{equation}
where $|i\rangle=|i_{2^N-1}\rangle\otimes\cdots\otimes|i_0\rangle$ is an outer product of single qubit computational states, with $i_{2^N-1}i_{2^N-2}\cdots i_1i_0$ the $N$-digit binary representation of the integer $i\in\{0,1,\dots,2^N-1\}$. These computational states are found from the reduction of the unperturbed single-qubit Hamiltonians. If the local reduction method is used for this, the rotation matrix $\mathbf{R}_2$ is guaranteed to be orthogonal.
\par Note that, although the effective Hamiltonian calculated with this method has the correct spectrum, the procedure is based on the approximate equality \eqref{eq:approx}, which is not often satisfied, particularly in the case of relatively large interactions. (This can be seen by considering, once again, the perturbative expansion \eqref{eq:perturb}.) Additionally, the previous derivation implicitly assumes that the circuit Hamiltonian is real, so that all the eigenstates and computational states can be chosen to have only real components. This ensures that $\mathbf{R}_1,\mathbf{R}_2\in\textrm{SO}(2^N)$. Some circuits, however, may have an efficient matrix representation of the Hamiltonian which is complex. In this case the definition of the two rotations would lead to the presence of arbitrary complex phases in their elements, which would need to be somehow taken care of. (Notice that even in the real case the scalar products defining the elements of $\mathbf{R}_1$ and $\mathbf{R}_2$ are only defined up to an arbitrary sign.)
\subsubsection{Diagonal Hamiltonian method}
\label{diag}
A second method of determining the effective Hamiltonian of a multi-qubit system is presented in a recent work by Melanson et al.\cite{melanson2019tunable}. This method works under the more restrictive assumption that the effective Hamiltonian is diagonal in the computational basis. In this case the lowest $2^N$ eigenstates of the circuit are also eigenstates of the single-qubit operators $\hat{O}_i$ specifying the computational basis and the corresponding eigenvalues can be calculated numerically as the expectation values $\langle E_n|\hat{O}_i|E_n\rangle$. Additionally the $2^N$ non-zero Pauli coefficients of the system can be expressed as a linear combination of its low-energy eigenvalues\cite{melanson2019tunable}. For instance, in the two-qubit case one has:
\begin{equation}
\begin{pmatrix}
E_{00}\\E_{01}\\E_{10}\\E_{11}
\end{pmatrix}=
\begin{pmatrix}
1&1&1&1\\
1&1&-1&-1\\
1&-1&1&-1\\
1&-1&-1&1
\end{pmatrix}\cdot\begin{pmatrix}
h_{II}\\h_{zI}\\h_{Iz}\\h_{zz}\end{pmatrix}:=\mathbf{M}\cdot\begin{pmatrix}
h_{II}\\h_{zI}\\h_{Iz}\\h_{zz}
\end{pmatrix},
\end{equation}
where $E_{ij},$ $i,j\in\{0,1\}$ is the eigenvalue of the circuit Hamiltonian corresponding to the computational state $|i\rangle|j\rangle$. We can therefore determine the Pauli coefficients of the two-qubit system by finding the lowest four energy eigenvalues of its circuit Hamiltonian, calculating the expectation value of the operators $\hat{O}_{1,2}$ on the each eigenstate to identify its corresponding computational state and by inverting the previous equation to get
\begin{equation}
\begin{pmatrix}
h_{II}\\h_{zI}\\h_{Iz}\\h_{zz}\end{pmatrix}=\mathbf{M}^{-1}\cdot\begin{pmatrix}
E_{00}\\E_{01}\\E_{10}\\E_{11}
\end{pmatrix}.
\end{equation}
The same procedure can be applied to systems with three or more qubits (plus eventual additional couplers).
\par In practice, an effective Hamiltonian diagonal in the computational basis is verified when the qubit tunnelling barriers are high (negligible transverse field $h_x$) and the qubits are coupled only through their \textit{z} degree of freedom (that is when the coupling is inductive between flux qubits or capacitive between charge qubits). A Hamiltonian of this form is however classical and cannot be sufficient for universal quantum computation\cite{cubitt2018universal}. This method can nevertheless still be useful when it is reasonable to assume that the different non-commuting terms of the qubit Hamiltonian can be turned on and off independently.
\subsubsection{Schrieffer-Wolff transformation method}
\label{SWT}
\par In this final subsection we introduce a new reduction protocol for multi-qubit systems which overcomes some of the limitations of the methods described above. In particular, this method does not require the mixing between the qubit subspace (Eq. \eqref{eq:multispace}) and its complement, resulting from the interactions, to be negligible, which is a crucial assumption of the approximate rotation reduction. Secondly, unlike the approximate rotation reduction, it can be applied directly to circuit Hamiltonians with complex elements, since the arbitrary phase choices made when numerically evaluating the Hamiltonian eigenvectors cancel out in all the necessary expressions. Thirdly, the reduction method introduced here can be applied to find arbitrary non-diagonal effective Hamiltonians. This is all made possible by the \textit{Schrieffer-Wolff transformation} (SWT), which by construction maps the total circuit Hamiltonian $\hat{H}_{e.m.}$ to a new Hermitian operator acting on the qubit subspace $\mathcal{H}_q$ and whose spectrum matches the low-energy spectrum of $\hat{H}_{e.m.}$, which is precisely what we expect from the effective qubit Hamiltonian\cite{bravyi2011schrieffer}.
\par The SWT relies on a single assumption regarding the form of the full system Hamiltonian, namely that the spectrum of the unperturbed part of the Hamiltonian (excluding the interactions) has a sufficiently large gap, as we will see below. For the purpose of this reduction method, we will replace this assumption with an equivalent pair of two distinct conditions. In order to state the first one, let us rewrite the unperturbed part of the \textit{N}-qubit \textit{M}-coupler system Hamiltonian \eqref{eq:csysham} as
\begin{equation}
\hat{H}_0=\hat{P}_0\hat{H}_0\hat{P}_0+\hat{Q}_0\hat{H}_0\hat{Q}_0,
\end{equation} where
\begin{equation}
\hat{P}_0=\sum_{i=0}^{2^N-1}|E_i^{(0)}\rangle\langle E_i^{(0)}|
\end{equation}
is the projector on the low-energy eigenspace $\mathcal{H}_{low}^{(0)}$, spanned by the eigenstates corresponding to the lowest $2^N$ eigenvalues of $\hat{H}_0$, and $\hat{Q}_0=\hat{\mathbb{I}}-\hat{P}_0$ projects on the complementary subspace $\mathcal{H}\setminus\hat{H}_{low}^{(0)}$. Notice that, given $E_{i}^{(0)}$, the \textit{i}-th eigenvalue of $\hat{H}_0$, the spectrum of $\hat{P}_0\hat{H}_0\hat{P}_0$ is by definition $\mathcal{S}_{low}^{(0)}=\{E_{0}^{(0)},E_{1}^{(0)},\dots,E_{2^N-1}^{(0)}\}$, whereas $\hat{Q}_0\hat{H}_0\hat{Q}_0$ has the set of eigenvalues $\mathcal{S}_{high}^{(0)}=\{E_{2^N}^{(0)},\dots\}$. The first assumption of our reduction is that $\mathcal{H}_q\equiv\hat{H}_{low}^{(0)}$, \textit{i.e.} that no additional excited state of the independent circuits is mixed in the low energy subspace of $\hat{H}_0$, and that the two sets $\mathcal{S}_{low}^{(0)}$ and $\mathcal{S}_{high}^{(0)}$ are separated by at least $\Delta>0$, that is $\vert E_{2^N}^{(0)}-E_{2^N-1}^{(0)}\vert\geq\Delta$. This composite condition can be written, more explicitly, in the following form:
\begin{equation}
\begin{gathered}
\bigg\vert \Delta E_{i,2}-\sum_{j=1}^N\Delta E_{j,1}\bigg\vert\geq\Delta,\; \forall i=1,\dots,N\\
\bigg\vert \Delta E_{ci,1}-\sum_{j=1}^N\Delta E_{j,1}\bigg\vert\geq\Delta,\; \forall i=1,\dots,M,
\end{gathered}
\label{eq:gapped}
\end{equation}
where we have introduced the notation $\Delta E_{i,j}=E_{i,j}-E_{i,0}$ and $\Delta E_{ci,j}=E_{ci,j}-E_{ci,0}$ (with $E_{i,j}$ ($E_{ci,j}$) again the \textit{j}-th eigenstate of the \textit{i}-th qubit (coupler) unperturbed Hamiltonian). Since the summations above grow linearly with the number of qubits in the system, this condition limits the size of the systems to which we can apply our reduction method. Intuitively this limit reflects the impossibility of finding any coherent description of the low-energy spectrum of a composite system in terms of interacting two-level subsystems, whenever the second excited state of one of these subsystems appears in the spectrum. Therefore if we are interested in characterising a very large circuit, we should first subdivide it into smaller connected subsystems for which the inequalities \eqref{eq:gapped} hold.
\par The second requirement is simply that the strength of the interaction Hamiltonian should be small compared to the \textit{spectral gap} of $\hat{H}_0$, $\Delta$. Namely:
\begin{equation}
\Vert\hat{H}_{int}\Vert_{op}<\frac{\Delta}{2},
\label{eq:cond2}
\end{equation}
where $\Vert\cdot\Vert_{op}$ is the \textit{operator norm}:
\begin{equation}
\Vert\hat{O}\Vert_{op}=\sup\{\Vert\hat{O}|\Psi\rangle\Vert:\Vert|\Psi\rangle\Vert=1\},
\end{equation}
with $\Vert\cdot\Vert$ the 2-norm $\sqrt{\langle\,\cdot\,|\,\cdot\,\rangle}$.
\par Since the addition of the interaction term $\hat{H}_{int}$ can shift the eigenvalues of $\hat{H}_0$ by at most $\Vert\hat{H}_{int}\Vert_{op}$, this second inequality implies that the spectrum of $\hat{H}_{e.m.}$ remains gapped. This in turn allows us to rewrite the total Hamiltonian in the block-diagonal form $\hat{H}_{e.m.}=\hat{P}\hat{H}_{e.m.}\hat{P}+\hat{Q}\hat{H}_{e.m.}\hat{Q}$, where $\hat{P}$ is the projector on the $2^N$-dimensional low-energy eigenspace of $\hat{H}_{e.m.}$, $\mathcal{H}_{low}$, and $\hat{Q}=\hat{\mathbb{I}}-\hat{P}$\cite{bravyi2011schrieffer}.
\par Additionally, according to \cite{bravyi2011schrieffer}, since $\mathcal{H}_{low}$ and $\mathcal{H}_q$ have the same dimension, they are connected by a \textit{direct rotation} $\hat{U}$ such that
\begin{equation}
\begin{gathered}
\hat{U}\hat{P}\hat{U}^\dagger=\hat{P}_0,\\
\hat{U}\hat{Q}\hat{U}^\dagger=\hat{Q}_0.
\end{gathered}
\label{eq:projectors}
\end{equation}
$\hat{U}$ is called the Schrieffer-Wolff transformation and can be written, in terms of the projectors, as\cite{bravyi2011schrieffer}:
\begin{equation}
\hat{U}=\sqrt{(2\hat{P}_0-\hat{\mathbb{I}})(2\hat{P}-\hat{\mathbb{I}})}.
\label{eq:SWT}
\end{equation}
The principal square root $\sqrt{\cdot}$ above is well-defined as long as
\begin{equation}
\Vert\hat{P}-\hat{P}_0\Vert_{op}<1,
\end{equation}
which in our case can be shown to be equivalent to \eqref{eq:cond2}\cite{bravyi2011schrieffer}.
\par Now the action of the SWT on $\hat{H}_{e.m.}$ is given by
\begin{equation}
\begin{aligned}
\hat{U}\hat{H}_{e.m.}\hat{U}^\dagger=&\hat{U}\hat{P}\hat{H}_{e.m.}\hat{P}\hat{U}^\dagger+\hat{U}\hat{Q}\hat{H}_{e.m.}\hat{Q}\hat{U}^\dagger\\=&\hat{P}_0\hat{U}\hat{H}_{e.m.}\hat{U}^\dagger\hat{P}_0+\hat{Q}_0\hat{U}\hat{H}_{e.m.}\hat{U}^\dagger\hat{Q}_0,
\end{aligned}
\label{eq:block}
\end{equation}
where we used the identities $\hat{U}\hat{P}=\hat{P}_0\hat{U}$ and $\hat{U}\hat{Q}=\hat{Q}_0\hat{U}$. According to equation \eqref{eq:block}, $\hat{U}\hat{H}_{e.m.}\hat{U}^\dagger$ is block-diagonal with respect to $\hat{P}_0$ and $\hat{Q}_0$. This finally leads us to the conclusion that
\begin{equation}
\hat{H}_q:=\hat{P}_0\hat{U}\hat{H}_{e.m.}\hat{U}^\dagger\hat{P}_0
\end{equation}
is an Hermitian operator, acting on $\mathcal{H}_q$, whose $2^N$ non-zero eigenvalues are the same as the lowest eigenvalues of the original interacting Hamiltonian $\hat{H}_{e.m.}$ (because the unitary $\hat{U}$ leaves the spectrum of $\hat{P}\hat{H}_{e.m.}\hat{P}$ unchanged)\cite{bravyi2011schrieffer}. $\hat{H}_q$ therefore represents our effective qubit Hamiltonian, from which we can directly extract the Pauli coefficients by rewriting equation \eqref{eq:paulis} as
\begin{equation}
h_{\vec{\eta}}=\frac{1}{2^N}\textrm{Tr}\left(\hat{H}_q\cdot\hat{\sigma}_{\vec{\eta}}\right).
\end{equation}
In this case, the Pauli operator $\hat{\sigma}_{\vec{\eta}}=\hat{\sigma}_{\eta_1}\otimes\cdots\otimes\hat{\sigma}_{\eta_N}\otimes\hat{P}_c$ is built from the single-qubit Pauli operators $\{\hat{\sigma}_{\eta_i}\}$, which, in turn, are obtained as in the single-qubit case, starting from the unperturbed Hamiltonian $\hat{H}_i$ of each qubit and the appropriate operator $\hat{O}_{p,i}$. The operator
\begin{equation}
\hat{P}_c=\bigotimes_{i=1}^M|E_{ci,0}\rangle\langle E_{ci,0}|
\end{equation}
represents the required identities acting on each of the ground-state energy subspaces of the coupler circuits.
\par Finally note that since both the approximate rotation reduction and the SWT reduction method determine an effective qubit Hamiltonian with the correct spectrum, the two results must be equivalent up to a unitary transformation. However, as mentioned before, the SWT reduction extends the range of applicability of the method to Hamiltonians with complex elements and does not involve the restrictive assumption \eqref{eq:approx}.
\section{Numerical results}
\label{numres}
In this section we present some numerical examples of Hamiltonian reduction for different SC qubit designs and interacting systems. For concreteness, we will focus on qubits of the flux-type (\cite{orlando1999superconducting,yan2016flux,friedman2000quantum,shimazu2009four}) and we will consider circuits and physical parameters from works in the recent literature.
\par For these simulations the circuit Hamiltonians and all other circuit operators were represented in matrix form by projection on a truncated orthonormal basis. The approximate Krylov-Schur method, implemented by the MATLAB${}^\textrm{\textcopyright}$ function \textit{eigs}\cite{matlab2}, was used to determine the relevant subsets of the operator eigenvalue-eigenvector pairs. This approach can be much faster than the complete diagonalisation of the operator, especially when it is very large and sparse, as is usually true for SC qubit Hamiltonians\cite{dempster2014understanding}.
\subsection{Single qubits}
\begin{figure}[b!]
	\centering
	\includegraphics[height=0.35\textwidth]{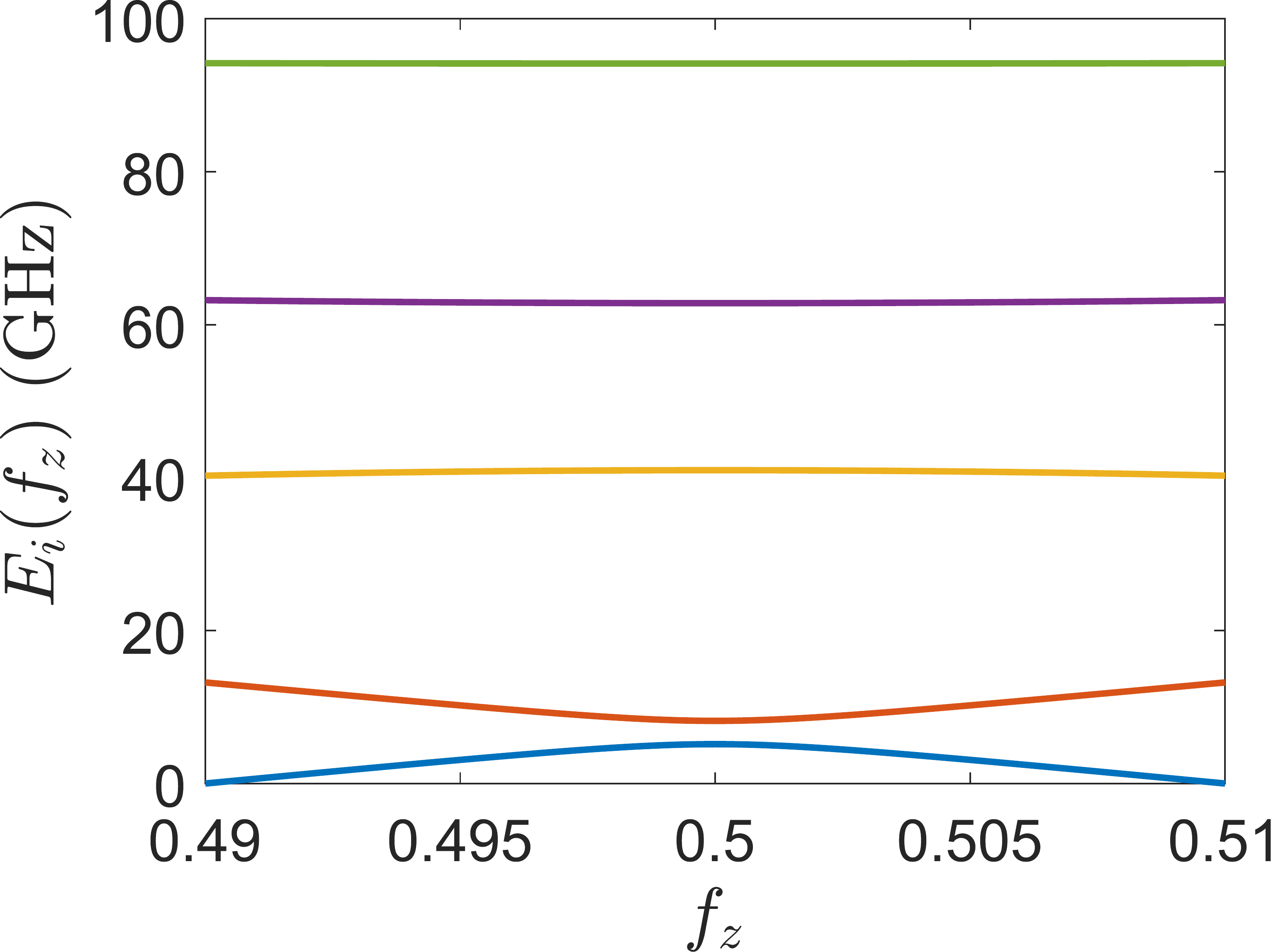}
	\caption{Low energy spectrum of the rf-SQUID flux qubit, as a function of the normalised magnetic flux $f_z$ applied to the superconducting loop.}
	\label{fig:rfsquidb}
\end{figure}
\subsubsection{rf-SQUID flux qubit}
\begin{figure*}[b!]
	\centering
	\subfloat[][Pauli coefficients as a function of the reduced magnetic flux bias $f_z$. Lines: local reduction, empty dots: perturbative reduction, crosses: instanton method.]{\includegraphics[height=0.35\textwidth]{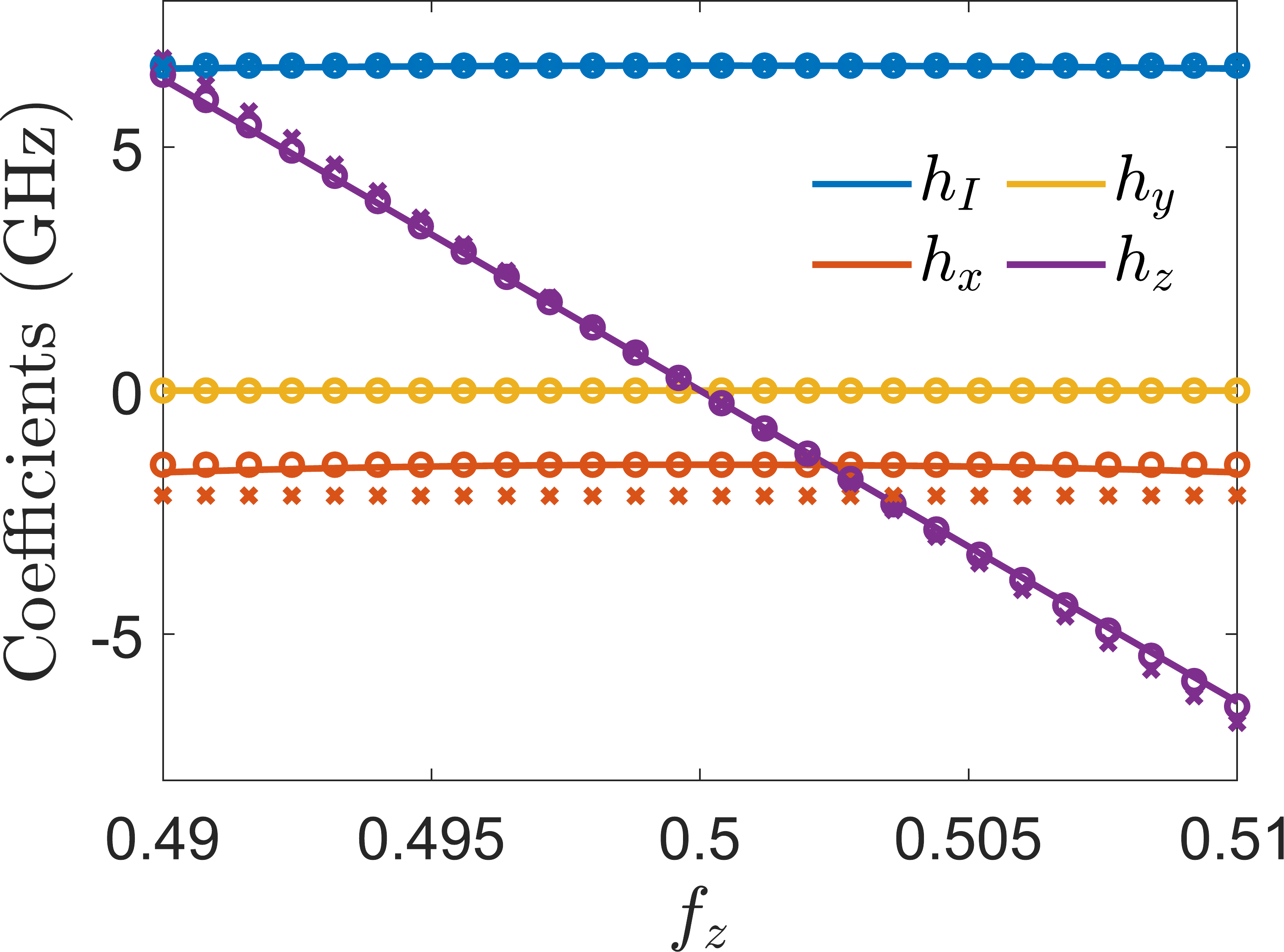}\label{fig:squbits1a}}\quad
	\subfloat[][Tunnelling energies (left \textit{y}-axis) at the symmetric bias point $f_z=0.5$ and the corresponding semi-classical potential barrier height (right \textit{y}-axis), as a function of the qubit loop inductance.]{\includegraphics[height=0.35\textwidth]{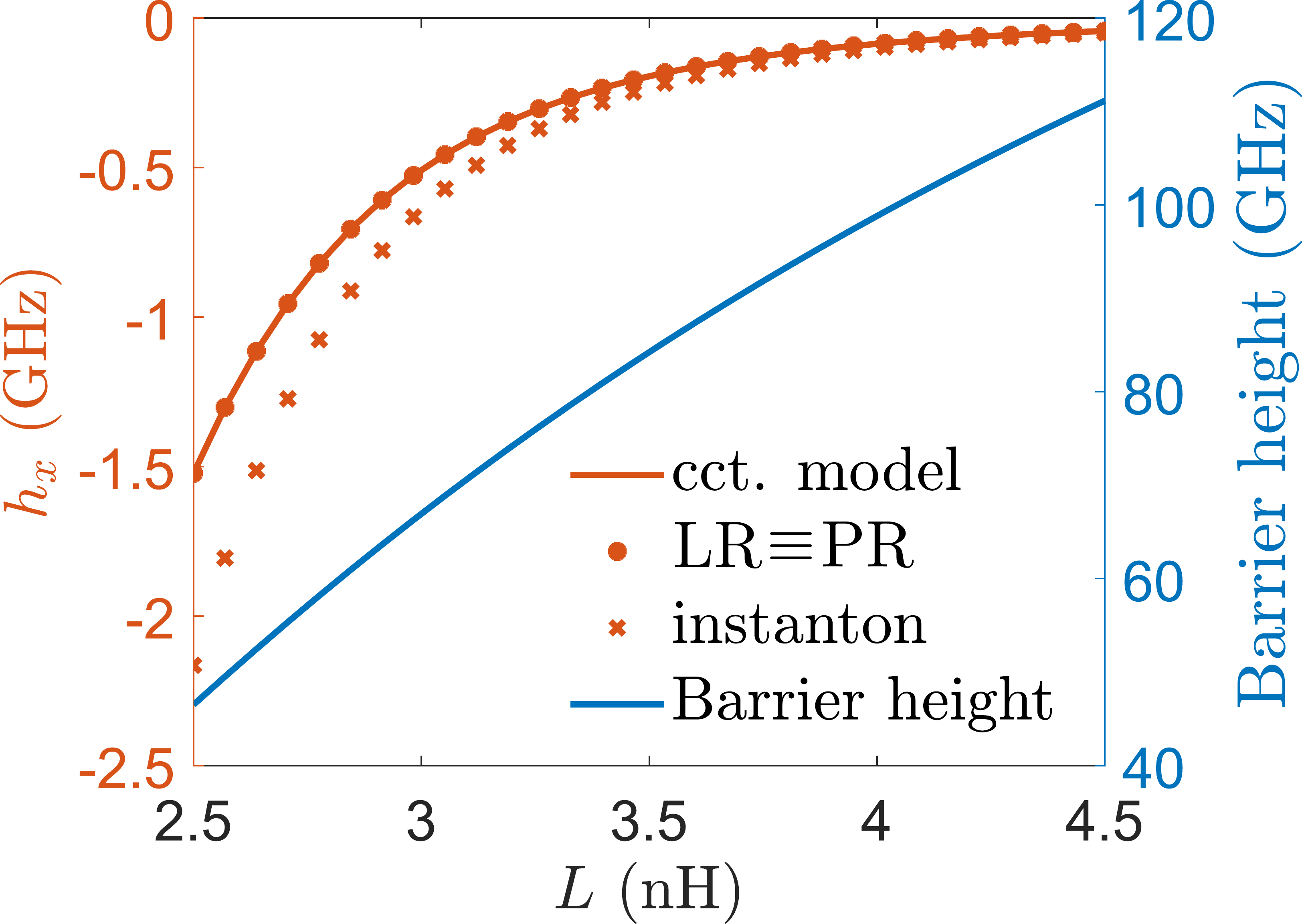}\label{fig:squbits1b}}\\
	\subfloat[][Comparison of the two lowest circuit levels, as a function of $f_z$, with the result of different reduced two-level system models. At this scale the LR and PR results overlap.]{\includegraphics[height=0.35\textwidth]{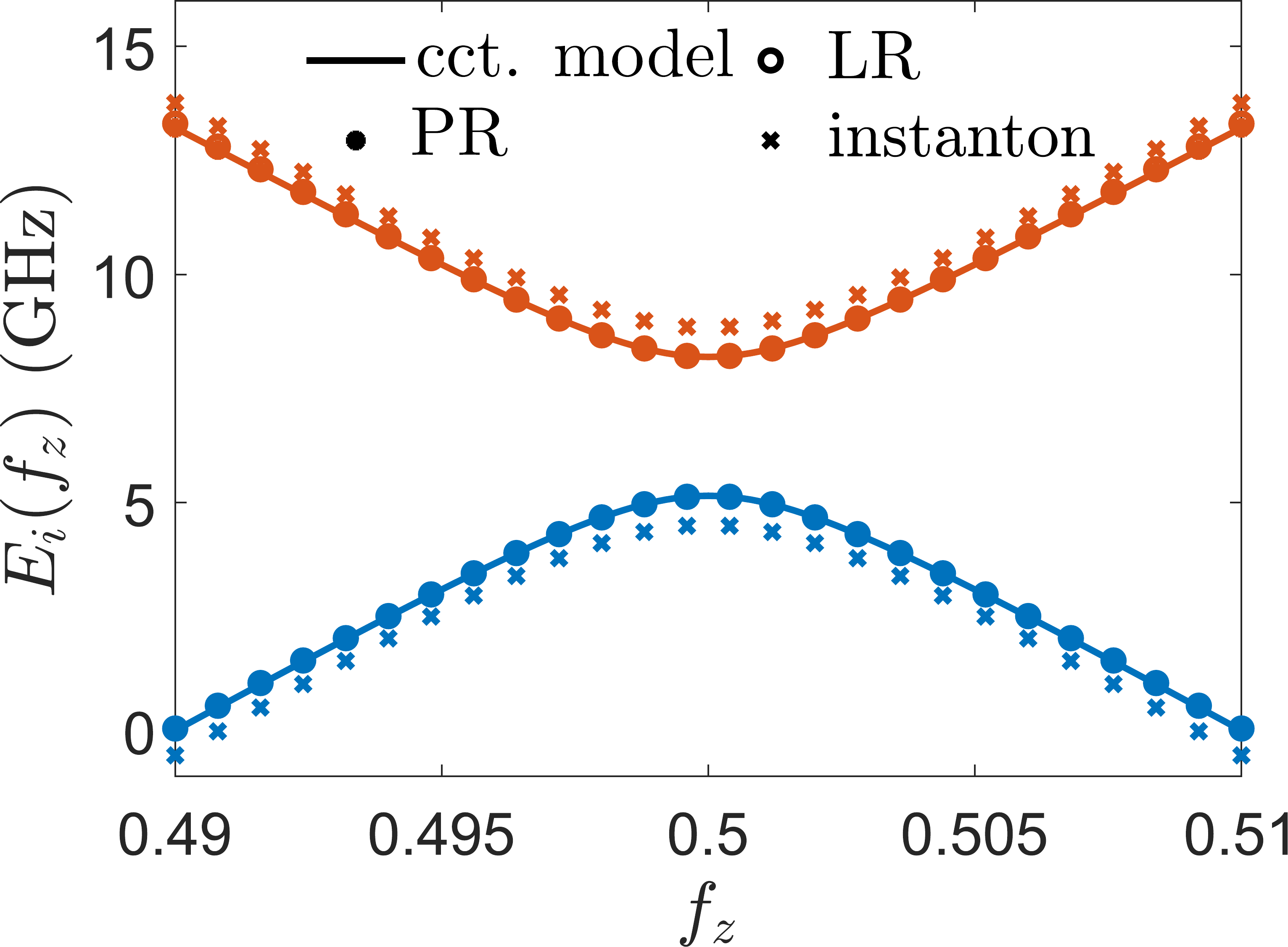}\label{fig:squbits1c}}\quad
	\subfloat[][Pauli coefficients calculated with LR over a broader range of $f_z$ (left \textit{y}-axis). Also shown are the lowest three eigenenergies of the system (dashed lines, values on the right \textit{y}-axis).]{\includegraphics[height=0.35\textwidth]{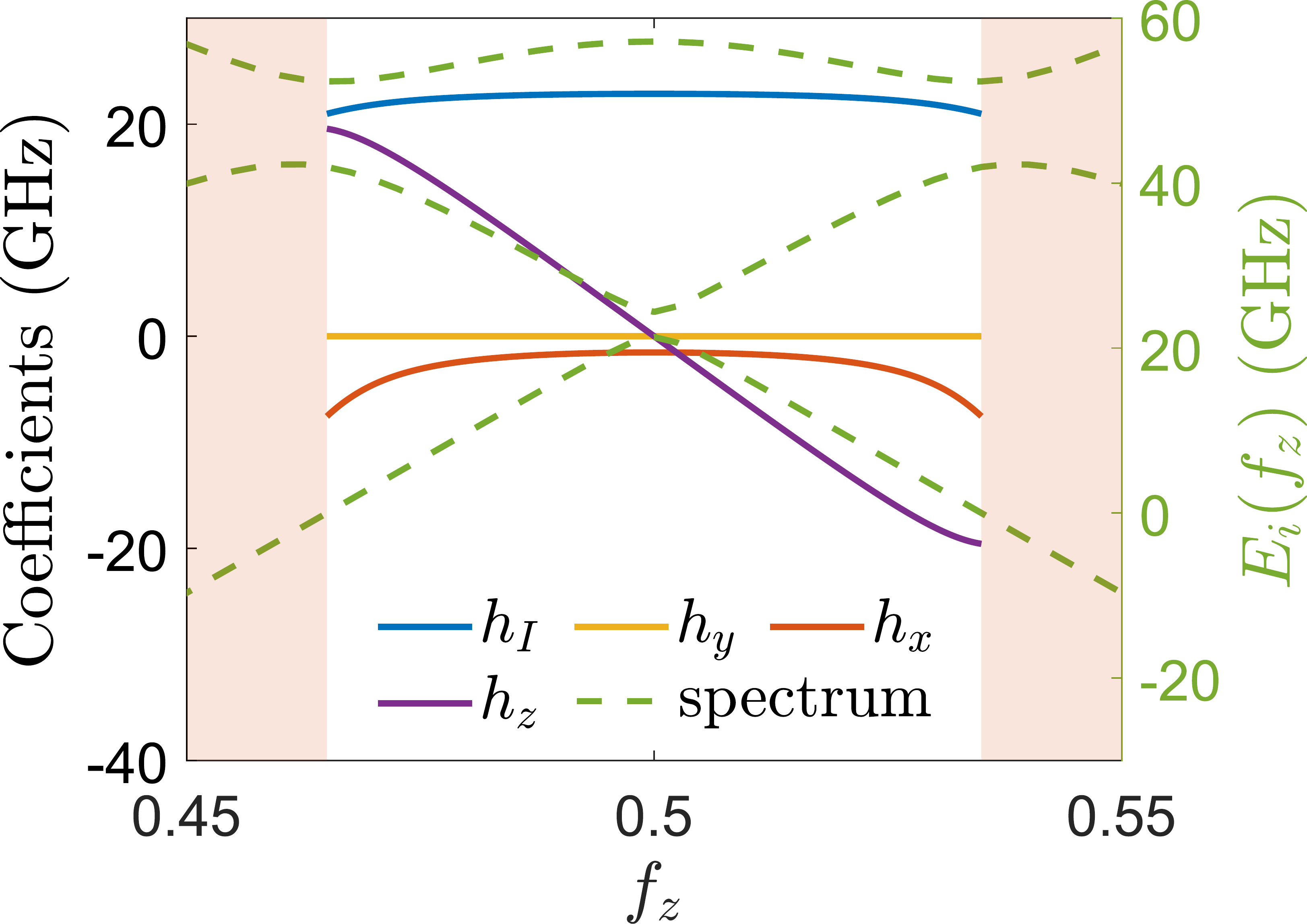}\label{fig:squbits1d}}
	\caption{Numerical results for the rf-SQUID circuit of figure \ref{fig:rfsquida}.}
	\label{fig:squbits1}
\end{figure*}
\label{singlequbits}
We start by considering the simplest example of a flux qubit, \textit{i.e.} the rf-SQUID circuit.
As shown in figure \ref{fig:rfsquida}, this consists of a Josephson junction, with tunnelling energy $E_J$, shunted by a superconducting inductive loop with self-inductance \textit{L} and in parallel with its intrinsic capacitance $C_J$\cite{wendin2017quantum}.
\par Figure \ref{fig:rfsquidb} shows the lowest five energy eigenvalues of the circuit, calculated as a function of the dimensionless external magnetic flux $f_z=\Phi_z/\Phi_0\equiv\Phi_{01}^{ext}/\Phi_0$. (Note that the constant offset $E_0(f_z=0.49)$ has been subtracted from all the energies.) The parameters used for the simulations are $E_J=125$ GHz, $C_J=5$ fF and $L=2.5$ nH, which are typical for this type of device\cite{harris2010experimental}. In this case, the Hamiltonian was represented in a basis of harmonic oscillator occupation number states, truncated at a maximum occupation number of 40, which ensured the convergence of the low energy spectrum (cf. appendix \ref{appendix3})\cite{zhu2013circuit}.
\par As we can see from the graph in figure \ref{fig:rfsquidb}, the lowest two energy levels of the system (i.e. the qubit states), vary approximately linearly with the flux $f_z$, except around the \textit{symmetry point} $f_z=0.5$, where they show a characteristic avoided crossing. In fact, as we saw previously, for small values of $|\delta f_z|=|f_z-0.5|$ the rf-SQUID Hamiltonian is well approximated by its first order expansion in $\delta f_z$. This maps to an effective qubit Hamiltonian of the form (see Eq. \eqref{eq:qham})
\begin{equation}
\mathbf{H}_q(f_z)=\frac{\Delta}{2}\pmb{\sigma}_x+\frac{\varepsilon(f_z)}{2}\pmb{\sigma}_x,
\end{equation}
where we have neglected the term proportional to the identity, $\Delta=(E_1(f_z)-E_0(f_z))_{f_z=0.5}$ and $\varepsilon(f_z)=2\Phi_0I_p\delta f_z$. The lowest two energy levels of the circuit are therefore approximately $E_{0,1}=const.\mp\sqrt{\Delta^2+4\Phi_0^2I_p^2|\delta f_z|^2}$, which become linear in $f_z$ for larger values of $|\delta f_z|$.
\par Figure \ref{fig:squbits1a} shows the values of the system Pauli coefficients as a function of $f_z$, calculated using equation \eqref{eq:paulicoeffs}. The solid lines correspond to values obtained by defining the Pauli operators according to the local reduction (LR) method introduced here (subsection \ref{localb}). These are compared with the result of the perturbative (PR, empty circles) and instanton (crosses) methods. As we can see, the three reduction methods produce largely compatible results for this circuit. In particular, away from the symmetry point the LR method finds a 10\% increase in the transverse field $h_x$ at the boundary of the flux interval considered, compared to its centre. The result of PR is instead independent of $f_z$, in agreement with Eq. \eqref{eq:qham}. The values of $h_z$ and $h_I$ calculated with the LR and PR methods are compatible to 1\% over the whole flux bias range. This implies that the definition of the computational basis in the LR method coincides, as it should, with that of the standard PR method in the limit in which the series expansion \eqref{eq:seriesexp} and perturbation theory apply. As for the semi-classical calculations, these appear to over-estimate both the longitudinal field (by $\simeq40$\%) and the transverse field (by up to 6\%), compared to the other two reduction methods.
\par Since the \textit{semi-classical approximation} applies in the limit where $\hbar$ is much smaller than the actions at play in the system, i.e. $S\gg\hbar$, and since the tunnelling energy $h_x$ decreases exponentially with the tunnelling action, $h_x\propto e^{-S/\hbar}$ (see appendix \ref{appendix4}), we expect the result of the instanton calculations to be more accurate in the limit where $h_x$ is small\cite{garg2000tunnel,landau2013quantum}. To verify this, we determined the qubit transverse field in the case of biasing at the symmetry point $f_z=0.5$ for increasing values of the loop inductance $L$. As we can see in figure \ref{fig:squbits1b}, increasing $L$ causes the barrier between the two semi-classical potential wells (blue line) to rise, therefore suppressing the tunnelling $h_x$ (data in red). Since $f_z=0.5$, the perturbative and local reduction methods coincide, and they both determine the correct value of the tunnelling energy: $h_x=-\Delta E/2$, where $\Delta E$ is the energy separation between the ground and first excited state of the circuit (cf. dots and solid line in Fig. \ref{fig:rfsquidb}). As expected, the instanton method result (crosses in Fig. \ref{fig:rfsquidb}) closely approaches that of the Hamiltonian reduction only as $L$ increases and $|h_x|$ becomes smaller.
\par At this point, as a consistency check, we can calculate the spectrum of the reduced qubit Hamiltonian simply as $E_{0,1}=h_I\mp\sqrt{h_x^2+h_y^2+h_z^2}$ and compare it with that obtained from the full circuit model. PR and LR do a good job in reproducing the low-energy spectrum of the rf-SQUID qubit, as we can see from the plot in figure \ref{fig:squbits1c}. This also shows the spectrum derived from the semi-classical model (crosses), which does not agree with the correct circuit spectrum as well.
\begin{figure}[t!]
	\centering
	\includegraphics[height=0.35\textwidth]{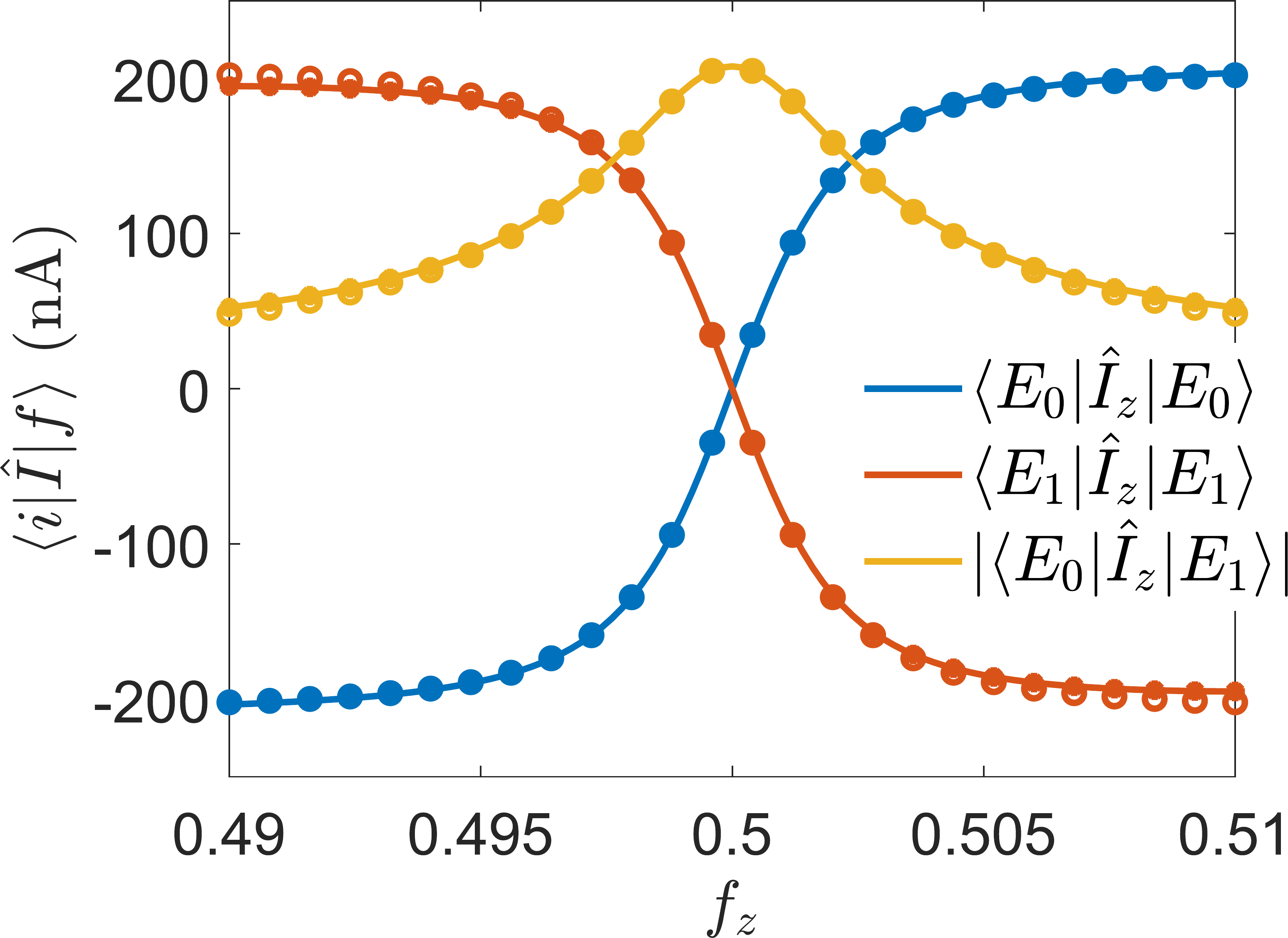}
	\caption{Expectation values of the current operator between the two rf-SQUID qubit eigenstates, as a function of $f_z$. Lines: circuit model, filled dots: LR, empty dots: PR.}
	\label{fig:squbits2a}
\end{figure}
\par Notice that LR is guaranteed to exactly reproduce the circuit levels as long as $f_z\simeq0.5$. As mentioned in the previous section, the LR protocol only fails when, as $|f_z-0.5|$ increases, the two eigenvalues of $\hat{I}_p(f_z)=\hat{P}_0(f_z)\hat{I}\hat{P}_0(f_z)$ begin to have the same sign, meaning that no measurement distinguishing two qubit states with opposite persistent current is possible at the given bias. For the particular rf-SQUID circuit considered here, the local reduction method breaks down for $|f_z-0.5|\gtrsim0.035$, as shown in figure \ref{fig:squbits1d} (region shaded in red). As we can see in this plot, as we approach this region the behaviour of the Pauli coefficients starts changing. In particular the transverse field increases considerably in magnitude, while the longitudinal field saturates. The green dotted lines in figure \ref{fig:squbits1d} show the circuit energy levels. We see that at the boundary of the unshaded region the second excited state starts mixing with the first, leading to an avoided crossing. This mixing means that, at this point, the two-level approximation does not hold any more, which leads to the failure of the LR.
\par Finally we might want to consider how well the reduced Hamiltonians are able to reproduce the correct expectation values of some circuit operator $\hat{O}$, \textit{i.e.} whether the following relationship holds
\begin{equation}
\langle E_i|\hat{O}|E_j\rangle=\langle \mathcal{E}_i|\hat{O}_p|\mathcal{E}_j\rangle,\;\forall i,j\in\{0,1\},
\end{equation}
where $\{|E_{i}\rangle\}_{i=0,1}$ are energy eigenstates of $\hat{H}_{e.m.}$ and $\{|\mathcal{E}_i\rangle\}_{i=0,1}$ are the eigenstates of the corresponding effective qubit Hamiltonian. $\hat{O}_p$ is defined locally as $\hat{P}_0(f_z)\hat{O}\hat{P}_0(f_z)$ (where $\hat{P}_0(f_z)$ is the projector on the two-dimensional low-energy subspace of $\hat{H}_{e.m.}(f_z)$) in the LR method case, and is defined globally as $\hat{P}_0(0.5)\hat{O}\hat{P}_0(0.5)$ in the PR case.
Figure \ref{fig:squbits2a} shows the matrix elements of the loop current operator $\hat{I}$ between qubit states, calculated with both the full and the reduced operators. We observe that LR is ensured to give the exact result, while PR produces a reasonable result.
\par We have shown here that the approximations inherent in the perturbative reduction method are valid and sufficient for for determining the reduced Hamiltonian in the case of the simple rf-SQUID qubit of Fig. \ref{fig:rfsquida}. We will see in the next subsection, however, that this is not true in the general case and that the local reduction method has a wider range of validity.
\subsubsection{C-shunt flux qubit}
\begin{figure}[t!]
	\centering
	\includegraphics[width=.45\textwidth]{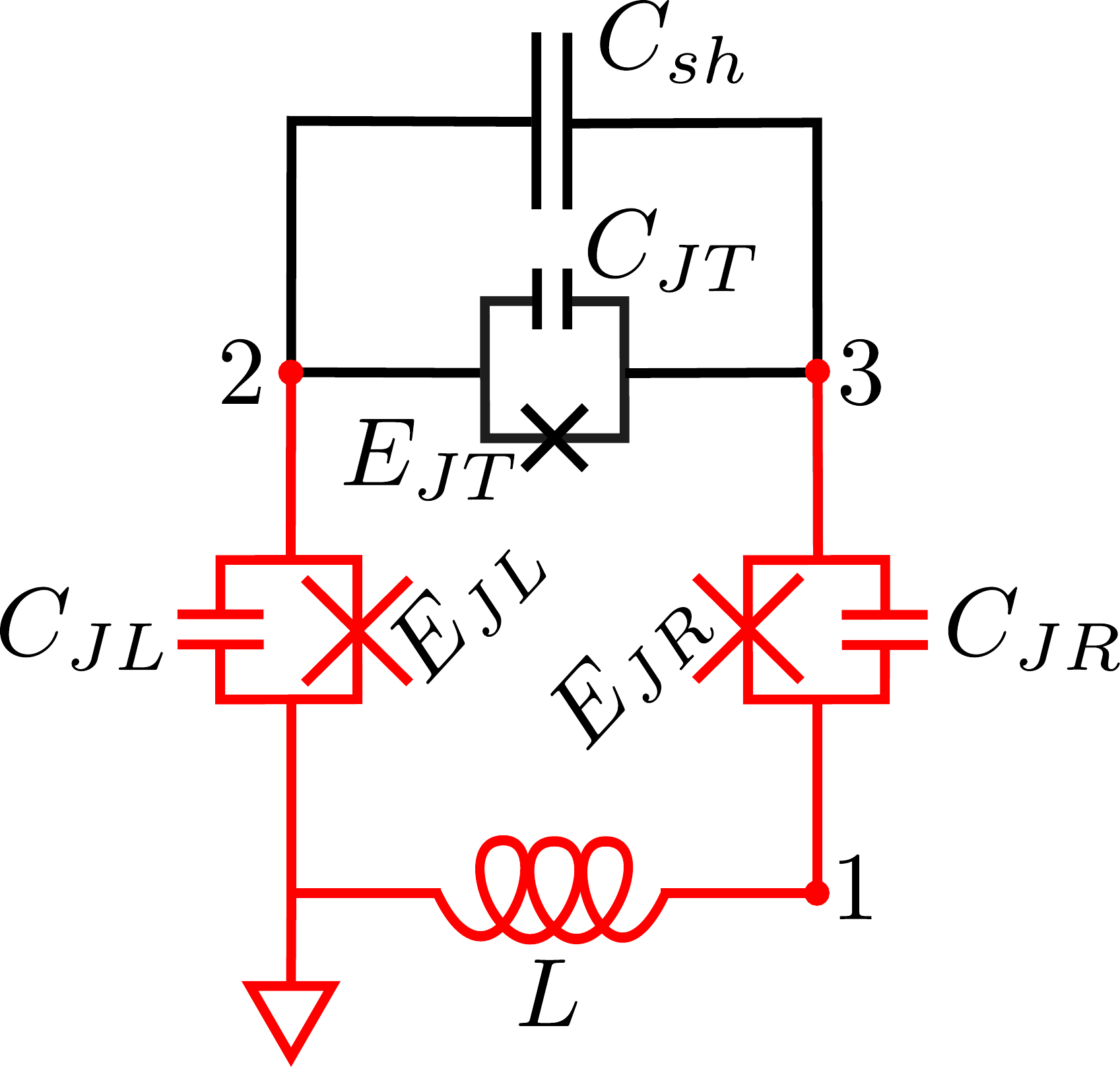}
	\caption{Equivalent lumped-element circuit of a capacitively-shunted flux qubit (as described in Ref. \cite{yan2016flux}). A possible choice of the spanning tree is highlighted in red.}
	\label{fig:3JJ}
\end{figure}
\begin{figure*}[t!]
	\centering
	\subfloat[][Low energy spectrum of the circuit as a function of the reduced magnetic flux bias $f_z$.]{\includegraphics[height=0.35\textwidth]{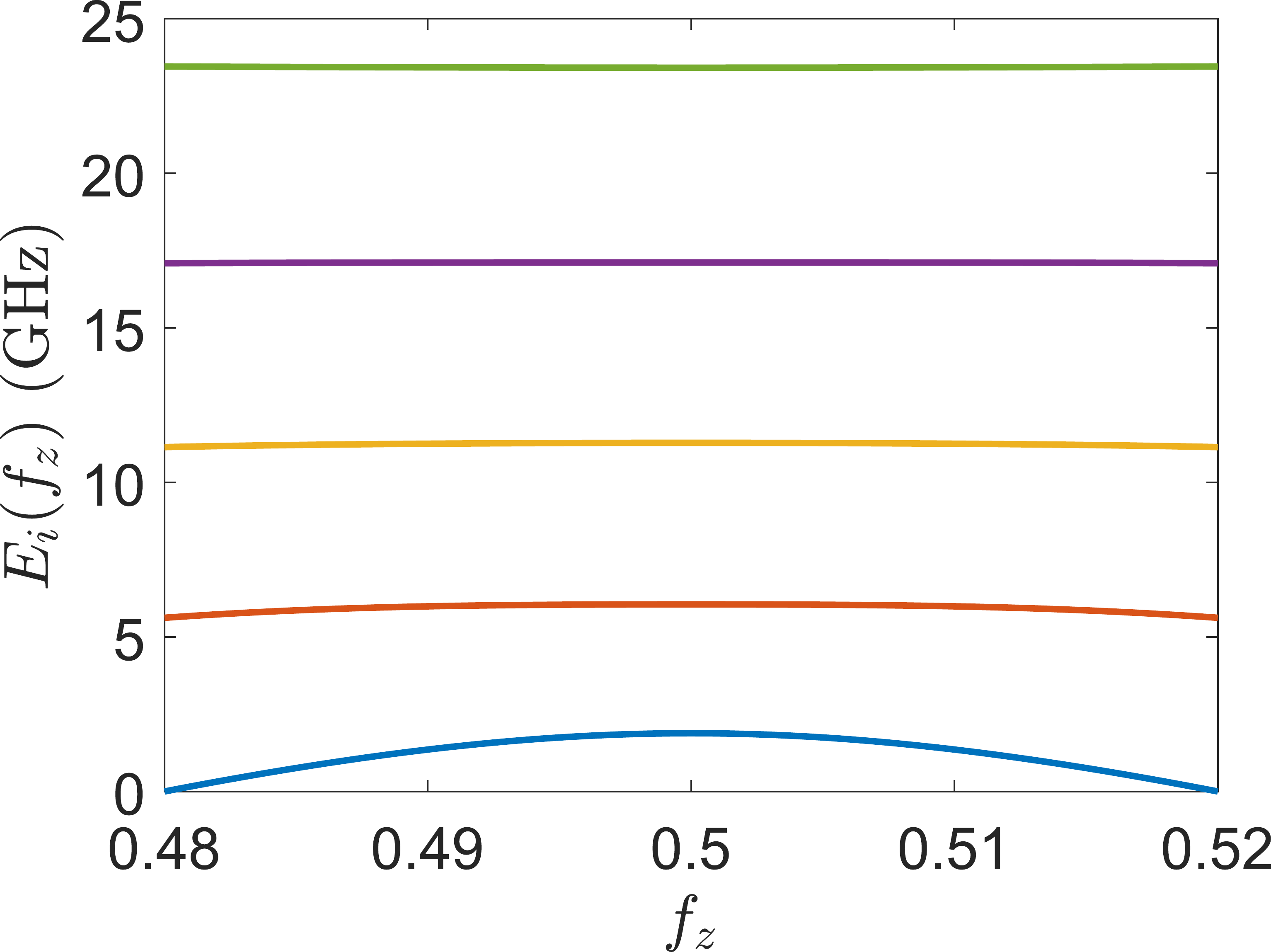}\label{fig:squbits3a}}\quad
	\subfloat[][Pauli coefficients as a function of $f_z$. Lines: local reduction, empty dots: perturbative reduction.]{\includegraphics[height=0.35\textwidth]{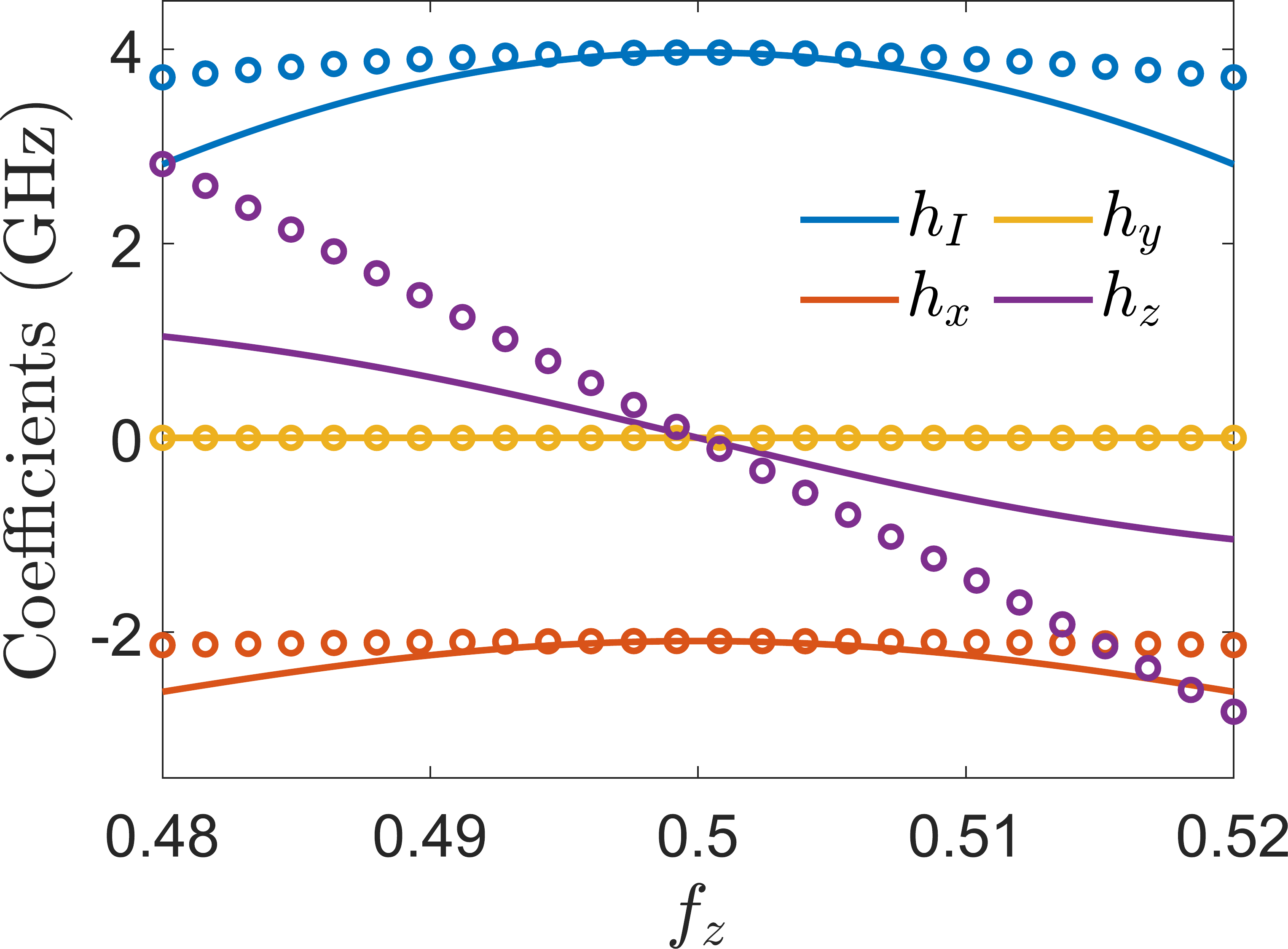}\label{fig:squbits3b}}\\
	\subfloat[][Comparison of the two lowest circuit levels, as a function of $f_z$, with those of the PR and LR qubit Hamiltonians.]{\includegraphics[height=0.35\textwidth]{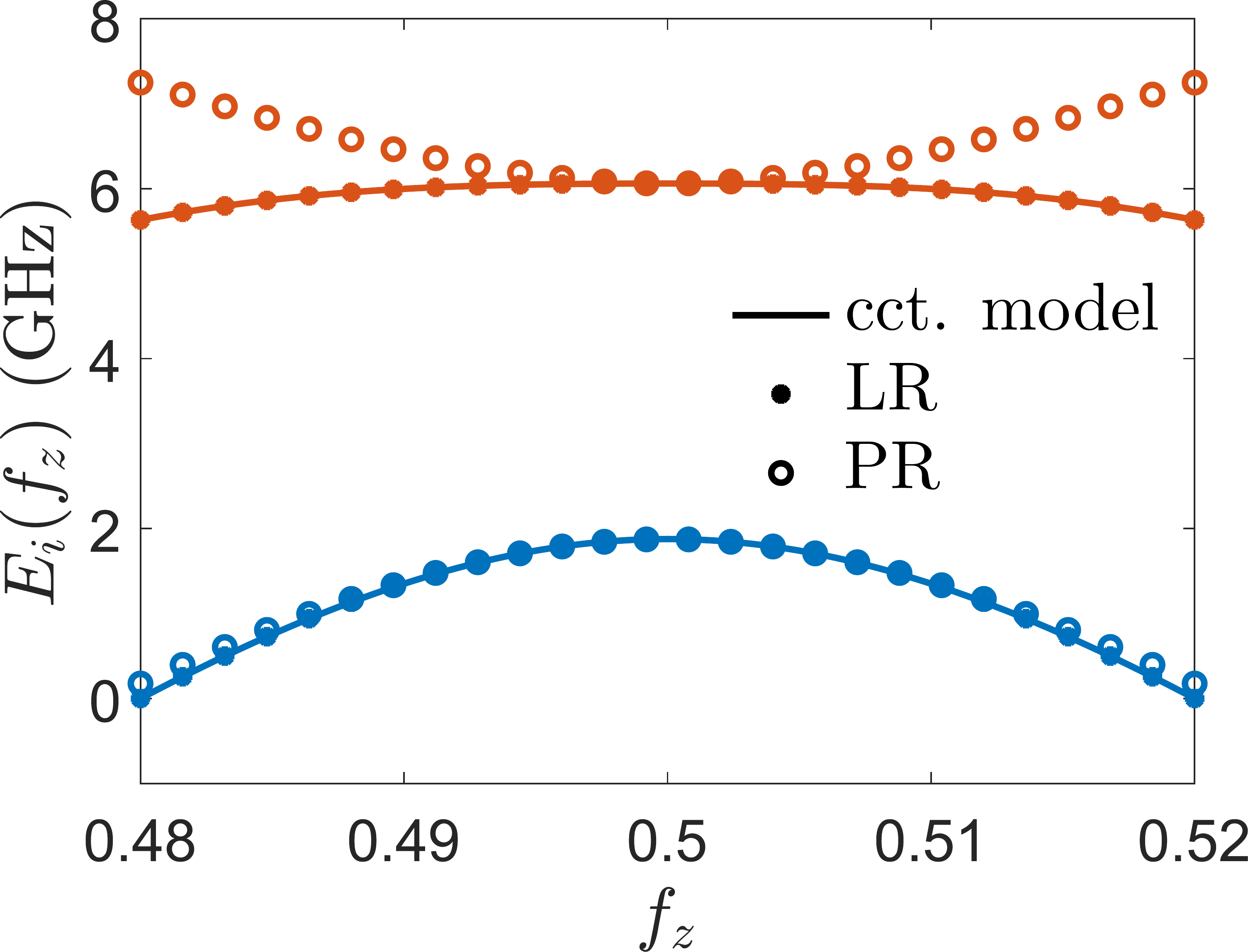}\label{fig:squbits3c}}\quad
	\subfloat[][Expectation values of the current operator between the two qubit energy eigenstates as a function of $f_z$. Lines: circuit model, filled dots: LR, empty dots: PR.]{\includegraphics[height=0.35\textwidth]{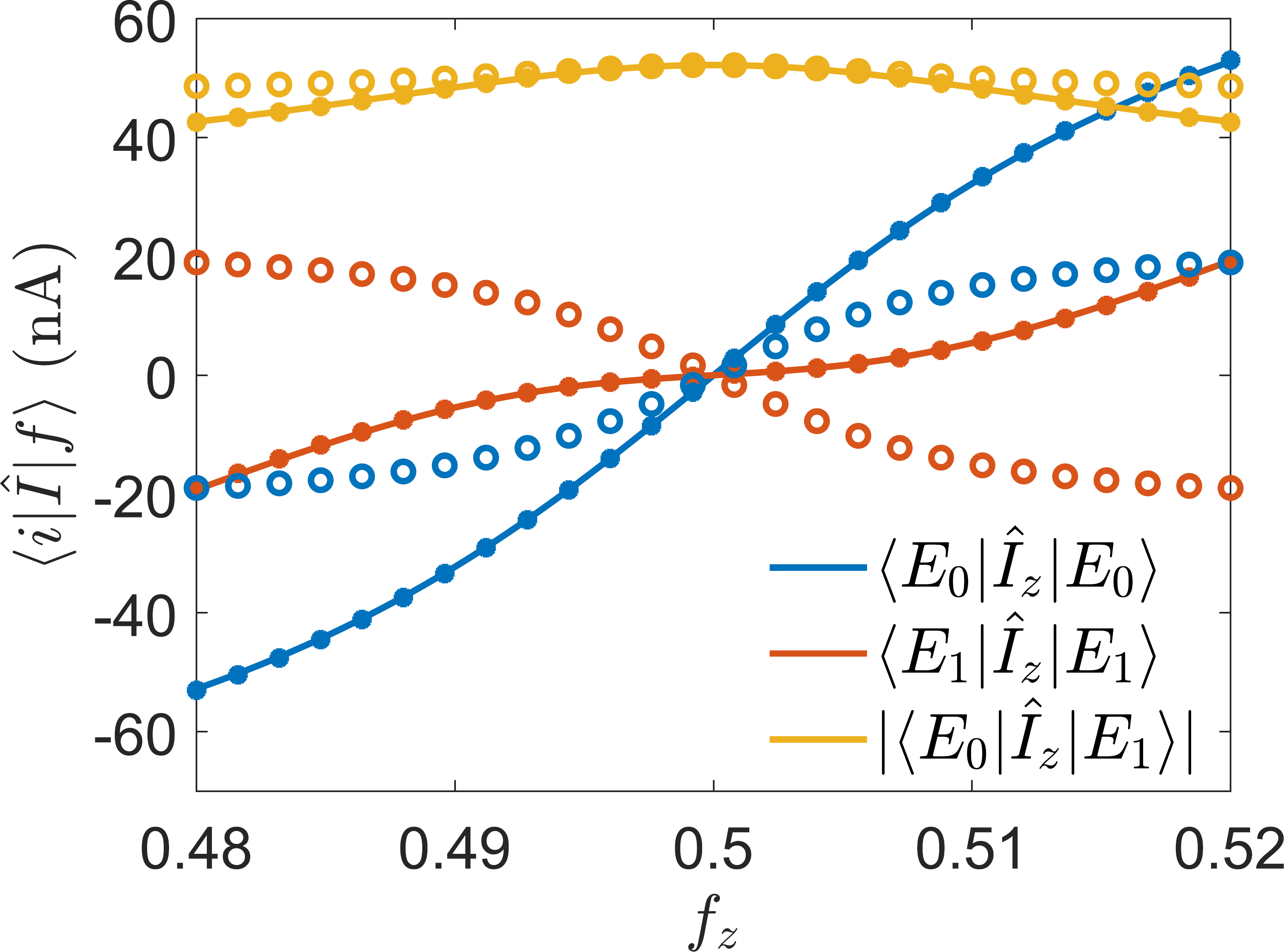}\label{fig:squbits3d}}
	\caption{Numerical results for the C-shunt flux circuit.}
	\label{fig:squbits3}
\end{figure*}
\par The accuracy of the perturbative reduction method deteriorates when we consider other flux qubit designs, particularly those with reduced anharmonicity like the capacitively-shunted flux qubit shown in figure \ref{fig:3JJ}. This consists of a superconducting loop interrupted by three Josephson junctions. The area of one junction is a factor $\alpha<1$ smaller than that of the other two and is shunted by a relatively large capacitor $C_{sh}\gg C_{JT}$. The capacitive shunt reduces the qubit sensitivity to charge noise, while improving the device reproducibility (by compensating for the fabrication variability of the junction size, which affects $C_{JT}$). At the same time, the effect of flux noise is mitigated by choosing small values of $\alpha$ (typically $0.125<\alpha<0.5$), which reduce the magnitude of the persistent current and therefore the magnetic dipole moment of the circuit\cite{yan2016flux}. The result is superconducting qubits with typical measured relaxation times $T_1$ in excess of 40\textmu s (three orders of magnitude longer than the standard rf-SQUID $T_1$) and decoherence times approaching the relaxation limit $T_2=2T_1$\cite{yan2016flux}.
\par This substantial coherence enhancement comes at the cost of a decrease in the spectrum anharmonicity. We can see this by looking at figure \ref{fig:squbits3a}, which shows the calculated low energy spectrum of a C-shunt qubit circuit as a function of $f_z=\Phi_{z}/\Phi_0=\Phi_{23}^{ext}/\Phi_0$, and comparing it with Fig. \ref{fig:rfsquidb}. The physical parameters used for the simulation are shown in table \ref{table0} (cf. Fig. \ref{fig:3JJ} for the meaning of the symbols). For the two lower junctions we used $E_{JL}=E_{JR}=E_{JT}/\alpha$ and $C_{JL}=C_{JR}=C_{JT}/\alpha$. These parameters are compatible with those reported in the experiments in Ref. \cite{yan2016flux}.
\begin{table}[h]
	\centering
	\begin{tabular}{r|l}
		Parameter	&  Value \\
		\hline
		$E_{JT}$ & 45 GHz \\ 
		\hline
		$C_{JT}$ & 1.8 fF \\
		\hline
		$\alpha$ & 0.43 \\
		\hline
		$C_{sh}$ & 50 fF \\
		\hline
		$L$ & 100 pH
	\end{tabular}
	\caption{}
	\label{table0}
\end{table}
In this case, the Hamiltonian was represented numerically by projecting on a finite basis consisting of harmonic oscillator states for the mode associated with the circuit node 1 and charge number states for the modes associated with nodes 2 and 3 (cf. appendix \ref{appendix3})\cite{zhu2013circuit,bouchiat1998quantum,kerman2018unpublished}.
\par As we can see from figure \ref{fig:squbits3a}, the two dispersion relations $E_{0,1}(f_z)$ have first derivatives with the same sign everywhere. Since
\begin{equation}
\langle\hat{I}\rangle_{0,1}:=-\langle\frac{\partial\hat{H}_{e.m.}}{\partial \Phi_z}\rangle\simeq\frac{\partial E_{0,1}}{\partial \Phi_z},
\end{equation}
this means that the average persistent currents in the two energy eigenstates have equal sign (cf. Fig. \ref{fig:squbits3d}). This is in contrast with the rf-SQUID flux qubit\cite{yan2016flux}, but does not preclude the possibility to find two current eigenstates with opposite sign in the qubit subspace.
\par Figure \ref{fig:squbits3b} shows the Pauli coefficients obtained by the perturbative (circles) and local (lines) reduction methods. As anticipated, there is a clear discrepancy between the two results. In fact, owing to the much smaller anharmonicity of this circuit compared to the rf-SQUID, the two low-energy eigenstates of the circuit Hamiltonian at $f_z=0.5$ are not a good approximation for those away from $f_z=0.5$. This implies that projecting $\hat{H}_{e.m.}(f_z)$ on the states \eqref{eq:computs} does not preserve its low-energy spectrum and does not lead to the correct reduction. From the numerical results we see that the slope of $h_z(f_z)$ in the local reduction case is smaller than in the perturbative reduction and further decreases away from $f_z=0.5$. Additionally, the transverse field $h_x(f_z)$ shows a clear negative curvature in the LR results, whereas it is roughly constant in $f_z$ in the PR case (as in the rf-SQUID). The strong dependence of the transverse field on $f_z$ is a known distinguishing feature of the C-shunt flux qubit design when compared to more standard flux qubit circuits like the rf-SQUID\cite{orlando1999superconducting,yan2016flux}.
\par Calculating the spectra of the two reduced Hamiltonians leads to the result shown in figure \ref{fig:squbits3c}. The local reduction result (filled dots) again reproduces the circuit ground and first excited states (lines) exactly, while the perturbative reduction fails to accurately predict the first excited state. Finally figure \ref{fig:squbits3d} shows the matrix elements of the current operator between the qubit energy eigenstates, calculated using the full circuit model (lines) and the two reduced two-level models (circles). The PR (empty circles) gives incorrect expectation values, which are opposite in sign for the two states.
\begin{figure}[b!]
	\centering
	\includegraphics[width=0.49\textwidth]{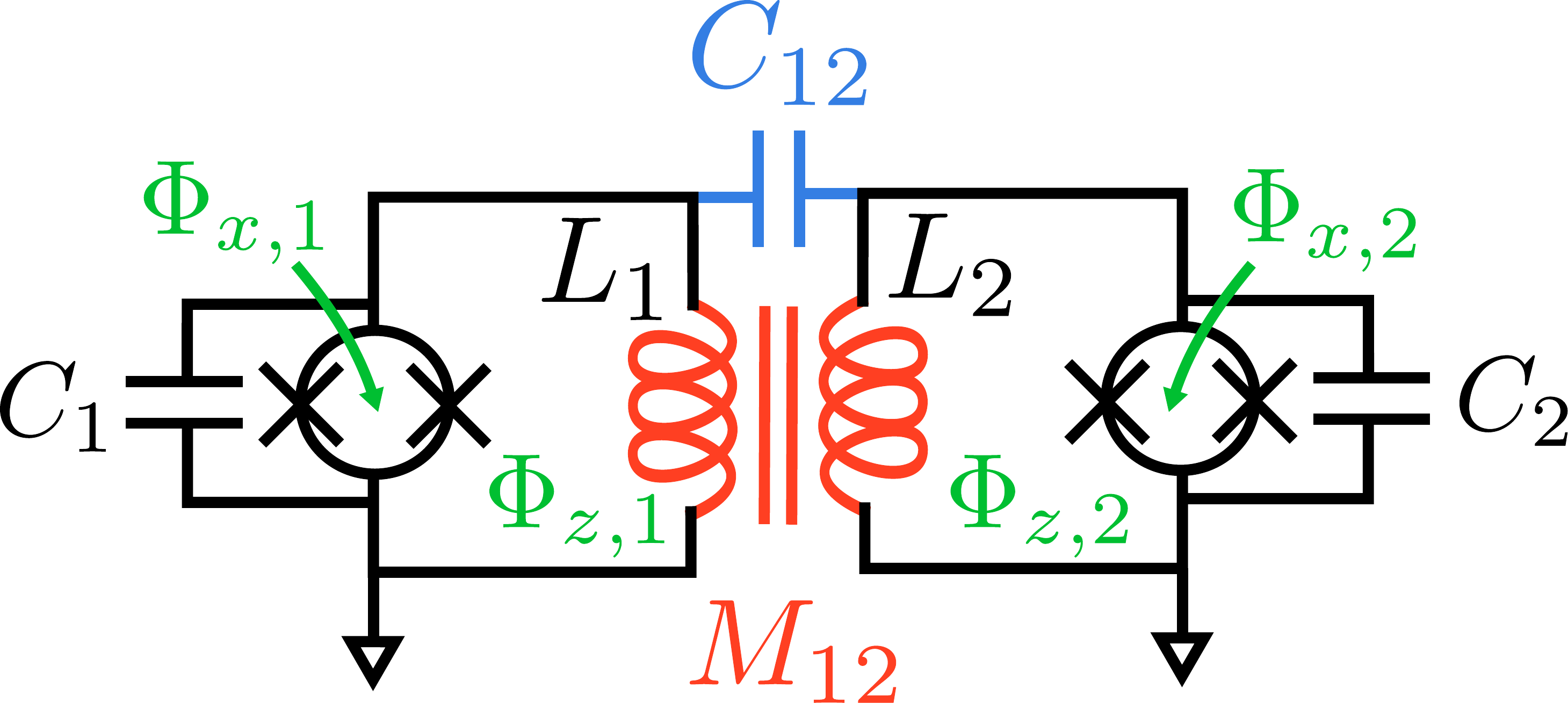}
	\caption{Circuit diagram of the system of two interacting qubits studied in \cite{ozfidan2019demonstration}. Highlighted in different colours are the coupling elements and the magnetic bias fluxes.}
	\label{fig:dwave1}
\end{figure}
\subsection{Multiple qubits}
\subsubsection{ZZ plus XX coupling}
\begin{figure*}[b!]
	\centering
	\subfloat[][Low energy circuit spectrum, relative to the ground state, of the circuit in figure \ref{fig:dwave1}, as a function of the mutual inductance $M_{12}$.]{\includegraphics[height=0.35\textwidth]{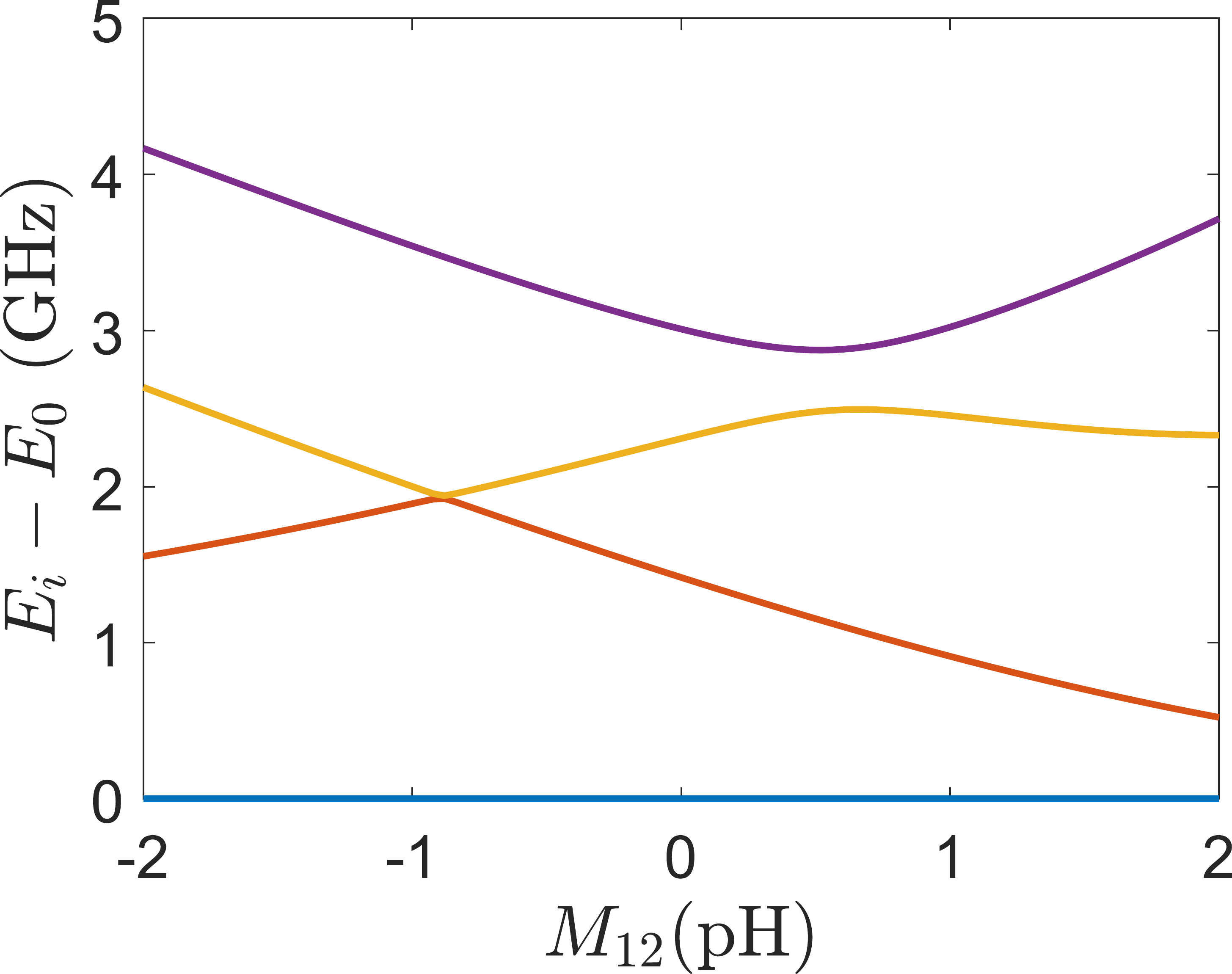}\label{fig:dwave2}}\quad
	\subfloat[][One-local Pauli coefficients calculated, as a function of $M_{12}$, by applying the Schrieffer-Wolff transformation reduction method to the full (solid lines) and the unperturbed Hamiltonian (dashed lines) of the circuit in Fig. \ref{fig:dwave1} (notice that the solid and dashed lines for $h_{zI}$ and $h_{Iz}$ all overlap at this scale). Circles: same coefficients, calculated using the approximate rotation reduction of \cite{ozfidan2019demonstration}.]{\includegraphics[height=0.35\textwidth]{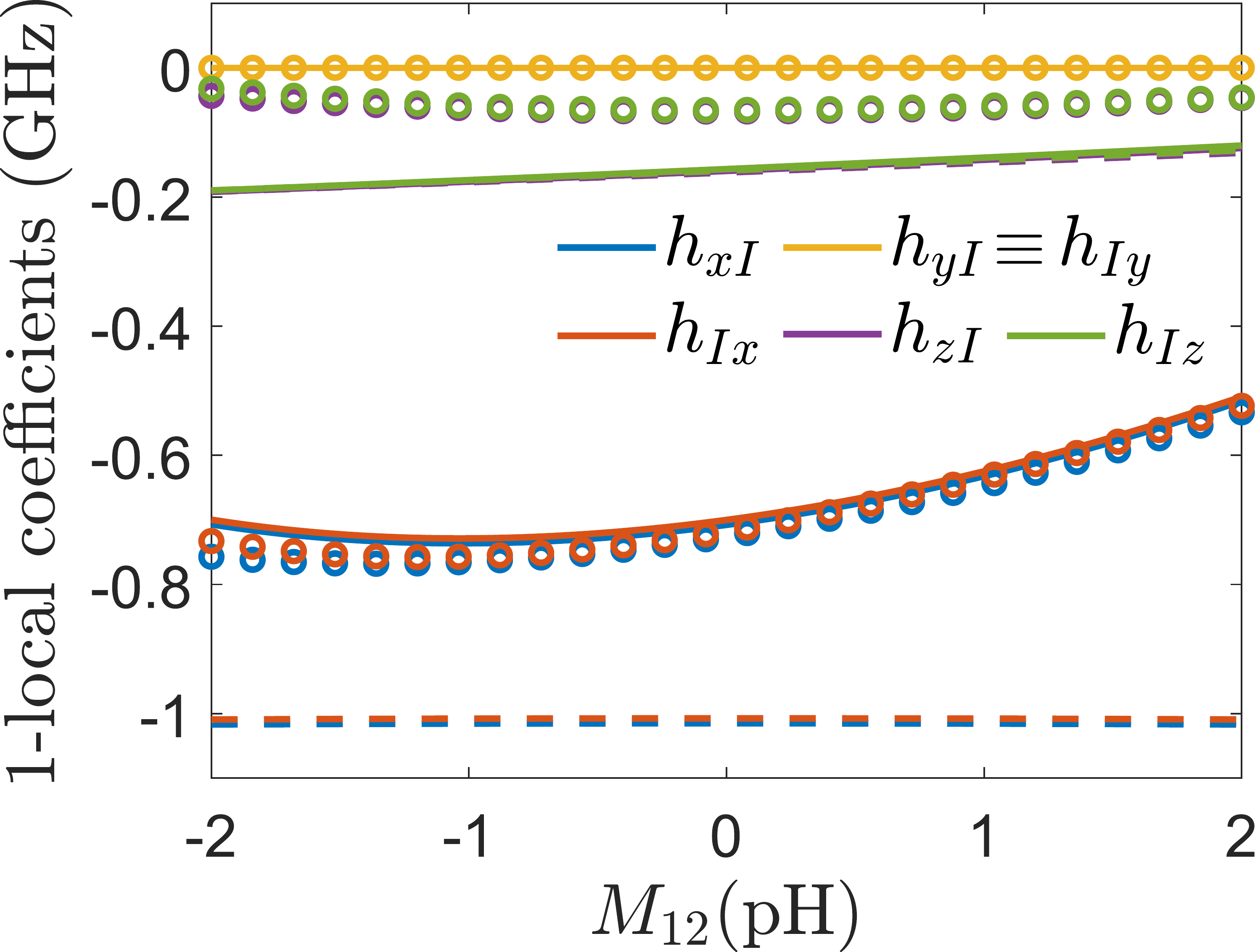}\label{fig:dwave3}}
	\caption{}
	\label{fig:dwavea}
\end{figure*}
We begin this subsection on coupled SC qubit systems by considering a simple two-qubit system, without any non-linear coupling element. As one such example we consider the system which Ozfidan et al. characterised experimentally in \cite{ozfidan2019demonstration}. This is composed of two compound-Josephson-junction rf-SQUID qubits (where the single Josephson junction is replaced by two junctions in parallel, forming a dc-SQUID) coupled both inductively and capacitively, as shown in figure \ref{fig:dwave1}. Assuming that the dc-SQUID loop is very small (such that its inductance is much smaller than both the main loop inductance and the Josephson inductance $(\Phi_0/2\pi)^2/E_J$), we can effectively describe it as a single junction whose Josephson energy depends on the flux $\Phi_x$ threading the dc-SQUID\cite{harris2009compound}:
\begin{equation}
E_J(\Phi_x)=E_{J0}\cos\left(\pi\frac{\Phi_x}{\Phi_0}\right).
\end{equation}
$E_{J0}=E_{J0,1}+E_{J0,2}$ here is the sum of the energies of the two junctions in parallel, which we are assuming to be equal.
\par Within this approximation, the Hamiltonian describing our circuit is:
\begin{equation}
\begin{gathered}
\hat{H}_{e.m.}=\sum_{i=1}^2\hat{H}_i+\hat{U}_C+\hat{U}_M=\\=\sum_{i=1}^2\bigg[\frac{\hat{Q}_i^2}{2\widetilde{C}_i}+\frac{\hat{\Phi}_i^2}{2\widetilde{L}_i}+\\-E_{J,i}(\Phi_{x,i})\cos\left(2\pi\frac{\hat{\Phi}_i-\Phi_{z,i}}{\Phi_0}\right)\bigg]+\\+\frac{C_{12}\hat{Q}_1\hat{Q}_2}{C_1C_2+(C_1+C_2)C_{12}}+\frac{M_{12}\hat{\Phi}_1\hat{\Phi}_2}{L_1L_2-M_{12}^2},
\end{gathered}
\label{eq:twoq}
\end{equation}
where $\widetilde{C}_{1(2)}=C_{1(2)}+C_{12}C_{2(1)}/(C_{2(1)}+C_{12})$ and $\widetilde{L}_{1(2)}=L_{1(2)}-M_{12}^2/L_{2(1)}$\cite{ozfidan2019demonstration}.
\par Using the physical parameters given in Ref. \cite{ozfidan2019demonstration}, i.e. $C_{12}=132$fF and those in table \ref{table},
\begin{table}[b]
	\centering
	\resizebox{0.48\textwidth}{!}{\begin{tabular}{c|c|c|c|c|c}
			Qubit	&  $E_{J0,i}$ (GHz) & $C_i$ (fF) & $L_i$ (pH) & $\Phi_{x,i}/\Phi_0$ & $\Phi_{z,i}/\Phi_0$ \\
			\hline
			Q1 & $1.603\cdot10^3$ & 119.5 & 231.9 & $-0.6538$ & $1\cdot10^{-4}$ \\ 
			\hline
			Q2 & $1.568\cdot10^3$ & 116.4 & 239 & $-0.6526$ & $1\cdot10^{-4}$
	\end{tabular}}
	\caption{}
	\label{table}
\end{table}
\begin{figure*}[b!]
	\centering
	\subfloat[][Two-local Pauli coefficients calculated, as a function of $M_{12}$, using the SWT method (lines) and approximate rotation method (circles). Note that the purple line lies flat at zero and overlaps with the green one. ]{\includegraphics[width=0.48\textwidth]{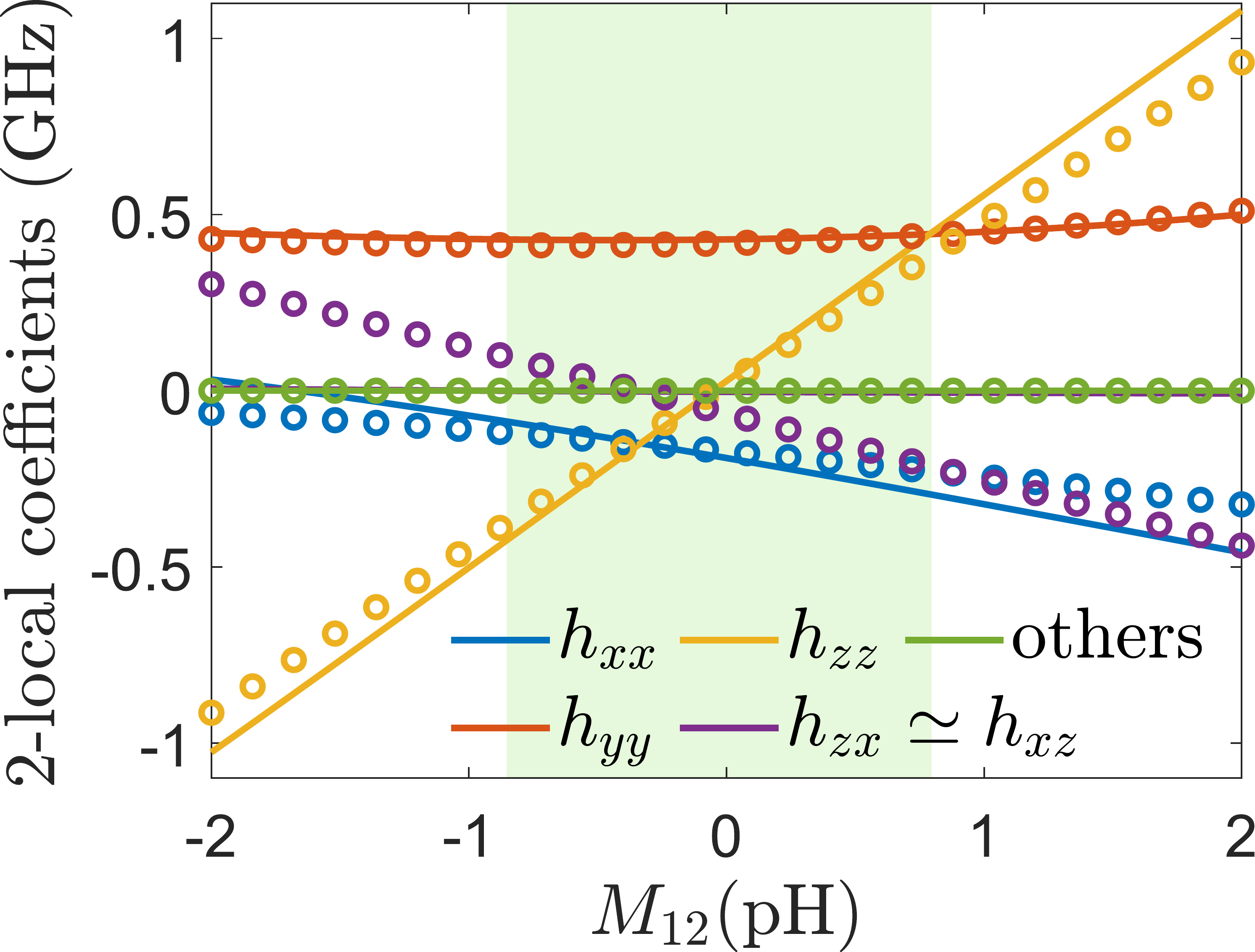}\label{fig:dwave4}}\quad
	\subfloat[][One and two-local Pauli coefficients determined with the approximate rotation method, after the application of the local rotation removing $XZ$ and $ZX$ terms (circles), compared against the ones calculated with the SWT method (solid lines). Note that the local Pauli coefficients for the two qubits overlap almost completely at this scale.]{\includegraphics[width=0.48\textwidth]{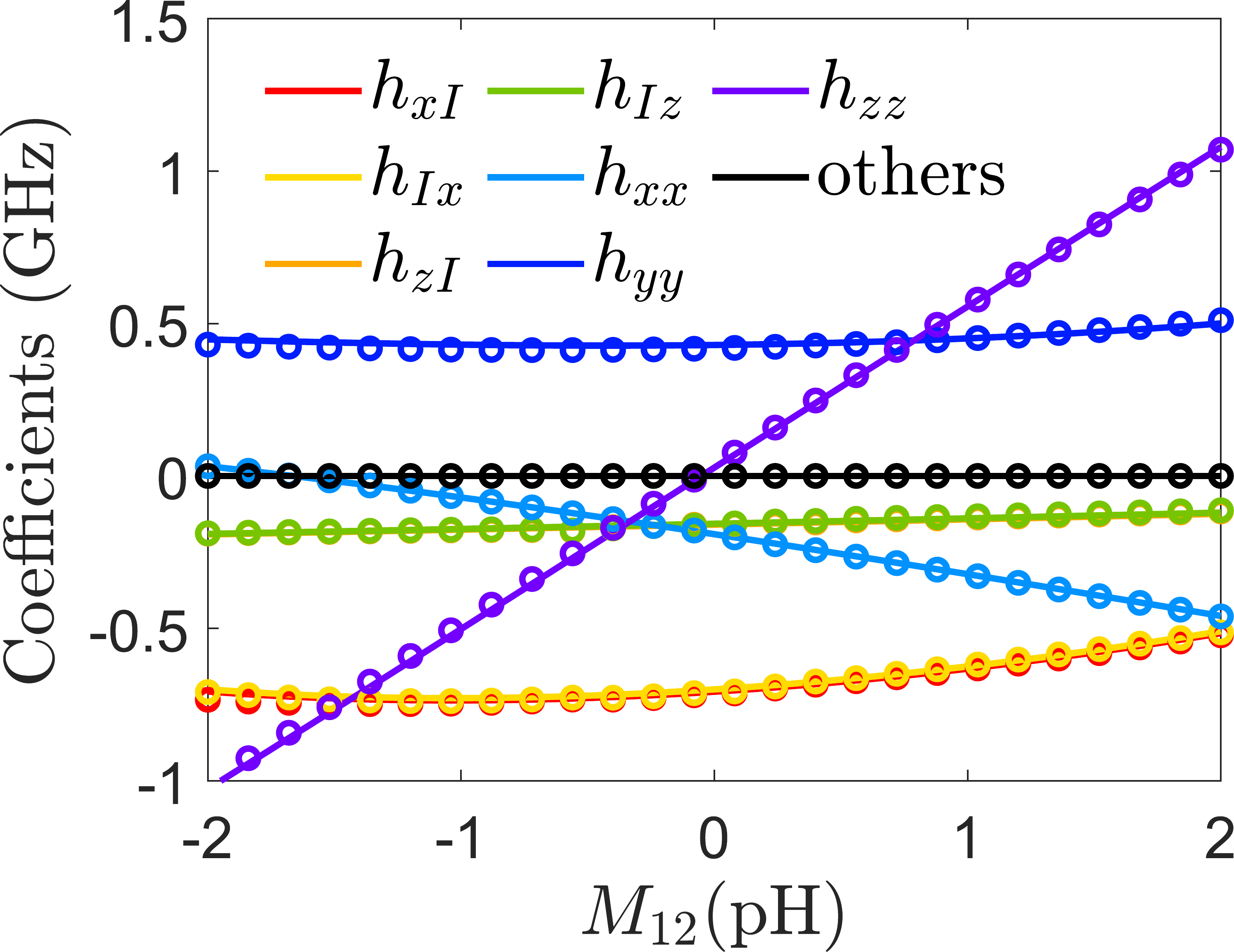}\label{fig:dwavec}}
	\caption{}
	\label{fig:dwaveb}
\end{figure*}
and calculating the lowest four eigenvalues of our Hamiltonian for different values of mutual inductance in the range $-2\textrm{pH}<M_{12}<2\textrm{pH}$, we obtained the graph shown in figure \ref{fig:dwave2}. This graph matches well with the corresponding one present in Fig. 3c of Ref. \cite{ozfidan2019demonstration}. The avoided level-crossing at $M_{12}\simeq0.7$pH is proportional to the capacitive coupling $C_{12}$ and only occurs at finite longitudinal fields, \textit{i.e.} $\Phi_{z,i}\neq0$\cite{ozfidan2019demonstration}. (Notice that when $-1<\Phi_{x,i}/\Phi_0<0$, the effective Josephson energy $E_{J,i}(\Phi_{x,i})$ is negative and the symmetry point where $h_z=0$ is displaced from $\Phi_{z,i}=\Phi_0/2$ to $\Phi_{z,i}=0$\cite{harris2009compound}.)
\par It is worth noting that, in order to efficiently represent a composite circuit Hamiltonian like \eqref{eq:twoq}, we cannot retain the representation of the circuit operators that we used for single circuits. In that case, the size of the total Hamiltonian matrix would equal the product of the sizes of all the individual circuit Hamiltonians, and would rapidly become unmanageable. Since we are, once again, only interested in the low-energy properties of the system, a good alternative basis choice is that of the outer products of some small number $N_i$ of low-energy eigenstates of each unperturbed (\textit{i.e.} non-interacting) circuit Hamiltonian $\hat{H}_i/\hat{H}_{c,i}$. In this case, for example, we can write:
\begin{equation}
\begin{aligned}
&\hat{H}_{e.m.}\simeq\\&\sum_{i,j=0}^{N_1-1}\sum_{k,l=0}^{N_2-1}\langle E_{1,i} E_{2,k}|\hat{H}_{e.m.}|E_{1,j} E_{2,l}\rangle|E_{1,i} E_{2,k}\rangle\langle E_{1,j} E_{2,l}|,
\end{aligned}
\end{equation}
with the meanings of the symbols introduced before. To ensure the convergence of our results, we first used 40 harmonic oscillator number states to represent the single qubit Hamiltonians and then projected onto their $N_1=N_2=10$ lowest-energy eigenstates.
\par Now that we have determined the low energy spectrum of the system, we can apply some reduction method to calculate the effective qubit Hamiltonian. We begin with the Schrieffer-Wolff transformation method, introduced in section \ref{SWT}. After verifying that the hypotheses of its construction are satisfied, in particular observing that $\Vert\hat{P}-\hat{P}_0\Vert_{op}\lesssim0.5$ in the whole range of $M_{12}$, we extracted the Pauli coefficients. These were calculated by defining the computational states and the Pauli operators locally for each qubit and then projecting the effective qubit Hamiltonian on them, as shown in section \ref{SWT}. The six one-local coefficients are shown in figure \ref{fig:dwave3} by solid lines. (We do not consider the coefficient $h_{II}=\textrm{Tr}(\hat{H}_q)$ here since we are focusing on relative energies.) The dashed lines represent the same coefficients obtained by applying the SWT reduction to the non-interacting part of the circuit Hamiltonian, \textit{i.e.} to the sum of the Hamiltonians of the isolated qubits (corrected for the \textit{static} inductive and capacitive loading). Since for $h_{zI}$ and $h_{Iz}$ the solid and the dashed lines overlap, the values of the longitudinal fields of the coupled system are completely determined by the static loading of the unperturbed Hamiltonians. This effect appears approximately linear in $M_{12}$. The values of the transverse fields for the coupled system, instead, are $\sim25\%$ lower in magnitude than those resulting from the loaded single-qubit Hamiltonians. The interaction with the other qubit, then, has an additional effect, which we call \textit{dynamic loading}. The change in transverse field appears approximately quadratic in $M_{12}$ and is not centred around $M_{12}=0$ due to the presence of the capacitive coupling (as we verified by comparing against the case $C_{12}=0$). As usual, the components of the local field along the \textit{y} direction have been removed by making the appropriate local gauge transformation. (Actually, the circuit Hamiltonian in this case is completely real, so that no imaginary terms can appear in the reduced Hamiltonian; the gauge transformation only ensures that the signs of different coefficients are consistent across the range of $M_{12}$.)
\par The empty circles in figure \ref{fig:dwave3} are the one-local Pauli coefficients determined with the approximate rotation method, introduced in \cite{ozfidan2019demonstration} and reviewed in section \ref{approt}.
Comparing with the previous results, we can see that we obtain qualitatively similar, but quantitatively different results. In particular the values for the transverse fields are close to those obtained with the SWT reduction, while the new longitudinal fields are everywhere smaller in magnitude, and, in this case, do not agree with their unperturbed values (dashed lines). 
\par Figure \ref{fig:dwave4} shows the coefficients of the nine effective qubit Hamiltonian two-local terms. According to the reduction based on the SWT (lines), the only non-negligible terms in the Hamiltonian are those proportional to $\pmb{\sigma}_{z,1}\pmb{\sigma}_{z,2}$, $\pmb{\sigma}_{x,1}\pmb{\sigma}_{x,2}$ and $\pmb{\sigma}_{y,1}\pmb{\sigma}_{y,2}$. The first term represents the inductive interaction, $\hat{U}_M\propto M_{12}\hat{\Phi}_1\hat{\Phi}_2$, the flux being our \textit{z} degree of freedom, and it indeed scales linearly with $M_{12}$. Since we have chosen to identify a flux degree of freedom with the real operator $\pmb{\sigma}_z$, the canonically conjugate charge operator must be complex (since $[\hat{\Phi},\hat{Q}]=i\hbar$), and therefore must be identified with $\pmb{\sigma}_y$. The \textit{YY} term, then, describes the capacitive interaction and, in fact, appears to be largely independent of $M_{12}$. Finally the $XX$ term is a result of the presence of the higher excited states of the system\cite{ozfidan2019demonstration}. It is related to both the inductive and the capacitive Hamiltonian terms and appears to scale linearly with $M_{12}$.
\par According to reference \cite{klassen2019two}, a two-local two-qubit Hamiltonian of the form
\begin{equation}
\begin{gathered}
\mathbf{H}=h_{xI}\pmb{\sigma}_{x,1}+h_{Ix}\pmb{\sigma}_{x,2}+h_{zI}\pmb{\sigma}_{z,1}+h_{Iz}\pmb{\sigma}_{z,2}+\\+h_{xx}\pmb{\sigma}_{x,1}\pmb{\sigma}_{x,2}+h_{yy}\pmb{\sigma}_{y,1}\pmb{\sigma}_{y,2}+h_{zz}\pmb{\sigma}_{z,1}\pmb{\sigma}_{z,2}
\end{gathered}
\end{equation}
is \textit{non-stoquastic}, and remains such after arbitrary local rotations, as long as $h_{xI},h_{Ix},h_{zI},h_{Iz}\neq0$ and $|h_{yy}|>|h_{xx}|,|h_{zz}|$. The region where this condition is satisfied is highlighted in green in figure \ref{fig:dwave4}. Non-stoquastic two-local catalyst Hamiltonians are know to provide an exponential speed-up to the convergence of quantum adiabatic optimisation, at least with specific problem classes, including the ferromagnetic \textit{p}-spin model\cite{nishimori2017exponential}. For this reason, they might be key to establish a quantum advantage over classical optimisation routines such as Quantum Monte Carlo\cite{ozfidan2019demonstration,albash2019role}.
\par Again, our implementation of the approximate rotation reduction produces qualitatively similar results to the SWT reduction for the two-local Pauli coefficients (see hollow circles in figure \ref{fig:dwave4}), except for $h_{xz}\simeq h_{zx}$ (purple circles), which are now of the same order of magnitude as the other coefficients. As we mentioned in section \ref{SWT}, the approximate rotation and the SWT reduction methods actually find equivalent effective qubit Hamiltonians, modulo a unitary. This was in fact verified by showing that both sets of coefficients lead to qubit Hamiltonians with the same spectrum.
\par Notice that Ref. \cite{ozfidan2019demonstration} actually reports the two $h_{xz}\simeq h_{zx}$ coefficients to be negligible, which we ascribe to the fact that the authors used a different form for the circuit Hamiltonian, and potentially a different definition of the computational basis, and hence of $\mathbf{R}_2$, as defined in section \ref{approt}\cite{ozfidan2019demonstration}. (In our case the computational basis was defined locally as shown in section \ref{localb}.) In fact, any mixed two-local term, involving different Pauli operators acting on the two qubits, can be eliminated from a two-qubit Hamiltonian by performing a local change of basis\cite{klassen2019two}. Applying this transformation produces a new set of coefficients which are within 5\% of those found by the SWT reduction method (see Fig. \ref{fig:dwavec}). In this case, then, the unitary mapping between the two is a local transformation.
\par We conclude the subsection on this two-qubit system by briefly considering how, in analogy to what we had in the single-qubit case, the reduced Hamiltonian not only contains information about the system low-energy spectrum, but also about state probabilities (as well as operator matrix elements). For instance, when we set $M_{12}=2$ pH, the SWT reduction produces the following effective qubit Hamiltonian:
\begin{equation}
\begin{aligned}
\mathbf{H}_q=&-0.125\pmb{\sigma}_{z,1}-0.121\pmb{\sigma}_{z,2}-0.516\pmb{\sigma}_{x,1}-0.509\pmb{\sigma}_{x,2}+\\
&-0.459\pmb{\sigma}_{x,1}\pmb{\sigma}_{x,2}+0.500\pmb{\sigma}_{y,1}\pmb{\sigma}_{y,2}+1.079\pmb{\sigma}_{z,1}\pmb{\sigma}_{z,2}.
\end{aligned}
\end{equation}
One can easily find that the first excited state of this Hamiltonian is $\ket{\mathcal{E}_1}=-0.0046\ket{00}+0.7041\ket{01}-0.7101\ket{10}-0.0012\ket{11}$, i.e. an entangled state where the two qubits are in opposite computational states with probability approximately one (i.e. $p(q_1=0|q_2=1)=\dots\simeq1$). This should translate to the fact that, at the circuit level, there is a high probability of measuring currents of opposite sign on the two qubits, when the system is in its first excited state. In other words, if the persistent current of one of the qubits is measured to be positive, the other qubit is projected on its negative persistent current state, and vice versa. We can verify that this is actually the case by using the projectors on the positive and negative subspaces of the qubit current operators and calculating their expectation value on the first excited state $\ket{E_1}$ of the circuit Hamiltonian. Table \ref{table1} gives the probabilities of measuring the different computational states on $\ket{\mathcal{E}_1}$ and different current sign combinations on $\ket{E_1}$. The two results are in good agreement.
\begin{table}[h]
	\centering
	\begin{tabular}{r|c|c|c|c}
		\multirow{2}{*}{Model}  & \multicolumn{4}{c}{Probabilities}\\
		 &  $p_{00}/p_{++}$ & $p_{01}/p_{+-}$ & $p_{10}/p_{-+}$ & $p_{11}/p_{--}$ \\
		\hline
		Qubit & $2\cdot10^{-5}$ & 0.50 & 0.50 & $1.3\cdot10^{-6}$  \\ 
		\hline
		Circuit & 0.06 & 0.44 & 0.45 & 0.05
	\end{tabular}
	\caption{}
	\label{table1}
\end{table}
%
\begin{figure*}[t!]
	\centering
	\subfloat[][Circuit diagram of three flux qubits and the three-local $ZZZ$ interaction circuit, described in \cite{melanson2019tunable}, consisting of two compound-Josephson-junction rf-SQUID tunable magnetic couplers, one of which, $c_2$, has a twist in the main loop.]{\includegraphics[height=0.35\textwidth]{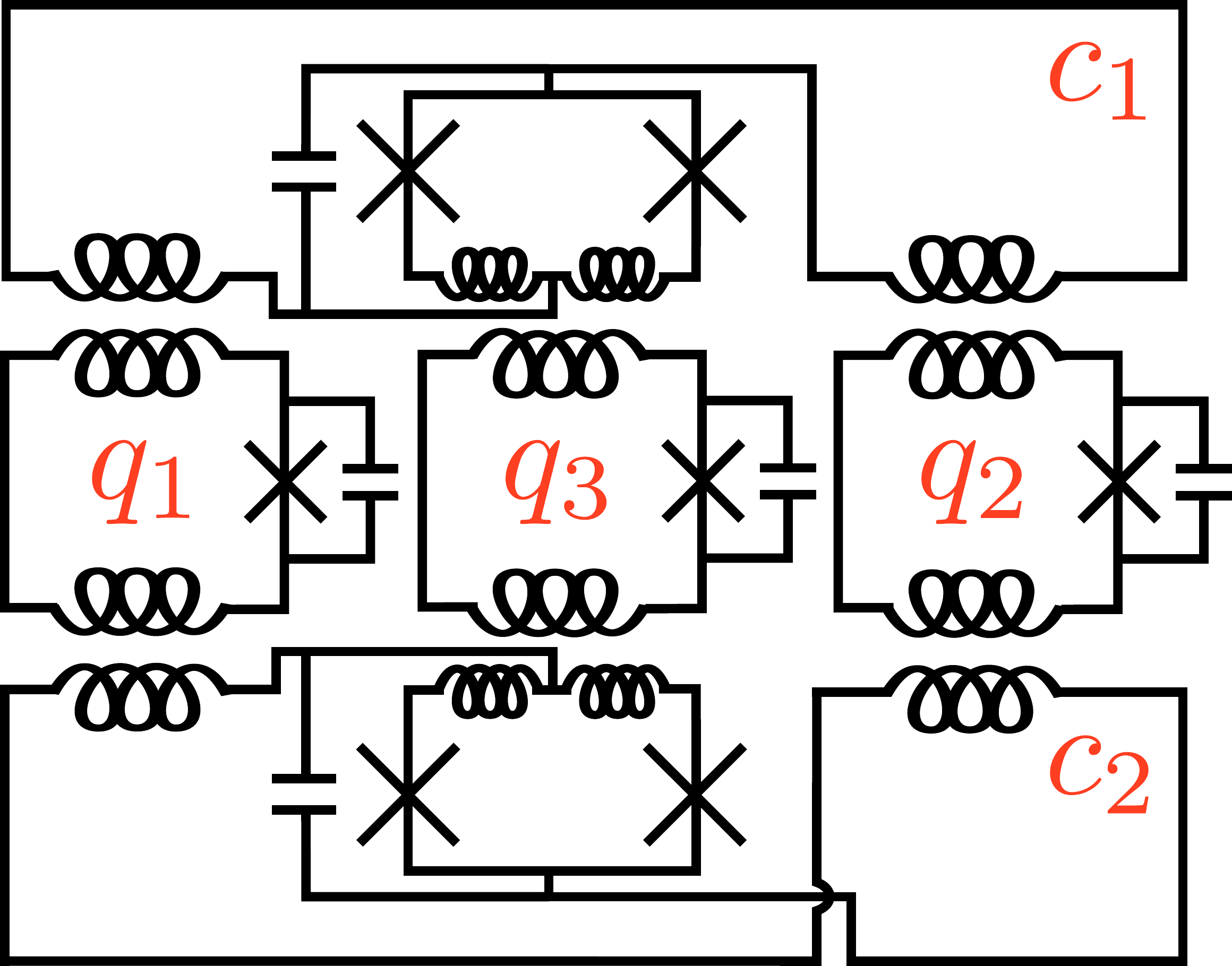}\label{fig:UW1}}\quad
	\subfloat[][Solid lines: Pauli coefficients extracted, using the SWT-based reduction method, for the system of three qubits and coupler $c_1$, as a function of $f_{x,c1}$. (Note that $c_2$ is absent here.) Filled circles: same coefficients, extracted using the diagonal Hamiltonian reduction method.]{\includegraphics[height=0.35\textwidth]{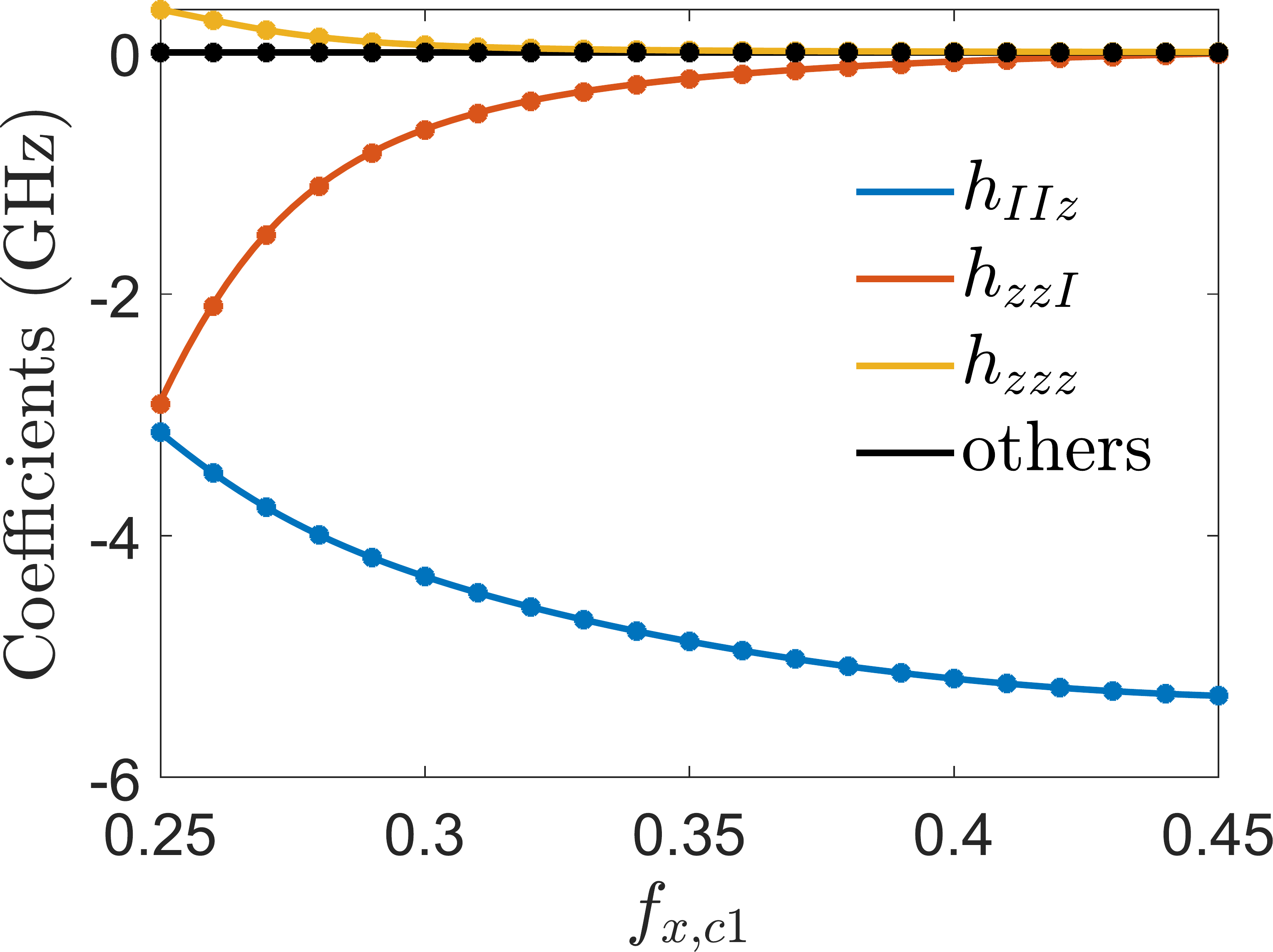}\label{fig:UW2}}\\
	\subfloat[][Pauli coefficients numerically extracted for the circuit in Fig. \ref{fig:UW1}. Solid lines: SWT-based reduction method. Filled circles: diagonal Hamiltonian reduction method.]{\includegraphics[height=0.35\textwidth]{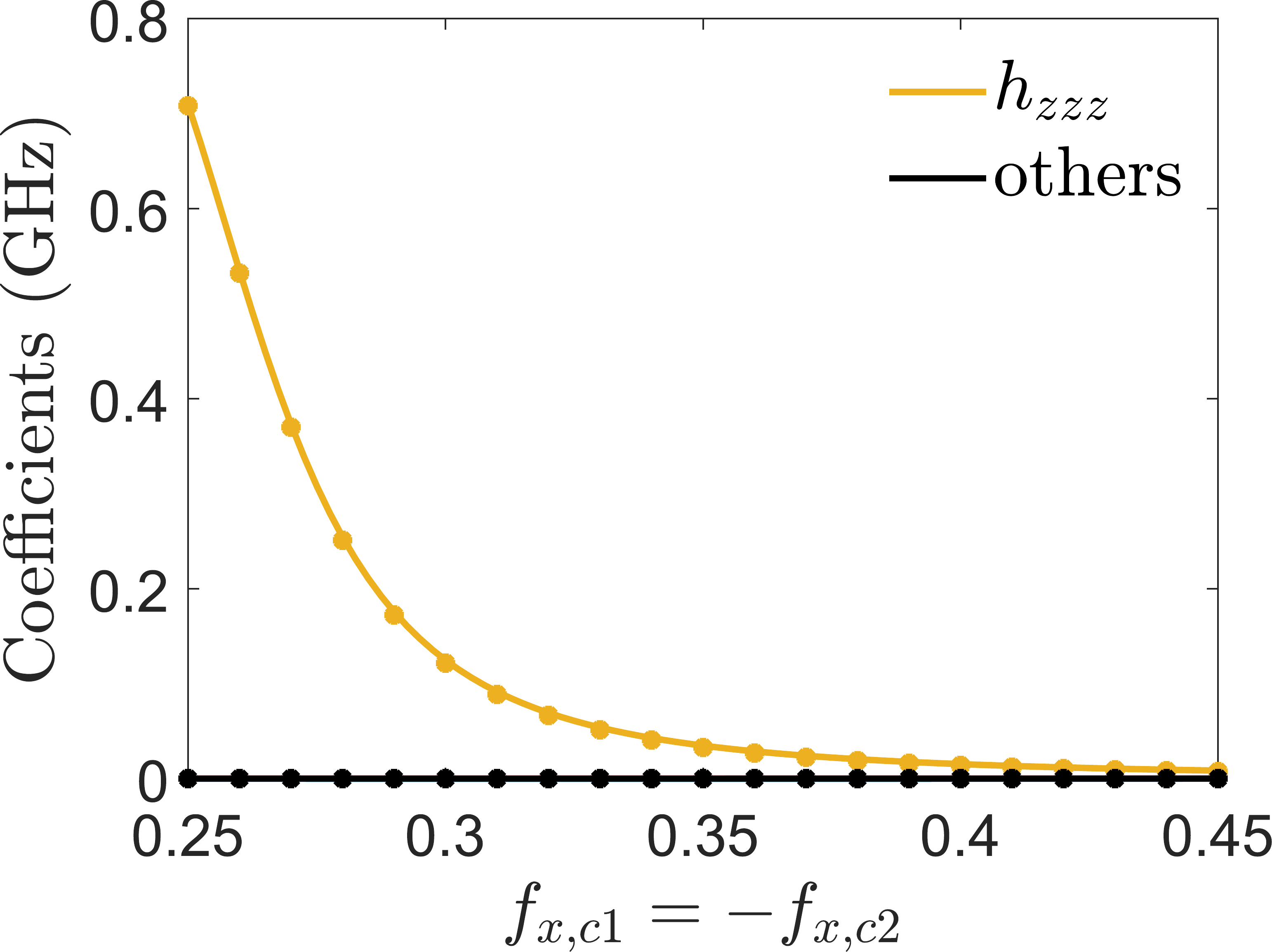}\label{fig:UWb1}}\quad
	\subfloat[][Energy spectrum, relative to the ground state, of the circuit in Fig. \ref{fig:UW1}. Solid lines: circuit Hamiltonian. Filled circles: effective qubit Hamiltonian. ]{\includegraphics[height=0.35\textwidth]{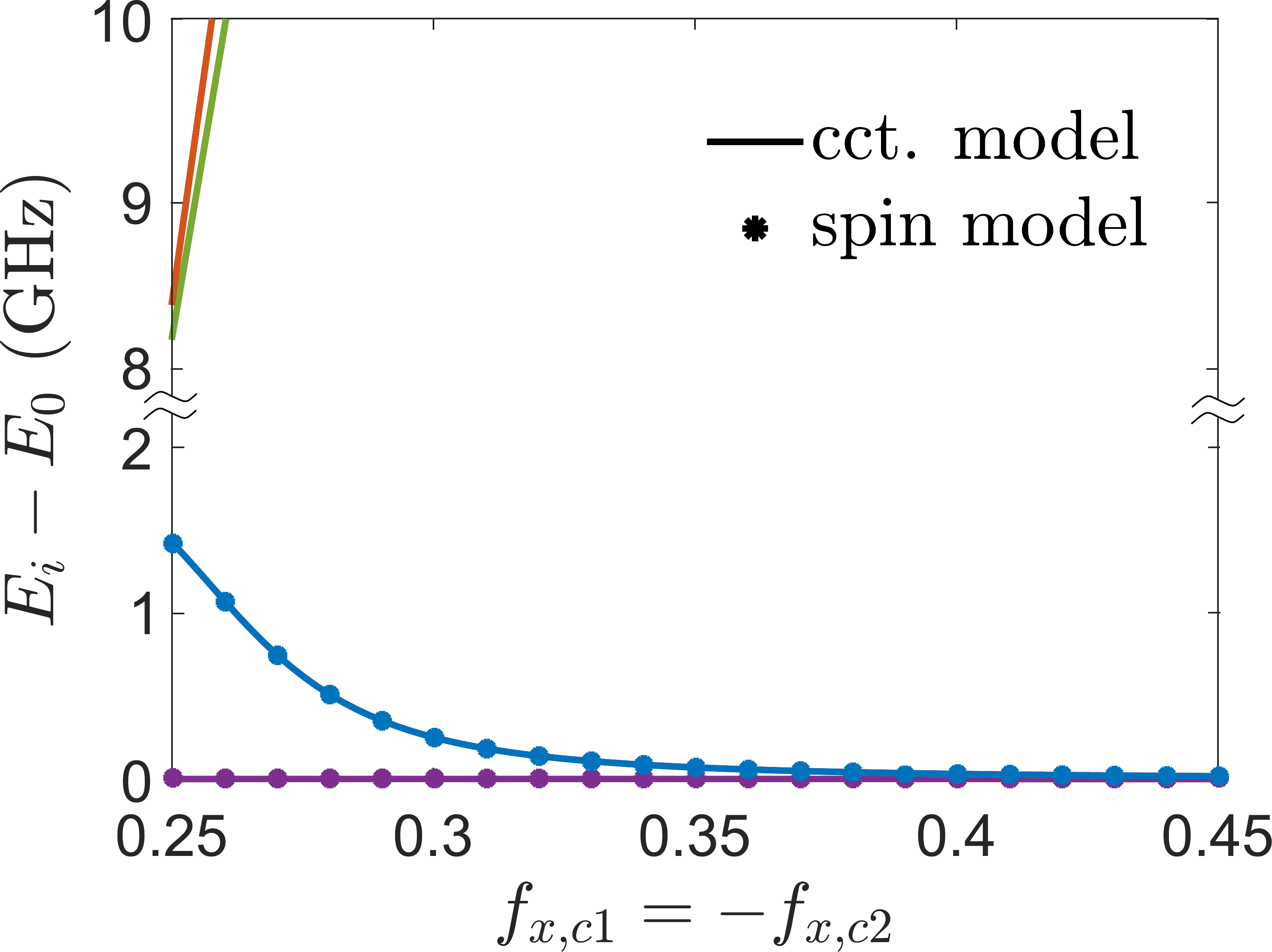}\label{fig:UWb2}}
	\caption{}
	\label{fig:UWb}
\end{figure*}
\subsubsection{ZZZ coupling}
\par As the final example we consider a proposed circuit implementing a three-local $ZZZ$ interaction between three flux qubits, presented in \cite{melanson2019tunable}. The circuit diagram is shown in figure \ref{fig:UW1} and consists of the three flux qubits (in this case rf-SQUID qubits) and two compound-Josephson-junction rf-SQUID couplers. The main loops of the two couplers, one of which contains a twist, mediate a magnetic interaction between the superconducting loops of qubits $q_1$ and $q_2$ (see Fig. \ref{fig:UW1}). If the flux applied to the coupler main loop, $\Phi_{z,ci}$ is kept constant, the flux applied to its dc-SQUID loop, $\Phi_{x,ci}$, controls the effective mutual inductance between the qubits and therefore the magnitude and sign of the effective $ZZ$ interaction\cite{harris2009compound}. By magnetically coupling the current loop of qubit $q_3$ to the coupler dc-SQUID loop, one can control the two local interaction between $q_1$ and $q_2$ with the current state of $q_3$, therefore obtaining a three-local $h_{zzz}\hat{\sigma}_{z1}\hat{\sigma}_{z2}\hat{\sigma}_{z3}$ interaction\cite{melanson2019tunable}.
\par The solid lines in figure \ref{fig:UW2} show the effective Hamiltonian coefficients for the system consisting of the three flux qubits and the single coupler $c_1$, extracted using the SWT reduction method. The main loop of the coupler and those of the three qubits are all biased at $\Phi_{z,c1}=\Phi_{z,i}=\Phi_0/2$, such that the qubit longitudinal fields are all zero. The transverse fields are also zero with the physical parameters considered (which are given below). As expected, we find a three-local interaction term $\propto h_{zzz}$, in addition to a residual two-local interaction between qubits $q_1$ and $q_2$, $\propto h_{zzI}$ and a large longitudinal field $h_{IIz}$ on qubit $q_3$.
\par The parameters used in the simulations are as follows: all qubits ($i=1,2,3$) have $E_{J,i}=99.3$GHz, $L_{q,i}=4.5$nH and a large shunting capacitance $C_{sh,i}=45$fF; the two coupler junction Josephson energies are $E_{J1,c1}=E_{J2,c1}=233.4$GHz, the coupler main loop inductance is $L_{z,c1}=550$pH, while the small loop has an inductance of $L_{x,c1}=170$pH and is shunted by a capacitance $C_{sh,c1}=10$fF; all mutual inductances are 50pH. As in the previous simulations, the rf-SQUID qubit Hamiltonians have been expressed in  a basis of 40 occupation number states. The three degrees of freedom of the coupler are expressed using 20 occupation number states for the small \textit{plasma frequency} mode and 7 for the higher plasma frequency modes. The total Hamiltonian is projected on the lowest 8 unperturbed eigenstates of each qubit and on the lowest 5 unperturbed coupler eigenstates. Since the effective Hamiltonian here is diagonal in the computational basis, its coefficients can also be calculated with the method used in \cite{melanson2019tunable} and reviewed in section \ref{diag}. The result of this reduction is represented by the filled dots in figure \ref{fig:UW2} and matches very well with the result of the SWT reduction.
\par Introducing a twist in the coupler, for instance changing the mutual inductance between the coupler and qubit $q_2$ from 50pH to $-50$pH (as in coupler $c_2$), and changing the sign of the coupler \textit{x}-bias, reverses the sign not only of the two-local coefficient $h_{zzI}$, but also of $h_{IIz}$. The three-local interaction coefficient, however, remains of the same sign. Therefore attaching both couplers $c_1$ and $c_2$ to the qubits leaves us with a purely three-local Hamiltonian. The numerical simulation of the full system agrees with this picture. The Pauli coefficients extracted, as a function of $f_{x,c1}=-f_{x,c2}$, are shown in figure \ref{fig:UWb1}, with the solid lines and the dots being the result of the SWT and the diagonal Hamiltonian reduction method, respectively. Coupler $c_2$ shares the same physical parameters as $c_1$ and is also biased at $f_{z,c2}=0.5$. Its lowest 5 unperturbed eigenstates are kept for representing the full system Hamiltonian. As we can see, the size of the three-local $ZZZ$ interaction can be changed from zero to as much as 700MHz in the range of fluxes considered. Its sign can also be changed to negative by biasing at $f_{x,c1}'=-f_{x,c2}'=2-f_{x,c1}$\cite{melanson2019tunable}.
\par Finally we can check that the reduced Hamiltonian has the correct spectrum. This is shown in figure \ref{fig:UWb2}, where the filled dots represent the effective qubit Hamiltonian transition energies and the solid lines those of the circuit Hamiltonian. The levels are grouped in two manifolds each of four degenerate levels, separated by an energy of $2\vert h_{zzz}\vert$. In the ground state manifold the expectation value of the product of the qubit currents, $\langle\hat{I}_1\hat{I}_2\hat{I}_3\rangle$, and therefore  $\langle\hat{\sigma}_{z1}\hat{\sigma}_{z2}\hat{\sigma}_{z3}\rangle$ in the reduced model, is negative, while it is positive in the excited manifold states. At energies above 8GHz we see the additional states of the system, specifically the first excited states of the couplers. As we can see, the interaction does not close the spectral gap of the Hamiltonian, which allows us to use the Schrieffer-Wolff transformation reduction method.
\section{Conclusions}
We have developed a systematic numerical method for determining the effective spin Hamiltonian, written in the appropriate computational basis, describing a system of interacting superconducting circuits. Our starting point was a numerical representation of the circuit Hamiltonian, in which each component is described as a lumped-element circuit, with potential magnetic and electrostatic biases, and interacts with the other components through mutual inductive or electrostatic interactions.
\par Comparison with other reduction approaches in the literature and self-consistency checks on the system spectrum allowed us to demonstrate the validity of our reduced model. At the same time, our approach is based on more general assumptions than other reduction methods in the literature. Therefore, in the case of isolated superconducting qubits we have seen that choosing the local computational basis with explicit reference to the measurement operator improves the accuracy of the reduced Hamiltonian, in terms of both the spectrum and expectation values of circuit operators. This is especially true for qubit designs with reduced anharmonicity, such as the capacitively-shunted flux qubit. In the multiple-qubit case, the Schrieffer-Wolff transformation theory provided the basis for calculating the effective spin Hamiltonian, the only requirement for its application being that the size of the spectral gap of the unperturbed Hamiltonian should be larger than the size of the interaction. In principle this limitation can be circumvented, as long as one is able to partition the system in smaller units, and as long as the qubits in each unit display sufficient anharmonicity. Numerical calculations of the effective multiple-qubit Hamiltonians provided results in good agreement with the existing reduction methods, when these were used within their range of applicability.
\par This reduction method should prove useful in different areas of applied quantum computation, where complex systems of continuous variable circuits are described in terms of interacting two-level systems. In practice one could start by fitting the parameters in the circuit model to some preliminary data, then extract the effective qubit Hamiltonian as a function of the control biases. The reduced model could then be verified with additional experiments, for instance spectroscopic or state population oscillation measurements, and successively be employed as the reference model for the operation of the system\cite{ozfidan2019demonstration}. In the context of circuit design this method can be used to model the interplay between different qubit Hamiltonian terms, for instance the effect of the coupler bias on the qubit transverse fields\cite{harris2010experimental} (i.e. dynamic inductive loading), or to predict the size of non-Ising terms like non-stoquastic or many-body interactions (as well as of Ising terms like the transverse fields, beyond the instanton approximation).\\
\linebreak
{\Large \textbf{Acknowledgements:}}
\\We thank T. Albash, M. Amin, A. Ciani, D. Ferguson, L. Fry-Bouriaux, I. Hen, A. J. Kerman, M. Khezri, J. Klassen, A. Lupascu, M. Marvian, D. Melanson, T. Menke, S. Novikov, I. Ozfidan, K. E. Porsch, M. Sch\"{o}ndorf and P. Schuhmacher for useful input. This material is based upon work supported by the Intelligence Advanced Research Projects Activity (IARPA) and the Army Research Office (ARO) under Contract No. W911NF-17-C-0050. Any opinions, findings and conclusions or recommendations expressed in this material are
those of the author(s) and do not necessarily reflect the views of the Intelligence Advanced Research Projects Activity (IARPA) and the Army Research Office (ARO). G.C. acknowledges the support of the EPSRC Centre for Doctoral Training in Delivering Quantum Technologies (grant ref: EP/L015242/1)
\bibliographystyle{unsrturl}
\bibliography{bibliography}

\begin{thebibliography}{10}

\bibitem{yurke1984quantum}
Bernard Yurke and John~S Denker.
\newblock Quantum network theory.
\newblock {\em Physical Review A}, 29(3):1419, 1984.
\newblock \href {http://dx.doi.org/10.1103/PhysRevA.29.1419}
  {\path{doi:10.1103/PhysRevA.29.1419}}.

\bibitem{reiter2012effective}
Florentin Reiter and Anders~S S{\o}rensen.
\newblock Effective operator formalism for open quantum systems.
\newblock {\em Physical Review A}, 85(3):032111, 2012.
\newblock \href {http://dx.doi.org/10.1103/PhysRevA.85.032111}
  {\path{doi:10.1103/PhysRevA.85.032111}}.

\bibitem{amin2012approximate}
Mohammad~H Amin, Anatoly~Yu Smirnov, Neil~G Dickson, and Marshall Drew-Brook.
\newblock Approximate diagonalization method for large-scale hamiltonians.
\newblock {\em Physical Review A}, 86(5):052314, 2012.
\newblock \href {http://dx.doi.org/10.1103/PhysRevA.86.052314}
  {\path{doi:10.1103/PhysRevA.86.052314}}.

\bibitem{wendin2017quantum}
G~Wendin.
\newblock Quantum information processing with superconducting circuits: a
  review.
\newblock {\em Reports on Progress in Physics}, 80(10):106001, 2017.
\newblock URL:
  \url{http://iopscience.iop.org/article/10.1088/1361-6633/aa7e1a}.

\bibitem{devoret2013superconducting}
Michel~H Devoret and Robert~J Schoelkopf.
\newblock Superconducting circuits for quantum information: an outlook.
\newblock {\em Science}, 339(6124):1169--1174, 2013.
\newblock \href {http://dx.doi.org/10.1126/science.1231930}
  {\path{doi:10.1126/science.1231930}}.

\bibitem{clarke2008superconducting}
John Clarke and Frank~K Wilhelm.
\newblock Superconducting quantum bits.
\newblock {\em Nature}, 453(7198):1031, 2008.
\newblock \href {http://dx.doi.org/10.1038/nature07128}
  {\path{doi:10.1038/nature07128}}.

\bibitem{krantz2019quantum}
Philip Krantz, Morten Kjaergaard, Fei Yan, Terry~P Orlando, Simon Gustavsson,
  and William~D Oliver.
\newblock A quantum engineer's guide to superconducting qubits.
\newblock {\em Applied Physics Reviews}, 6(2):021318, 2019.
\newblock \href {http://dx.doi.org/10.1063/1.5089550}
  {\path{doi:10.1063/1.5089550}}.

\bibitem{paraoanu2014recent}
GS~Paraoanu.
\newblock Recent progress in quantum simulation using superconducting circuits.
\newblock {\em Journal of Low Temperature Physics}, 175(5-6):633--654, 2014.
\newblock \href {http://dx.doi.org/10.1007/s10909-014-1175-8}
  {\path{doi:10.1007/s10909-014-1175-8}}.

\bibitem{orlando1999superconducting}
TP~Orlando, JE~Mooij, Lin Tian, Caspar~H van~der Wal, LS~Levitov, Seth Lloyd,
  and JJ~Mazo.
\newblock Superconducting persistent-current qubit.
\newblock {\em Physical Review B}, 60(22):15398, 1999.
\newblock \href {http://dx.doi.org/10.1103/PhysRevB.60.15398}
  {\path{doi:10.1103/PhysRevB.60.15398}}.

\bibitem{arute2019quantum}
Frank Arute, Kunal Arya, Ryan Babbush, Dave Bacon, Joseph~C Bardin, Rami
  Barends, Rupak Biswas, Sergio Boixo, Fernando~GSL Brandao, David~A Buell,
  et~al.
\newblock Quantum supremacy using a programmable superconducting processor.
\newblock {\em Nature}, 574(7779):505--510, 2019.
\newblock \href {http://dx.doi.org/10.1038/s41586-019-1666-5}
  {\path{doi:10.1038/s41586-019-1666-5}}.

\bibitem{hauke2019perspectives}
Philipp Hauke, Helmut~G Katzgraber, Wolfgang Lechner, Hidetoshi Nishimori, and
  William~D Oliver.
\newblock Perspectives of quantum annealing: Methods and implementations.
\newblock {\em arXiv preprint arXiv:1903.06559}, 2019.
\newblock URL: \url{https://arxiv.org/abs/1903.06559}.

\bibitem{koch2007charge}
Jens Koch, Terri~M. Yu, Jay Gambetta, A.~A. Houck, D.~I. Schuster, J.~Majer,
  Alexandre Blais, M.~H. Devoret, S.~M. Girvin, and R.~J. Schoelkopf.
\newblock Charge-insensitive qubit design derived from the cooper pair box.
\newblock {\em Phys. Rev. A}, 76:042319, 2007.
\newblock \href {http://dx.doi.org/10.1103/PhysRevA.76.042319}
  {\path{doi:10.1103/PhysRevA.76.042319}}.

\bibitem{yan2016flux}
Fei Yan, Simon Gustavsson, Archana Kamal, Jeffrey Birenbaum, Adam~P Sears,
  David Hover, Ted~J Gudmundsen, Danna Rosenberg, Gabriel Samach, Steven Weber,
  et~al.
\newblock The flux qubit revisited to enhance coherence and reproducibility.
\newblock {\em Nature communications}, 7:12964, 2016.
\newblock \href {http://dx.doi.org/10.1038/ncomms12964}
  {\path{doi:10.1038/ncomms12964}}.

\bibitem{harris2010experimental}
R.~Harris, J.~Johansson, A.~J. Berkley, M.~W. Johnson, T.~Lanting, Siyuan Han,
  P.~Bunyk, E.~Ladizinsky, T.~Oh, I.~Perminov, E.~Tolkacheva, S.~Uchaikin,
  E.~M. Chapple, C.~Enderud, C.~Rich, M.~Thom, J.~Wang, B.~Wilson, and G.~Rose.
\newblock Experimental demonstration of a robust and scalable flux qubit.
\newblock {\em Phys. Rev. B}, 81:134510, Apr 2010.
\newblock \href {http://dx.doi.org/10.1103/PhysRevB.81.134510}
  {\path{doi:10.1103/PhysRevB.81.134510}}.

\bibitem{nakamura1999coherent}
Yu~Nakamura, Yu~A Pashkin, and JS~Tsai.
\newblock Coherent control of macroscopic quantum states in a
  single-cooper-pair box.
\newblock {\em nature}, 398(6730):786, 1999.
\newblock URL: \url{https://www.nature.com/articles/19718}.

\bibitem{makhlin2001quantum}
Yuriy Makhlin, Gerd Sch{\"o}n, and Alexander Shnirman.
\newblock Quantum-state engineering with josephson-junction devices.
\newblock {\em Reviews of modern physics}, 73(2):357, 2001.
\newblock \href {http://dx.doi.org/10.1103/RevModPhys.73.357}
  {\path{doi:10.1103/RevModPhys.73.357}}.

\bibitem{boixo2016computational}
Sergio Boixo, Vadim~N Smelyanskiy, Alireza Shabani, Sergei~V Isakov, Mark
  Dykman, Vasil~S Denchev, Mohammad~H Amin, Anatoly~Yu Smirnov, Masoud Mohseni,
  and Hartmut Neven.
\newblock Computational multiqubit tunnelling in programmable quantum
  annealers.
\newblock {\em Nature communications}, 7:10327, 2016.
\newblock \href {http://dx.doi.org/10.1038/ncomms10327}
  {\path{doi:10.1038/ncomms10327}}.

\bibitem{ozfidan2019demonstration}
Isil Ozfidan, Chunqing Deng, AY~Smirnov, T~Lanting, R~Harris, L~Swenson,
  J~Whittaker, F~Altomare, M~Babcock, C~Baron, et~al.
\newblock Demonstration of nonstoquastic hamiltonian in coupled superconducting
  flux qubits.
\newblock {\em arXiv preprint arXiv:1903.06139}, 2019.
\newblock URL: \url{https://arxiv.org/abs/1903.06139}.

\bibitem{albash2018adiabatic}
Tameem Albash and Daniel~A Lidar.
\newblock Adiabatic quantum computation.
\newblock {\em Reviews of Modern Physics}, 90(1):015002, 2018.
\newblock \href {http://dx.doi.org/10.1103/RevModPhys.90.015002}
  {\path{doi:10.1103/RevModPhys.90.015002}}.

\bibitem{jordan2006error}
Stephen~P Jordan, Edward Farhi, and Peter~W Shor.
\newblock Error-correcting codes for adiabatic quantum computation.
\newblock {\em Physical Review A}, 74(5):052322, 2006.
\newblock \href {http://dx.doi.org/10.1103/PhysRevA.74.052322}
  {\path{doi:10.1103/PhysRevA.74.052322}}.

\bibitem{friedman2002aharonov}
Jonathan~R Friedman and Dmitri~V Averin.
\newblock Aharonov-casher-effect suppression of macroscopic tunneling of
  magnetic flux.
\newblock {\em Physical review letters}, 88(5):050403, 2002.
\newblock \href {http://dx.doi.org/10.1103/PhysRevLett.88.050403}
  {\path{doi:10.1103/PhysRevLett.88.050403}}.

\bibitem{rajaraman1989solitons}
Ramamurti Rajaraman.
\newblock {\em Solitons and Instantons: An Introduction to Solitons and
  Instantons in Quantum Field Theory}.
\newblock Elsevier, 1982.

\bibitem{bravyi2011schrieffer}
Sergey Bravyi, David~P DiVincenzo, and Daniel Loss.
\newblock Schrieffer--wolff transformation for quantum many-body systems.
\newblock {\em Annals of physics}, 326(10):2793--2826, 2011.
\newblock \href {http://dx.doi.org/10.1016/j.aop.2011.06.004}
  {\path{doi:10.1016/j.aop.2011.06.004}}.

\bibitem{zagoskin2011quantum}
Alexandre~M Zagoskin.
\newblock {\em Quantum engineering: theory and design of quantum coherent
  structures}.
\newblock Cambridge University Press, 2011.

\bibitem{devoret2004superconducting}
Michel~H Devoret, Andreas Wallraff, and John~M Martinis.
\newblock Superconducting qubits: A short review.
\newblock {\em arXiv preprint cond-mat/0411174}, 2004.
\newblock URL: \url{https://arxiv.org/abs/cond-mat/0411174}.

\bibitem{wendin2005superconducting}
G{\"o}ran Wendin and VS~Shumeiko.
\newblock Superconducting quantum circuits, qubits and computing.
\newblock {\em arXiv preprint cond-mat/0508729}, 2005.
\newblock URL: \url{https://arxiv.org/pdf/cond-mat/0508729.pdf}.

\bibitem{breuer2002theory}
Heinz-Peter Breuer, Francesco Petruccione, et~al.
\newblock {\em The theory of open quantum systems}.
\newblock Oxford University Press on Demand, 2002.

\bibitem{cubitt2018universal}
Toby~S Cubitt, Ashley Montanaro, and Stephen Piddock.
\newblock Universal quantum hamiltonians.
\newblock {\em Proceedings of the National Academy of Sciences},
  115(38):9497--9502, 2018.
\newblock \href {http://dx.doi.org/10.1073/pnas.1804949115}
  {\path{doi:10.1073/pnas.1804949115}}.

\bibitem{friedman2000quantum}
Jonathan~R Friedman, Vijay Patel, Wei Chen, SK~Tolpygo, and James~E Lukens.
\newblock Quantum superposition of distinct macroscopic states.
\newblock {\em nature}, 406(6791):43, 2000.
\newblock \href {http://dx.doi.org/10.1038/35017505}
  {\path{doi:10.1038/35017505}}.

\bibitem{shimazu2009four}
Y~Shimazu, Y~Saito, and Z~Wada.
\newblock Four-josephson-junction flux qubit with controllable energy gap.
\newblock In {\em Journal of Physics: Conference Series}, volume 150, page
  022075. IOP Publishing, 2009.
\newblock \href {http://dx.doi.org/10.1088/1742-6596/150/2/022075}
  {\path{doi:10.1088/1742-6596/150/2/022075}}.

\bibitem{vinci2017non}
Walter Vinci and Daniel~A Lidar.
\newblock Non-stoquastic hamiltonians in quantum annealing via geometric
  phases.
\newblock {\em npj Quantum Information}, 3(1):38, 2017.
\newblock \href {http://dx.doi.org/10.1038/s41534-017-0037-z}
  {\path{doi:10.1038/s41534-017-0037-z}}.

\bibitem{shankar2012principles}
Ramamurti Shankar.
\newblock {\em Principles of quantum mechanics}.
\newblock Springer Science \& Business Media, 2012.

\bibitem{garg2000tunnel}
Anupam Garg.
\newblock Tunnel splittings for one-dimensional potential wells revisited.
\newblock {\em American Journal of Physics}, 68(5):430--437, 2000.
\newblock \href {http://dx.doi.org/10.1119/1.19458}
  {\path{doi:10.1119/1.19458}}.

\bibitem{landau2013quantum}
Lev~Davidovich Landau and Evgenii~Mikhailovich Lifshitz.
\newblock {\em Quantum mechanics: non-relativistic theory}, volume~3.
\newblock Elsevier, 2013.

\bibitem{bouchiat1998quantum}
Vincent Bouchiat, D~Vion, Ph~Joyez, D~Esteve, and MH~Devoret.
\newblock Quantum coherence with a single cooper pair.
\newblock {\em Physica Scripta}, 1998(T76):165, 1998.
\newblock \href {http://dx.doi.org/10.1238/physica.topical.076a00165}
  {\path{doi:10.1238/physica.topical.076a00165}}.

\bibitem{hutter2006tunable}
Carsten Hutter, Alexander Shnirman, Yuriy Makhlin, and Gerd Sch{\"o}n.
\newblock Tunable coupling of qubits: Nonadiabatic corrections.
\newblock {\em EPL (Europhysics Letters)}, 74(6):1088, 2006.
\newblock \href {http://dx.doi.org/10.1209/epl/i2006-10054-4}
  {\path{doi:10.1209/epl/i2006-10054-4}}.

\bibitem{melanson2019tunable}
Denis Melanson, Antonio~J Martinez, Salil Bedkihal, and Adrian Lupascu.
\newblock Tunable three-body coupler for superconducting flux qubits.
\newblock {\em arXiv preprint arXiv:1909.02091}, 2019.
\newblock URL: \url{https://arxiv.org/pdf/1909.02091.pdf}.

\bibitem{matlab2}
Matlab online documentation.
\newblock URL: \url{https://uk.mathworks.com/help/matlab/ref/eigs.html}.

\bibitem{dempster2014understanding}
Joshua~M Dempster, Bo~Fu, David~G Ferguson, DI~Schuster, and Jens Koch.
\newblock Understanding degenerate ground states of a protected quantum circuit
  in the presence of disorder.
\newblock {\em Physical Review B}, 90(9):094518, 2014.
\newblock \href {http://dx.doi.org/10.1103/PhysRevB.90.094518}
  {\path{doi:10.1103/PhysRevB.90.094518}}.

\bibitem{zhu2013circuit}
Guanyu Zhu, David~G Ferguson, Vladimir~E Manucharyan, and Jens Koch.
\newblock Circuit qed with fluxonium qubits: Theory of the dispersive regime.
\newblock {\em Physical Review B}, 87(2):024510, 2013.
\newblock \href {http://dx.doi.org/10.1103/PhysRevB.87.024510}
  {\path{doi:10.1103/PhysRevB.87.024510}}.

\bibitem{kerman2018unpublished}
Andrew~Jamie Kerman.
\newblock {\em Unpublished}, 2018.

\bibitem{harris2009compound}
R~Harris, T~Lanting, AJ~Berkley, J~Johansson, MW~Johnson, P~Bunyk,
  E~Ladizinsky, N~Ladizinsky, T~Oh, and Siyuan Han.
\newblock Compound josephson-junction coupler for flux qubits with minimal
  crosstalk.
\newblock {\em Physical Review B}, 80(5):052506, 2009.
\newblock \href {http://dx.doi.org/10.1103/PhysRevB.80.052506}
  {\path{doi:10.1103/PhysRevB.80.052506}}.

\bibitem{klassen2019two}
Joel Klassen and Barbara~M Terhal.
\newblock Two-local qubit hamiltonians: when are they stoquastic?
\newblock {\em Quantum}, 3:139, 2019.
\newblock URL: \url{https://quantum-journal.org/papers/q-2019-05-06-139/pdf/?}

\bibitem{nishimori2017exponential}
Hidetoshi Nishimori and Kabuki Takada.
\newblock Exponential enhancement of the efficiency of quantum annealing by
  non-stoquastic hamiltonians.
\newblock {\em Frontiers in ICT}, 4:2, 2017.
\newblock \href {http://dx.doi.org/10.3389/fict.2017.00002}
  {\path{doi:10.3389/fict.2017.00002}}.

\bibitem{albash2019role}
Tameem Albash.
\newblock Role of nonstoquastic catalysts in quantum adiabatic optimization.
\newblock {\em Physical Review A}, 99(4):042334, 2019.
\newblock \href {http://dx.doi.org/10.1103/PhysRevA.99.042334}
  {\path{doi:10.1103/PhysRevA.99.042334}}.

\bibitem{garashchuk2014calculation}
Sophya Garashchuk, Bing Gu, and James Mazzuca.
\newblock Calculation of the quantum-mechanical tunneling in bound potentials.
\newblock {\em Journal of Theoretical Chemistry}, 2014, 2014.
\newblock \href {http://dx.doi.org/10.1155/2014/240491}
  {\path{doi:10.1155/2014/240491}}.

\bibitem{vranivcar2000accuracy}
Marko Vrani{\v{c}}ar and Marko Robnik.
\newblock Accuracy of the wkb approximation: the case of general quartic
  potential.
\newblock {\em Progress of Theoretical Physics Supplement}, 139:214--233, 2000.
\newblock \href {http://dx.doi.org/10.1143/PTPS.139.214}
  {\path{doi:10.1143/PTPS.139.214}}.

\bibitem{rastelli2012semiclassical}
G.~Rastelli.
\newblock Semiclassical formula for quantum tunneling in asymmetric double-well
  potentials.
\newblock {\em Phys. Rev. A}, 86:012106, Jul 2012.
\newblock \href {http://dx.doi.org/10.1103/PhysRevA.86.012106}
  {\path{doi:10.1103/PhysRevA.86.012106}}.

\end{thebibliography}
\appendix
\section{Appendices}
\subsection{Capacitance and inverse inductance matrices}
\label{appendix1}
In this appendix we give the definition of the capacitance and inverse inductance matrices used to specify the linear part of the circuit Hamiltonian $\hat{H}_{LC}$.
\par For a circuit with \textit{N} nodes (ground node excluded), these are two symmetric $N\times N$ matrices. In the capacitance matrix, each diagonal element $(\mathbf{C})_{ii}$ represents the sum of the capacitances connected to the \textit{i}-th node, while, for every pair of nodes $i\neq j$, the off-diagonal element $(\mathbf{C})_{ij}$ equals minus the total capacitance between \textit{i} and \textit{j}. For the circuit in figure \ref{fig:3JJ}, for instance, the capacitance matrix is
\begin{equation}
\resizebox{0.45\textwidth}{!}{$
	\mathbf{C}=
	\begin{pmatrix}
	C_{JR}&0&-C_{JR}\\
	0&C_{JL}+C_{JT}+C_{sh}&-C_{JT}-C_{sh}\\
	-C_{JR}&-C_{JT}-C_{sh}&C_{JR}+C_{JT}+C_{sh}
	\end{pmatrix}$},
\end{equation}
whose inverse is
\begin{equation}
	\mathbf{C}^{-1}=
	\begin{pmatrix}
	\frac{C_{JL}+C_{JR}+C_{\parallel}}{C_{JL}C_{JR}C_{\parallel}}&\frac{1}{C_{JL}}&\frac{C_{JL}+C_{\parallel}}{C_{JL}C_{\parallel}}\\
	\frac{1}{C_{JL}}&\frac{1}{C_{JL}}&\frac{1}{C_{JL}}\\
	\frac{C_{JL}+C_{\parallel}}{C_{JL}C_{\parallel}}&\frac{1}{C_{JL}}&\frac{C_{JL}+C_{\parallel}}{C_{JL}C_{\parallel}}
	\end{pmatrix},
\end{equation}
where $C_{\parallel}=C_{JT}+C_{sh}$. Notice that $1/(\mathbf{C}^{-1})_{ii}$ corresponds to the effective capacitance between node \textit{i} and ground.
\par In analogy with $\mathbf{C}$, the inverse inductance matrix $\mathbf{L}^{-1}$ has, along the diagonal, the sums of the inverse inductances connected to each node and, in the off-diagonal elements, the total inverse inductance between pairs of nodes. The inverse inductance matrix for the circuit in figure \ref{fig:3JJ} is, for instance,
\begin{equation}
	\mathbf{L}^{-1}=
	\begin{pmatrix}
	\frac{1}{L}&0&0\\
	0&0&0\\
	0&0&0
	\end{pmatrix}.
\end{equation}
\subsection{Capacitance and inverse inductance matrices: interacting circuits case}
\label{appendix5}
In this appendix we show how to modify the capacitance and inverse inductance matrices of two circuits in order to take into account their interactions. The following definitions can easily be extended to the case of more than two interacting circuits.
\par Let $\mathbf{C}_1$ and $\mathbf{C}_2$ be the two original capacitance matrices of the two circuits (as defined in appendix \ref{appendix1}), and let their sizes be $N\times N$ and $M\times M$, respectively. Let $\mathbf{C}_{12}$ be the $N\times M$ matrix whose elements are the capacitances between pairs of nodes belonging to different circuits. Consider then the following $(N+M)\times(N+M)$ matrix:
\begin{equation}
\mathbf{C}=\begin{pmatrix}
\mathbf{C}'_1&-\mathbf{C}_{12}\\
-\mathbf{C}_{12}^T& \mathbf{C}'_2
\end{pmatrix},
\end{equation}
where the primed matrices include the additional capacitance attached to each node, i.e.:
\begin{equation}
\begin{gathered}
(\mathbf{C}'_1)_{kk}=(\mathbf{C}_1)_{kk}+\sum_{k'=1}^{M}(\mathbf{C}_{12})_{kk'},\forall k=1,\dots,N\\
(\mathbf{C}'_2)_{kk}=(\mathbf{C}_2)_{kk}+\sum_{k'=1}^N(\mathbf{C}_{12})_{k'k},\forall k=1,\dots,M.
\end{gathered}
\end{equation}
Notice that $\mathbf{C}$ is nothing but the capacitance matrix defined for the extended circuit including all the nodes of the two interacting circuits. By inverting it, we get:
\begin{equation}
\mathbf{C}^{-1}=\begin{pmatrix}
\widetilde{\mathbf{C}}^{-1}_1&\mathbf{C}^{-1}_{m}\\
(\mathbf{C}^{-1}_{m})^T& \widetilde{\mathbf{C}}^{-1}_2
\end{pmatrix},
\end{equation}
where $\widetilde{\mathbf{C}}^{-1}_1$ and $\widetilde{\mathbf{C}}^{-1}_2$ are the new inverse capacitance matrices of the two circuits (cf. Eq. \eqref{eq:crescale}) which include the effect of the external capacitive loading, and $\mathbf{C}^{-1}_{m}$ is the inverse mutual capacitance matrix, describing the interaction between the two circuits, which appears in equation \eqref{eq:cinteract}.
\par For the inductive interactions, these involve pairs of inductive branches belonging to different circuits, coupled by their mutual inductance. Let $N'$ and $M'$ be the number of branches in the two circuits and consider the following $(N'+M')\times(N'+M')$ matrix:
\begin{equation}
\mathbf{L}_b=\begin{pmatrix}
\mathbf{L}_{b1}&-\mathbf{M}\\
-\mathbf{M}^T& \mathbf{L}_{b2}
\end{pmatrix},
\end{equation}
where $\mathbf{L}_{bi}$ is the inductance matrix of circuit \textit{i} in the branch representation, having along the diagonal the self-inductance of each branch $(\mathbf{L}_{bi})_{kk}=L_{b_{i_k}}$ and zeros everywhere else, and $\mathbf{M}$ is the $N'\times M'$ matrix whose elements are the mutual inductances between pairs of inductive branches. Inverting $\mathbf{L}_b$, we obtain
\begin{equation}
\mathbf{L}_b^{-1}=\begin{pmatrix}
\mathbf{L}_{b1}^{-1}&\mathbf{M}^{-1}\\
(\mathbf{M}^{-1})^T& \mathbf{L}_{b2}^{-1}
\end{pmatrix},
\end{equation}
where $\mathbf{M}^{-1}$ is the matrix appearing in equation \eqref{eq:minteract}. $\mathbf{L}_{b1}^{-1}$ and $\mathbf{L}_{b2}^{-1}$ can be used to rescale the inverse inductance matrices of the two circuits (see Eq. \eqref{eq:lrescale}). This is accomplished by replacing each branch inductance $L_{b_{i_k}}$ appearing in the expression of $\mathbf{L}^{-1}_i$ with $1/(\mathbf{L}_{bi}^{-1})_{kk}$.
\subsection{Spectrum convergence}
\label{appendix3}
\begin{figure}[b!]
	\centering
	\includegraphics[width=.45\textwidth]{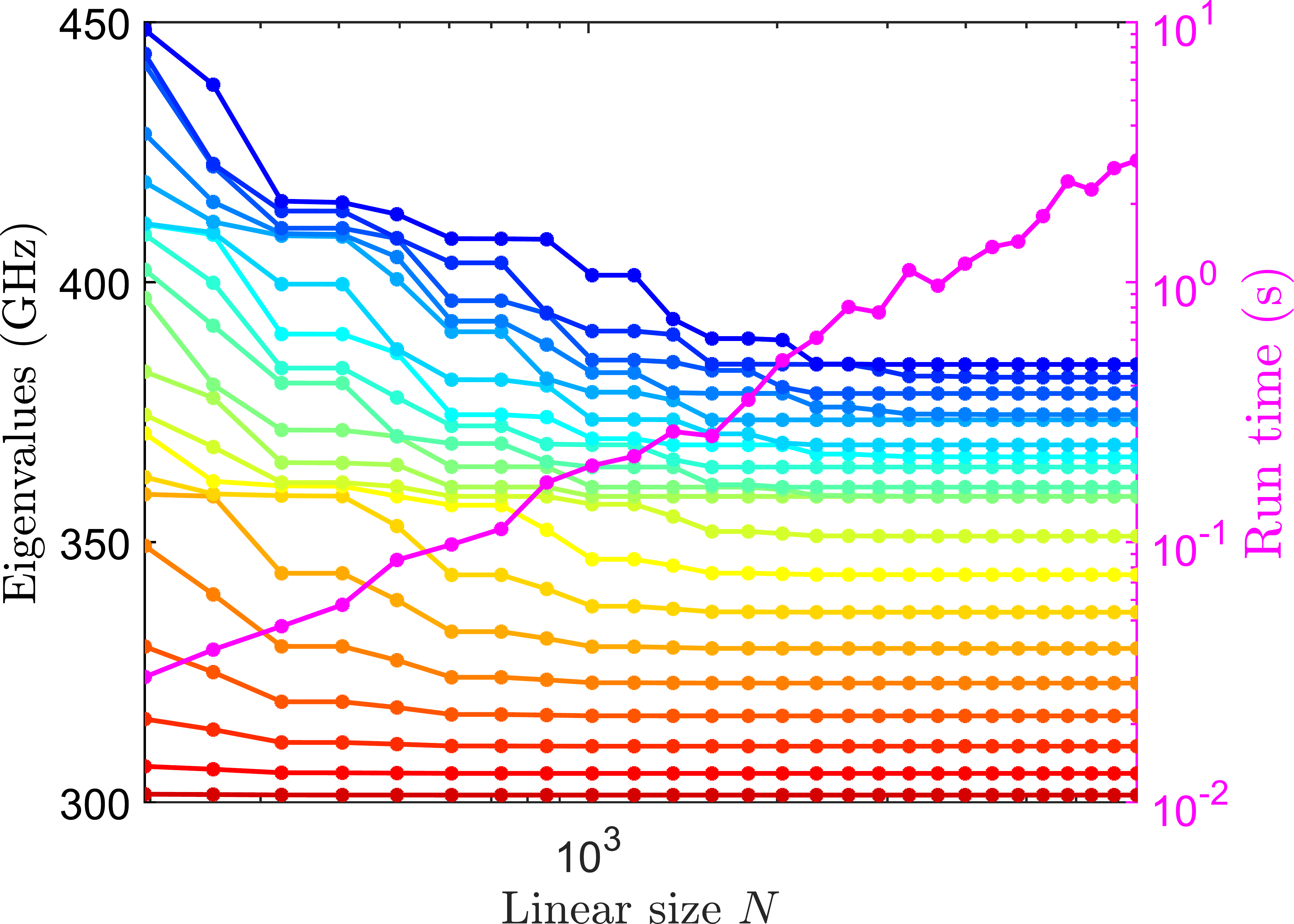}	
	\caption{Lowest 20 energy eigenvalues of a C-shunt flux qubit, as a function of the linear size  $N$ of its truncated circuit Hamiltonian. The pink line shows the time (indicated on the right vertical scale) required to numerically compute each set of 20 eigenvalues.}
	\label{fig:specvstr}
\end{figure}
In this section we consider the convergence of the numerical spectrum of a qubit circuit as a function of the number of states included in the basis used to describe each of its modes. We refer to this number as the (mode) truncation.
\par The circuit examined here is that of the capacitively-shunted flux qubit shown in Fig.\ref{fig:3JJ}. By inspecting its circuit Hamiltonian, we find that the mode associated with node 1 (\textit{O1}) is conveniently expressed in a basis of harmonic oscillator states below a certain occupation number $N_{O}^{max}$, while those associated with nodes 2 ($C1$) and 3 ($C2$) are better expressed in the charge number basis, keeping only integer charges lower in absolute value than $N_{C1}^{max}$ ($N_{C2}^{max}$)\cite{zhu2013circuit,bouchiat1998quantum,kerman2018unpublished}.
\par Figure \ref{fig:specvstr} shows the lowest 20 eigenvalues of the approximate circuit Hamiltonian $\mathbf{H}_{e.m.}^{(N)}$, as a function of its linear size $N=(N_{O1}^{max}+1)\cdot(2N_{C1}^{max}+1)\cdot(2N_{C2}^{max}+1)$, as well as the time required to evaluate them (shown by the pink line and indicated on the right vertical axis). The qubit is taken to be biased at the optimal point $f_z=\Phi_{23}^{ext}/\Phi_0=0.5$ and its other physical parameters are given in section \ref{singlequbits} of the main text.  In the graph the values of the truncations $N_{O1}^{max},N_{C1}^{max}$ and $N_{C2}^{max}$ are increased sequentially going from left to right, starting from the values $(N_{O1}^{max},N_{C1}^{max},N_{C2}^{max})=(2,3,3)$. As we can see, all of the 20 lowest eigenvalues have converged for the set of truncations $(9,10,10)$, corresponding to a Hamiltonian of linear size $N=4410$. As it turns out, the convergence is mainly determined by the Josephson modes, and the set $(3,10,10)$ ($N=1323$) is already sufficient to obtain the same eigenvalues. Also notice that the lowest three eigenvalues already converge for the set of truncations $(3,5,5)$ and $N=363$.
\par The eigenvalue evaluation times refer to the use of \textit{MATLAB}${}^\textrm{\textcopyright}$ \textit{eigs} algorithm\cite{matlab2}, run on a quad-core laptop CPU. As the pink line in the graph shows, the run time scales as a power law of the linear matrix size (notice the log-log scale), namely $t_{run}\simeq(1.1\cdot10^{-5}\textrm{s})\cdot N^{1.4}$, as results from a non-linear fit. 
\subsection{Tunnelling rates in the rf-SQUID qubit with the instanton method}
\label{appendix4}
\begin{figure}[b!]
	\centering
	\includegraphics[width=.45\textwidth]{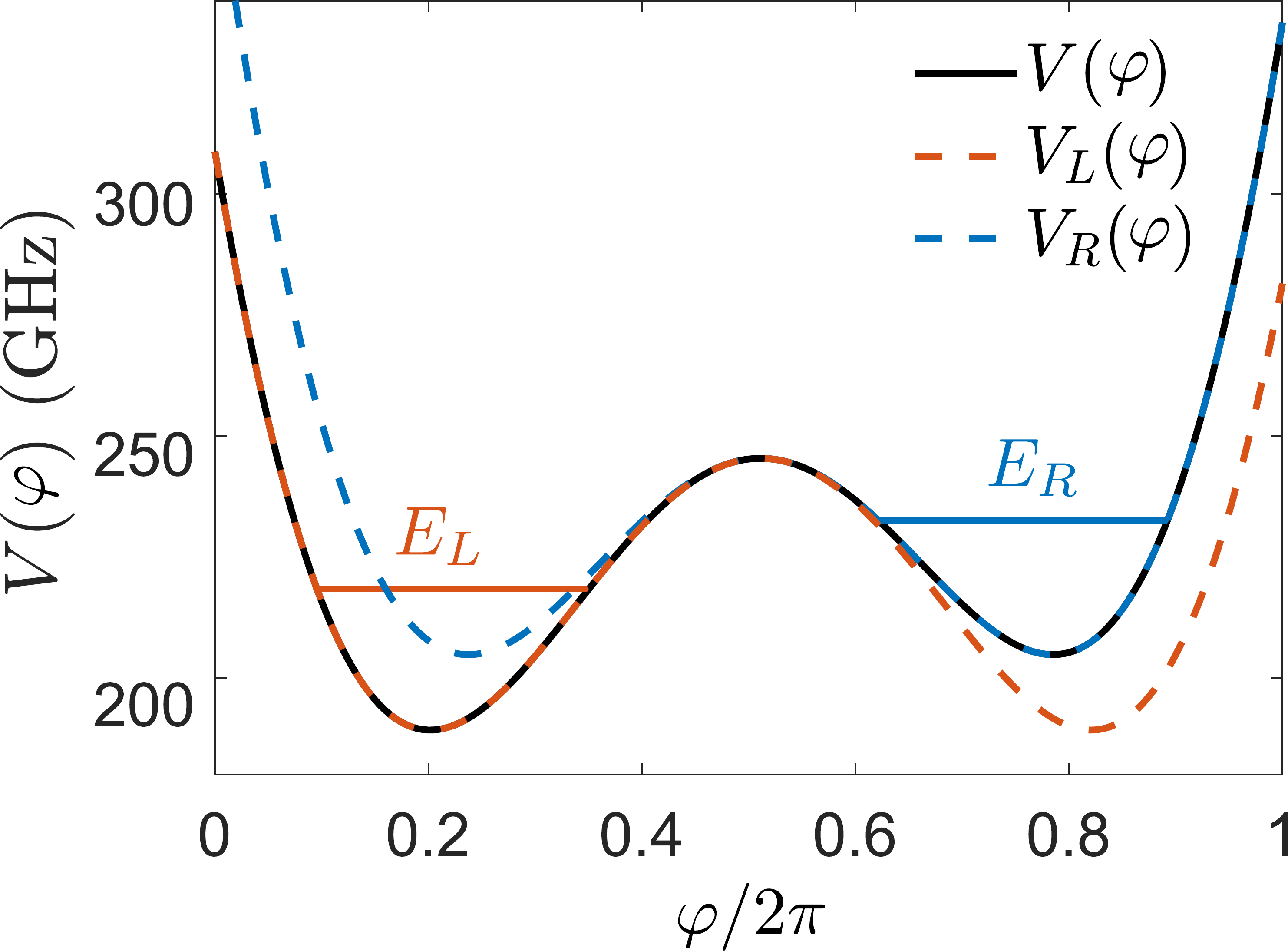}	
	\caption{rf-SQUID semi-classical potential (black line) for $\varphi_{ext}/2\pi=0.49$, and its two symmetrised versions (dashed lines). Also shown are the energies of the lowest bound states in the two wells.}
	\label{fig:tunnel}
\end{figure}
The semi-classical description of tunnelling through a potential barrier is a very well-known subject in quantum mechanics and is routinely used in many applications of chemistry and quantum physics\cite{garg2000tunnel,garashchuk2014calculation,vranivcar2000accuracy}. In order to describe the tunnelling between the two opposite persistent current states of the rf-SQUID qubit, we are going to use the formalism developed in \cite{rastelli2012semiclassical}, which applies to a generic, potentially asymmetric double-well potential. Let us first write the semi-classical potential of the circuit\cite{wendin2017quantum}:
\begin{equation}
V(\varphi)=U_L\cdot\left[\frac{(\varphi-\varphi_{ext})^2}{2}+\beta_L(1-\cos\varphi)\right],
\end{equation}
where $\varphi=2\pi\Phi/\Phi_0$ is the dimensionless total flux, $\varphi_{ext}$ is the externally applied flux, $U_L=(\Phi_0/2\pi)^2/L$ is the characteristic inductive energy and $\beta_L=E_JL(2\pi/\Phi_0)^2$ is called the \textit{screening parameter}. When $\beta_L\gtrsim1$ and $\varphi_{ext}/2\pi\simeq0.5$, this potential has three stationary points, given by the solutions of the transcendental equation
\begin{equation}
\beta_L\sin\varphi=\varphi_{ext}-\varphi.
\end{equation}
Two of the solutions, say $\varphi_L$ and $\varphi_R$, correspond to the minima of the left and right potential wells, respectively, while the third, $\varphi_M$, is the maximum of the barrier between them ($\varphi_L<\varphi_M<\varphi_R$). For instance, when $U_L=65$GHz, $\beta_L=1.9$ and $\varphi_{ext}/2\pi=0.49$, we obtain the potential profile shown in figure \ref{fig:tunnel} (solid black line, sitting below the dashed lines).
\par According to the semi-classical theory, the low energy behaviour of the rf-SQUID system can be described in terms of the tunnelling between the lowest bound states in its two potential wells, $\Psi_L(\varphi)$ and $\Psi_R(\varphi)$\cite{garg2000tunnel}. These represent the local solutions to the stationary Schr\"{o}dinger equation, in the limit where the two wells are completely isolated from each other (eg. $\varphi_L\ll\varphi_R$). One way to approximately identify these solutions is by considering the second-order series expansion of the potential around its minima:
\begin{equation}
V(\varphi)\simeq V(\varphi_i)+\frac{V''(\varphi_i)}{2}(\varphi-\varphi_i)^2, \; i=L,R.
\end{equation}
Then $\Psi_L(\varphi)$ and $\Psi_R(\varphi)$ approximately correspond to the vacuum states of two displaced harmonic oscillators, such that
\begin{equation}
\left[-\frac{(2e)^2}{2C}\frac{\partial^2}{\partial\varphi^2}+\frac{V''(\varphi_i)}{2}(\varphi-\varphi_i)^2\right]\Psi_i(\varphi)=E_i\Psi_i(\varphi),
\end{equation}
with $C$ the total capacitance across the Josephson junction, and
\begin{equation}
E_i=V(\varphi_i)+\frac{\hbar\omega_i}{2}.
\end{equation}
The oscillator frequency here is
\begin{equation}
\omega_i=\frac{2\pi}{\Phi_0}\sqrt{\frac{V''(\varphi_i)}{C}}=\sqrt{\frac{1+\beta_L\cos\varphi_i}{LC}}.
\end{equation}
Notice that these states have a phase expectation value of $\langle\hat{\varphi}\rangle_i=\varphi_i$ and an average persistent current of
\begin{equation}
\begin{gathered}
\langle\hat{I}\rangle_i:=-\langle\frac{\partial\hat{H}_{e.m.}}{\partial\Phi_{ext}}\rangle_i=\frac{2\pi U_L}{\Phi_0}\langle\hat{\varphi}-\varphi_{ext}\rangle_i=\\=\frac{\Phi_0}{2\pi}\frac{\varphi_i-\varphi_{ext}}{L}.
\end{gathered}
\end{equation}
Therefore, since $\varphi_L<\varphi_{ext}<\varphi_R$, the bound states also correspond to persistent current states of opposite sign, as expected.
\par Quantum tunnelling across the potential barrier couples the two bound states, leading to the repulsion between their energy levels. The resulting eigenstates of the system are determined by the following two-level Hamiltonian, expressed in the persistent current basis $\{|\Psi_R\rangle,|\Psi_L\rangle\}$:
\begin{equation}
\begin{gathered}
\mathbf{H}_q=\begin{pmatrix}
E_R &-\Delta\\
-\Delta &E_L
\end{pmatrix}=\frac{E_R+E_L}{2}\pmb{\sigma}_I+\\-\Delta\pmb{\sigma}_x+\frac{E_R-E_L}{2}\pmb{\sigma}_z,
\end{gathered}
\end{equation}
where $\Delta$ is the tunnelling energy. This represents the effective qubit Hamiltonian of the circuit, and is again in the standard form of Eq. \eqref{eq:qubithamgen}.
\par Finally, following reference \cite{rastelli2012semiclassical}, we can write the tunnelling energy explicitly as:
\begin{equation}
\Delta=A\cdot\sqrt{\Delta_L\Delta_R},
\end{equation}
where
\begin{equation}
A=\frac{1}{2}\left[\left(\frac{V_0-E_L}{V_0-E_R}\right)^{1/4}+\left(\frac{V_0-E_R}{V_0-E_L}\right)^{1/4}\right],
\end{equation}
with $V_0=V(\varphi_M)$, and where $\Delta_{L,R}$ is the tunnelling energy relative to the symmetric double-wells $V_L(\varphi)$ and $V_R(\varphi)$, obtained by reflecting $V(\varphi)$ about the local maximum $\varphi_M$ (cf. dashed lines in figure \ref{fig:tunnel}):
\begin{equation}
\begin{gathered}
V_L(\varphi)=V(\min(\varphi,2\varphi_M-\varphi)),\\
V_R(\varphi)=V(\max(\varphi,2\varphi_M-\varphi)).
\end{gathered}
\end{equation}
The instanton result for the symmetric double-well tunnelling energies reads:
\begin{equation}
\Delta_i=\hbar\omega_i e^{-\frac{S_i}{\hbar}},\;i=L,R
\end{equation}
with $S_i$ the \textit{tunnelling action}, given by:
\begin{equation}
S_i=\frac{\Phi_0}{2\pi}\int_{\varphi_{i,1}}^{\varphi_{i,2}}\sqrt{2C(V_i(\varphi')-E_i)}d\varphi',
\end{equation}
where $\varphi_{i,1}=2\varphi_M-\varphi_{i,2}$ are the two points at which the potential barrier intersects the energy level: $V_i(\varphi_{i,1})=V_i(\varphi_{i,2})=E_i$. This semi-classical formula holds when $S_i\gg\hbar$ and therefore in the limit of small tunnelling energies\cite{landau2013quantum}.
\end{document}